\documentclass{ieeeaccess}
\usepackage{graphicx}
\usepackage[utf8]{inputenc}
\usepackage{amsmath,amssymb,amsfonts}
\usepackage{bm}
\usepackage[caption=false]{subfig}
\usepackage{float}
\usepackage{stfloats}
\usepackage{enumitem}
\usepackage[super]{nth}
\usepackage{booktabs}
\usepackage{tabularx}
\usepackage{colortbl}
\usepackage{multirow}
\usepackage{makecell}
\usepackage[]{placeins}  
\usepackage{algpseudocode}
\usepackage{algorithm}
\usepackage{hyperref}

\renewcommand*\thetable{\Roman{table}}

\algrenewcommand{\algorithmiccomment}[2][.53\linewidth]{%
  \leavevmode\hfill\makebox[#1][l]{$\triangleright$~#2}}

\newcommand{\echo}{Echo}
\newcommand{\EPP}{EPP}
\newcommand{\ESP}{ESP}
\newcommand{\GP}{GP}
\newcommand{\LP}{LP}

\newcommand{\Classic}{Classic}
\newcommand{\Neural}{Neural}
\newcommand{\Poly}{Poly}
\newcommand{\NF}{Neural-fast}
\newcommand{\NS}{Neural-slow}
\newcommand{\PF}{Poly-fast}
\newcommand{\PS}{Poly-slow}
\newcommand{\real}[1]{\operatorname{Re}\{#1\}}
\newcommand{\imag}[1]{\operatorname{Im}\{#1\}}

\begin{document}

\title{Blind interactive learning of modulation schemes: Multi-agent cooperation without co-design}
\author{\uppercase{Anant Sahai}\authorrefmark{1}, \uppercase{Joshua Sanz}, \uppercase{Vignesh Subramanian}, \uppercase{Caryn Tran},  \uppercase{and Kailas Vodrahalli}}
\address[1]{Prof.~Anant Sahai was the PI for this project and all the authors, including him, are listed in alphabetical order. All authors had an equal contribution to this paper.}

\tfootnote{This work was supported in part by the UC Berkeley ML4Wireless center member companies, the DARPA Spectrum Collaboration Challenge, and NSF grants AST-144078 and ECCS-1343398.
 }

\history{This article has been accepted for publication in a future issue of IEEE Access. Citation information: DOI10.1109/ACCESS.2020.2984218, IEEE Access}
\doi{DOI10.1109/ACCESS.2020.2984218}


\markboth
{A. Sahai \headeretal: Blind interactive learning of modulation schemes}
{A. Sahai \headeretal: Blind interactive learning of modulation schemes}

\corresp{Corresponding author: Joshua K. Sanz (e-mail: jsanz@berkeley.edu).}

\begin{abstract}

We examine the problem of learning to cooperate in the context of wireless communication. In our setting, two agents must learn modulation schemes that enable them to communicate across a power-constrained additive white Gaussian noise channel. We investigate whether learning is possible under different levels of information sharing between distributed agents which are not necessarily co-designed.  
We employ the ``\echo{}'' protocol, a ``blind'' interactive learning protocol where an agent hears, understands, and repeats (echoes) back the message received from another agent, simultaneously training itself to communicate. 
To capture the idea of cooperation between ``not necessarily co-designed'' agents we use two different populations of function approximators --- neural networks and polynomials. We also include interactions between learning agents and non-learning agents with fixed modulation protocols such as QPSK and 16QAM.  We verify the universality of the \echo{} learning approach, showing it succeeds independent of the inner workings of the agents. In addition to matching the communication expectations of others, we show that two learning agents can collaboratively invent a successful communication approach from independent random initializations. We complement our simulations with an implementation of  the \echo{} protocol in software-defined radios. 
To explore the continuum of co-design, we study how learning is impacted by different levels of information sharing between agents, including sharing training symbols, losses, and full gradients. We find that co-design (increased information sharing) accelerates learning. Learning higher order modulation schemes is a more difficult task, and the beneficial effect of co-design becomes more pronounced as the task becomes harder.

\end{abstract}



\titlepgskip=-15pt

\maketitle

\begingroup
\newlength{\xfigwd}
\setlength{\xfigwd}{\textwidth}
\endgroup



\section{Introduction}

Machine learning is a technology and associated design paradigm that has recently seen a resurgence largely due to advances in computational capabilities.
Consequently, there has been increasingly active research in the areas of supervised and reinforcement learning, both in the underlying technology as well as in the development of design paradigms appropriate to using these technologies in diverse application contexts.
This paper\footnote{An earlier, limited, preliminary version of this work was presented at Allerton 2019 as an invited paper \cite{AllertonVersion}.
This paper is the full version (about four times longer) and all the key plots here are new.} is about seeing whether machine learning paradigms can be used to aid us with achieving interoperability in a wireless communication setting.
The established paradigm for interoperation is that of standards --- communication protocols are not only hand-crafted by individual humans, these hand-crafted protocols are standardized and certified by authorized committees of people.
Can we instead use machine learning techniques to learn how to communicate with minimal assumptions on shared information, and if so, how well can we learn? 

Communication is a fundamentally cooperative activity between at least two agents.
Consequently, communication itself can be viewed as both a special case of cooperation as well as a building block that can be leveraged to permit more effective cooperation.
The fundamental limits to learning how to cooperate with a stranger have been studied in an abstract theoretical setting in   \cite{tw:juba2008universal, tw:juba2008universal2, tw:KomargodskiKS15, tw:CanonneGMS14, tw:GoldreichJS12}.
By asking how two intelligent agents might understand and help each other without a common language, a basic theory of goal-oriented communication was developed in these papers.
The principal claim is that for two agents to robustly succeed in the task of learning to collaborate, the goal must be explicit, verifiable, and forgiving.
However the approach in these works took a fundamentally {\em semantic} perspective on cooperation.
As Shannon pointed out in \cite{shannon1948mathematical}, the arguably simpler cooperative problem of communicating messages can be understood in a way that is decoupled from the issue of what the messages mean.
To see whether existing machine learning paradigms can be adapted to achieve cooperation with strangers, we consider the concrete problem where two agents learn to communicate in the presence of a noisy channel.
Each agent consists of a modulator and demodulator and must learn compatible modulation schemes to communicate with each other.

This problem of learned communication has been tackled using learning techniques under different assumptions on the information that the agents are allowed to share and how tightly coordinated their interaction is.
Early work in this area \cite{cae:8054694}, \cite{cae:DBLP:journals/corr/OSheaKC16}, where gradients are shared among agents, demonstrated the success of training a channel auto-encoder using supervised learning when a stochastic model of the channel is known.
Subsequent works relax the assumption on the known channel model by learning a stochastic channel model by using GANs as in \cite{wcm:lcm:DBLP:journals/corr/abs-1903-02551}, \cite{wcm:lcm:DBLP:journals/corr/abs-1807-00447}, \cite{wcm:lcm:DBLP:journals/corr/abs-1805-06350} or by approximating gradients across the channel and using that for training.
However, these approaches cannot be said to represent communication with strangers, and instead represent a way of having co-designed systems learn to communicate.
If instead of sharing gradients we can only share scalar loss values then with access to a shared preamble, then reinforcement learning-style techniques must be used to train the system as demonstrated in \cite{wcm:de2018cooperative} and \cite{wcm:DBLP:journals/corr/abs-1804-02276} since they can work without access to an explicit channel model.

Moving closer to minimal co-design, if we further restrict ourselves to the case where the two agents only have access to a shared preamble, the ``\echo{}'' protocol, where an agent hears, understands, and repeats (echoes) back the message received from the other agent, as specified in \cite{wcm:de2018cooperative} has been shown to work.
By comparing the original message to the received echo, a learning agent can get feedback about how well the two agents understand each other\footnote{Round-trip stability is not by itself a sufficient condition to guarantee mutual comprehension.
After all, one agent might only be doing simple mimicry \textemdash~ repeating back the raw analog signal value received with no attempt to actually demodulate.
However, in \cite{wcm:de2018cooperative}, the key insight was that intelligent agents, though strangers, are believed to be cooperative and so wish to actually understand and communicate with each other.
They do not need to actually coordinate with another designer to realize that sheer mimicry would not necessarily advance their goal of cooperation.
Consequently, the \echo{} protocol can rely on good intentions to eliminate the possibility of agents just mirroring what has been heard instead of trying to understand what was sent and repeating it back.}.
The work in \cite{wcm:de2018cooperative} considered a neural network based modulator that was trained using reinforcement learning via policy gradients \cite{williams1992REINFORCE}, but the demodulator was nearest neighbors based and required no training --- it used small-sample-based supervised learning.
Our work in the present paper builds on this and studies the truly ``blind'' case where agents do not have access to a shared preamble.

We dub this ``blind interactive learning'' to acknowledge the motivational connection with the well known and traditional problems of blind equalization and blind system identification --- where we have to deal with a channel and implicitly learn a model for it without knowing the actual input to the channel.
(See, for example, the book \cite{li2001blind} for a survey of well understood approaches.) 
Such blind approaches are fundamentally motivated by the desire for universality and the resulting robust modularity.
Traditional blind approaches in signal processing are intellectually akin to what are called unsupervised learning approaches in machine learning.
Reinforcement learning has always occupied a middle ground between supervised and unsupervised learning, because in a sense, it is self-supervised and carried out via interaction with an environment.
For us, ``blind interactive learning'' involves agents interacting blindly via an environment --- where the individual agents might be self-supervised but there is no explicit joint self-supervision.
They are blind in the traditional sense of not really knowing what exactly went into the channel whose output they are observing, and in particular, not sharing a known training sequence.

It is important to differentiate here between our work on blind \textit{interactive} modulation learning, and automatic modulation recognition (AMR) as in \cite{amr:azzouz2013automatic}. 
AMR seeks to take an unknown signal present in the environment and classify as one of a set of known modulation schemes for subsequent demodulation. 
No interaction with the signal source occurs --- indeed, for surveillance applications interaction might destroy all value! 
By contrast, our work requires interaction to learn how to demodulate any possible signal, even ones never before seen or imagined, and further to learn to modulate in a similarly arbitrary way understandable by an agent on the other side of a communication channel.
This interaction takes the form of round-trip training so that both the modulation and demodulation functions can be updated.
Although AMR-based techniques could play a role in the demodulation half of this process by speeding up learning for known signals, they are insufficient on their own to complete the circle. 
Our work introduces a new problem, blindly learning an entire modulation and demodulation scheme, rather than introducing a new technique for classifying existing modulation schemes.

We also introduce the concept of ``alienness'' among agents.
After all, if our goal is to understand learning of communication between strangers, we need to be able to test with strangers. 
A natural question is what it means for two agents to be alien.
For now, we consider agents to be alien if they differ in their learning architecture or their hyperparameters.
In this work we examine modulators and demodulators represented using different types of function approximators such as neural networks and polynomials.

\textbf{Our main contribution is to show how to make the \echo{} protocol blind and investigate the extent to which it is universal}, i.e.
does it allow two agents to learn to communicate irrespective of their inner workings. To do this we  \textbf{study what level of information sharing is necessary for successful learning}.
Although we do not have a formal proof of universality yet, we provide some empirical evidence by pairing up agents with different levels of ``alienness'' based on the hyperparameters, architectures, and techniques used in their modulators and demodulators.
By doing so we wish to separate the effect of the of the agents' implementations, such as those owed to specific function approximators, from the meta protocols (specifically the \echo{} protocol) used to do the interactive learning.
To both connect to the literature and explore the spectrum between complete co-design and cooperative learning among strangers, we look at different levels of information sharing: shared gradients, shared loss information, shared preamble, and finally the case where only the overall protocol is shared.

Machine learning scholarship is notorious for producing results that are not easily reproducible, and failure to identify the source of and explain the reasoning behind performance gains \cite{jacob}.
Keeping this in mind, in order to evaluate the ease, speed, and robustness of the learning task under various levels of alienness and information sharing, we conduct repeated trials for each setting using different random seeds and multiple sets of hyperparameters.
We report the fraction of trials that succeeded as a function of the number of symbols exchanged, as well as aggregate statistics about the bit error rate achieved at different signal to noise ratios (SNR) by the learned modulation schemes. 
We compare our experimental bit error rates to those achieved by optimal modulation schemes for AWGN channels to provide a baseline for comparison.
The code used to generate the results is available in \cite{Repo}.
From our experiments we observe and conclude that the \echo{} protocol does enable two agents to learn a modulation scheme even under minimal shared information, and that as we decrease the amount of shared information the learning task becomes harder, i.e a lower fraction of trials succeed and the agents take longer to learn.

It appears that learning to communicate with ``alien'' agents is not necessarily more difficult than learning to communicate with agents of the same type.
 However, it is significantly easier to learn to communicate if one of the agents already uses a good modulation scheme, for example a hand-designed scheme like QPSK.
Finally, as we increase the modulation order for communication the learning task becomes harder, especially so for settings with little information sharing.

Although a majority of the results reported in this paper were performed purely in simulation, we replicate our main results using USRP radios and observe similar results --- two agents can learn to communicate in a decentralized fashion even using real hardware.



\section{Related Work}
\label{sec:related-work}

Deep learning has shown great success in tasks that historically  relied on multi-stage processing using a series of well-designed, hand-crafted features such as computer vision, natural language processing, and more recently robotics. 
Wireless communication is another area that historically uses hand-crafted schemes for various processing stages such as modulation, equalization, demodulation, and encoding and decoding using error correcting codes. 
Thus, as alluded to in \cite{ecc:2018arXiv180509317K} and \cite{cae:DBLP:journals/corr/OSheaH17}, one might believe that bringing deep learning into wireless communication is a worthwhile endeavor. 
In fact, learning and deep learning have been present in the sub-field of AMR since at least the 1980s. 
AMR has undergone a similar transition from hand-crafted features, such as phase difference and amplitude histograms as detailed in \cite{amr:azzouz2013automatic}, to modern deep learning techniques \cite{amr:dl:DBLP:journals/corr/OSheaC16}, \cite{amr:dl:r22:8645696}, \cite{amr:modstoimprove:r21:8978670}, \cite{amr:modstoimprove:r23:article}. 

Beyond modulation recognition, the pioneering work in \cite{cae:8054694}, \cite{cae:DBLP:journals/corr/OSheaKC16} demonstrated the promise of the channel auto-encoder model by using supervised learning techniques to learn an end-to-end communication scheme, including both transmission and reception.
This approach assumes the knowledge of an analytical (differentiable) model of a channel and the ability to share gradient information between the receiver and transmitter.
This approach was a natural first step given the known connection between auto-encoders and compression (see e.g. \cite{hinton2006reducing}) as well as the well-known duality between source-coding (compression) and channel-coding (communication) \cite{shannon1948mathematical}.
  
  Building on this foundation, other works deal with the case where the channel model is unknown, as is the case when we perform end-to-end training over the air. In \cite{wcm:DBLP:journals/corr/abs-1804-02276},  a stochastic model that allows backpropagation of gradients to approximate the channel is used with a two phase training strategy.  Phase one involves  auto-encoder-style training using a stochastic channel model, whereas phase two involves supervised fine-tuning of the receiver part of auto-encoder based on the labels of messages sent by transmitter and the IQ-samples recorded at the receiver. This approach relies on starting out with a good stochastic channel model. Use of Generative Adversarial Networks to learn such models is explored in \cite{wcm:lcm:DBLP:journals/corr/abs-1903-02551}, \cite{wcm:lcm:DBLP:journals/corr/abs-1807-00447}, \cite{wcm:lcm:DBLP:journals/corr/abs-1805-06350}.  In \cite{wcm:backpropthroughair18}, instead of estimating the channel model, stochastic approximation techniques are used to calculate the approximate gradients across the channel. The idea of approximating gradients at the transmitter has also been used in \cite{wcm:modelfree19} to successfully perform end-to-end training. 
  
  In the absence of a known channel model, reinforcement learning can also be used to train the transmitter as demonstrated in \cite{wcm:de2018cooperative} and  \cite{wcm:DBLP:journals/corr/abs-1804-02276}. In \cite{wcm:de2018cooperative}, the \echo{} protocol --- a learning protocol where an agent hears, understands, and repeats (echoes) back the message received from the other agent --- was used to obtain a scalar loss that was used to train the neural-network based transmitter using policy gradients. Here the receiver used a lightweight nearest-neighbor based scheme that was trained afresh in each communication round. This work assumed that the agents have access to a shared preamble so we dub it \textit{\echo{} with Shared Preamble} (\ESP{}). In \cite{wcm:DBLP:journals/corr/abs-1804-02276}
  both the transmitter and receiver were neural-network based. The receiver was trained using supervised-learning whereas the transmitter was trained using policy gradients by passing scalar loss values obtained at the receiver back to the transmitter. Reinforcement learning techniques have the added advantage of being implementable in software-defined radios to perform end-to-end learning over the air. To do this one must tackle the issue of time synchronization between the transmitted and received symbols as done in \cite{sync:2019arXiv190510468S} and \cite{cae:DLCommOverAir18}.  In \cite{sync:2016arXiv160500716O}, the general problem of synchronization in wireless networks is addressed via the use of attention models.

Other parts of the communication pipeline such as channel equalization and error correcting code encoding and decoding have also been studied using machine learning techniques.  The use of neural networks for equalization is studied in \cite{eq:bce:2018arXiv180301526C} and \cite{eq:lbe}. Construction and decoding of error correcting codes is considered in \cite{ecc:2018arXiv180509317K}, \cite{ecc:learncodes:2018arXiv181112707J}, \cite{ecc:mind:2019arXiv190302268J}, and \cite{ecc:ldc:DBLP:journals/corr/abs-1807-00592}. Joint source channel coding is an area where performance gains are possible through co-design as demonstrated in \cite{joint:jsc:DBLP:journals/corr/abs-1811-07557} for wireless communication, and in the application of wireless image transmission in  
\cite{joint:jsci:DBLP:journals/corr/abs-1809-01733}.
  End-to-end  auto-encoder style training continues to be an area of interest in wireless communication. There has been recent work demonstrating the success of 
 convolutional neural network based architectures and block based schemes in this setting in
 \cite{cae:2018arXiv180803242Z},
 \cite{cae:recent:cnn:8768327}, \cite{cae:recent:cnn:8755977}, and \cite{cae:recent:block:2019arXiv190606563M}. 
 This approach has also been used successfully in OFDM systems   \cite{ofdm:DBLP:journals/corr/abs-1803-05815} to learn the symbols transmitted over the sub-carriers. Deep learning techniques and auto-encoder style training have been used in the fields of fiber-optic \cite{fbo:li2018achievable}, \cite{fbo:karanov2018end} and molecular communication \cite{mol:lee2017machine}, \cite{mol:8715741} to model the channel and to leverage the channel model to learn communication schemes that achieve low  error rates.

 A theoretical analysis of the general learning to cooperate problem is done in the works \cite{tw:juba2008universal, tw:juba2008universal2, tw:KomargodskiKS15, tw:CanonneGMS14, tw:GoldreichJS12}. This body of work investigates the possibility for two intelligent beings to cooperate when a shared context is absent or limited. In particular, this work also does not presume a pre-existing communication protocol. In asking how two intelligent agents might understand each other without a common language, a theory of goal-oriented communication is developed. The principal claim is that for two agents to robustly succeed in the cooperative task, the goal must be explicit, verifiable, and forgiving. Agents should have feedback about whether the goal is achieved or not, and it should be possible for the agents to achieve the goal from any state that is reached after a finite set of actions. The works     \cite{cmrl:lobms:DBLP:journals/corr/abs-1802-09640},
    \cite{cmrl:gt:NIPS2017_7007}, 
    \cite{cmrl:fdccl:DBLP:journals/corr/abs-1802-08757}, 
    \cite{cmrl:NIPS2016_6042} bring about these ideas in a limited setting. 
 
 From a psychological perspective, developmental psychology \cite{children} provides a rich account of how human infants learn to communicate. How do babies come to understand sounds, words, and meaning? It begins in the development of `categorical perception of sound' which creates discrete categories of sound perception, not unlike the task of demodulation. Later on, other tasks emerge such as word segmentation, attributed to statistical learning, where in the child grows increasingly aware of sounds and words that belong together. Soon after, the child engages in babbling as an exploration of language production, investigating rhythm, sound, intonation, and meaning, a task similar to modulation. Important to all the above processes, is social interaction and exchange, most often between child and caretaker, which provides the rich information required for learning to be successful. 
 



\section{Overview}
\label{sec:procedure}
\subsection{Problem Formulation}
\label{subsec:problem-formulation}

We consider the setting where two agents communicate in the presence of a discrete-time additive white Gaussian noise (AWGN) channel. Each agent consists of an encoder (modulator) and  a decoder (demodulator). We treat the modulator as an abstract (black box) object that converts bit symbols to complex numbers, i.e. we treat it as a mapping $M: B \rightarrow \mathbb{C}$ where $B$ refers to the set of bit symbols and $\mathbb{C}$ refers to the set of complex numbers. Similarly we treat the demodulator as an abstract object that converts complex numbers to bit symbols, i.e. a mapping $D: \mathbb C \rightarrow B$. The set of bit symbols $B$, is specified by the modulation order (bits per symbol). For instance, when bits per symbol is 1, $B = \{0,1\}$  and when bits per symbol is 2, $B = \{00,01,10,11\}$. For the case where bits per symbol is 1, the classic\footnote{Here we use classic to refer to a modulation scheme that is fixed and specified identically for all communicating agents by a certain standard. One example of such a scheme is BPSK signaling as described here.} BPSK (binary phase shift keying) modulation scheme is given by:
\begin{align}
    M^{BPSK}(0) &= 1 + 0j,\\
    M^{BPSK}(1) &= -1 + 0j.
\end{align}
The corresponding demodulator performs the demodulation as,
\begin{align}
    D^{BPSK}(c) = \begin{cases} 0, & Re(c) \geq 0\\
                                1, & Re(c) < 0
    \end{cases}.
\end{align}
In addition to agents that use fixed modulation schemes we also consider `learning' agents. These agents use function approximators to learn the mappings performed by  modulator and demodulator, and we denote these as $M(\cdot;\theta)$ and $D(\cdot; \phi)$ where $\theta$ and $\phi$ denote the parameters of the underlying function approximators and are updated during training. The specifics of the learning agents and their update methods can be found in Appendix \ref{app:agent-details}.

 The main focus of our work is in learning modulation schemes, thus, to make it easier to conduct experimental simulations we make the following simplifying assumptions:
\begin{enumerate}

    \item There are at most two agents, and they engage in perfect turn-taking.
    \item The two agents are separated by a unit gain AWGN channel. There is no carrier frequency offset, timing offset or phase offset. 
    \item Both agents encode and decode data using the same, fixed number of bits per symbol (i.e., the modulation order is preset). Section~\ref{subsec:mod_order} describes the modulation orders and their reference modulation schemes used in this paper.
    \item The environment is stationary and non-adversarial during the learning process.
\end{enumerate}

\begin{figure*}[ht]
    \centering
    \includegraphics[width=0.8\linewidth]{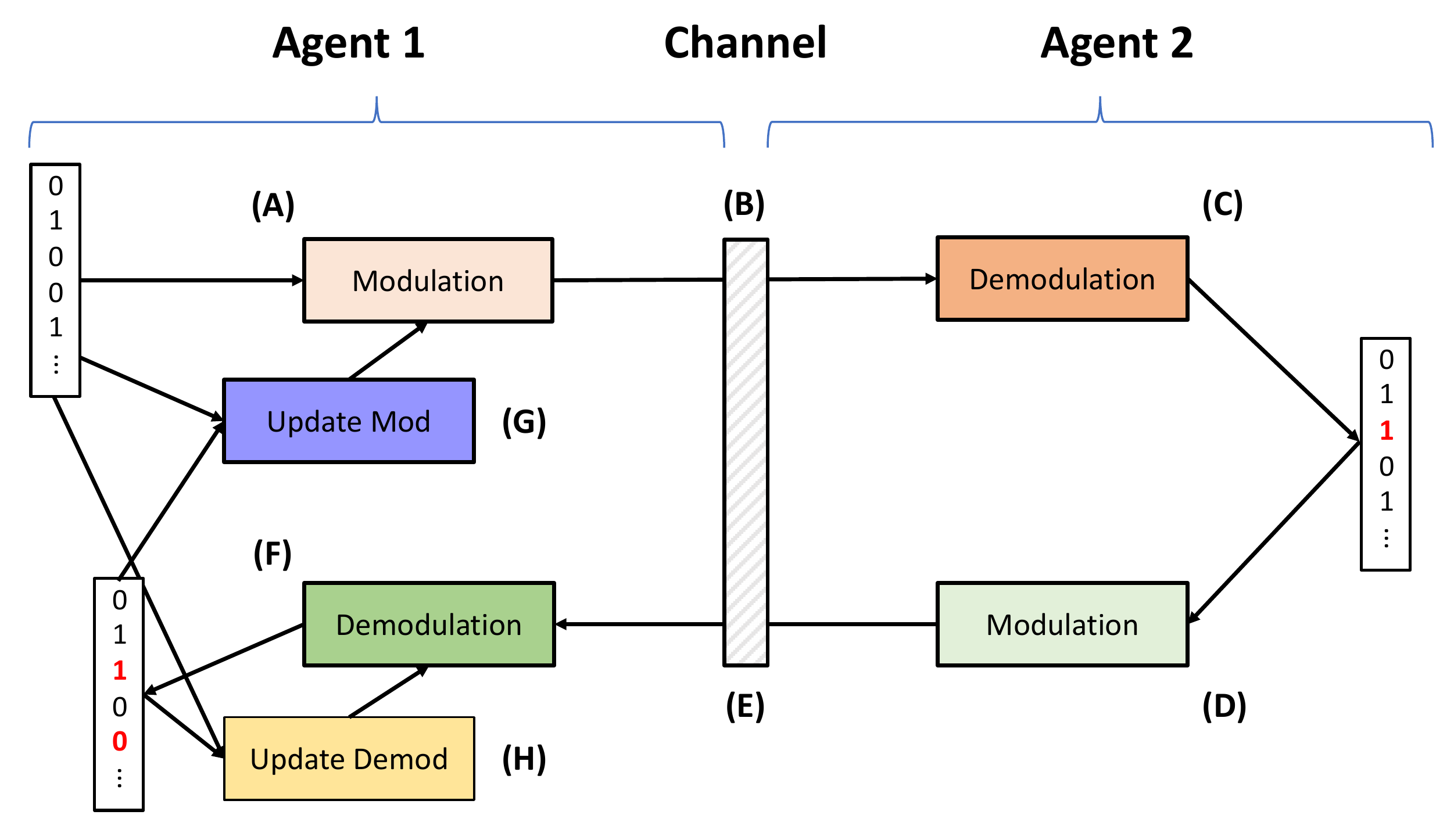}
    \caption{Visualization of the \echo{} protocol. (A) Speaker Agent (A1) modulates a bit sequence and (B) sends it across a (AWGN) channel. (C) Echoer Agent (A2) receives the sequence and demodulates it. (D) A2 then modulates the recovered sequence and (E) sends it back over the channel. (F) A1 receives this echoed version of its original sequence and demodulates it. (G, H) Then A1 uses the received echo to update its modulator and demodulator. The agents switch roles and repeat until convergence. Details of the protocol are elaborated in  Fig.~\ref{fig:private_preamble} and Sec.~\ref{subsec:relaxations-echo}.}
    \label{fig:the_echo_protocol}
\end{figure*}

\subsection{Motivation  and  Approach -- Echo with Private Preamble Protocol}
\label{subsec:procedure_metaprotocol}

The main objective of our work is to specify a robust communication-learning protocol that allows two independent agents to learn a modulation scheme under minimal assumptions on information sharing beyond shared knowledge of learning protocol and the ability to take turns. No other information is shared {\em a priori} or via a side channel during training. We name this protocol \textit{Echo with Private Preamble} (\EPP{}). Details about the \EPP{} protocol are provided in Section~\ref{subsec:relaxations-echo}.  The \EPP{} protocol is a special variant of the \echo{} protocol described in Fig.~\ref{fig:the_echo_protocol}.

The underlying premise of the \echo{} protocol is that an echo of the message---originating from one agent and repeated back to them by another agent---provides sufficient feedback for an agent to learn expressive modulation schemes. 
Under the \echo{} protocol, one agent (the ``speaker'') broadcasts a message and receives back an estimate of this message (preamble), an echo, from the other agent (the ``echoer''). The passage of the original message from the speaker to the echoer and back to the speaker as an echo is denoted as a  \textit{round-trip}. (A \textit{half-trip} goes only from speaker to echoer.) After a \textit{round-trip}, the speaker compares the original message to the echo and trains its  modulator and demodulator to minimize the difference (usually measured in bit-errors) between the two messages. The two agents then switch roles and repeat. When the difference between the original message and the demodulated echo is small, we infer that the agents can communicate with one another.

The ideas here are  similar to the approach to solving image-to-image mapping, or style transfer, popularized by CycleGAN \cite{cyclegan:2017arXiv170310593Z}.
Both works solve the problem of learning mappings between domains with only weak supervision by introducing a round-trip and defining `goodness' as how close the output of the round-trip is to the input.
Having round-trip feedback from either the other radio agent or the other GAN crucially enables performance measurement, and hence training.

By contrast, in the \textit{\echo{} with Shared Preamble} protocol from \cite{wcm:de2018cooperative},
the echo behavior is introduced only to train the modulator, and knowledge of a shared preamble between the two agents is assumed to facilitate direct supervised training of the demodulator after a half-trip exchange. In the \EPP{} protocol, the agents do not have access to a shared preamble and must learn to demodulate blindly, without knowing what was actually sent by the other agent. There is no shared preamble.

We believe that the \EPP{} protocol minimizes the information sharing assumptions for learning modulation schemes for two reasons. First, some sort of feedback is required for learning, and the echo provides this feedback.  Second, the \EPP{} protocol treats the environment as a regenerative channel, i.e. a channel that provides helpful feedback without requiring assumptions about the nature of the other communicating agent. As long as the other agent is cooperative (in the sense that it echoes back what is heard), then the environment behaves like a regenerative channel.

Next, we argue that the \EPP{} protocol is a plausible mechanism for learning modulation schemes when the channel is regenerative by considering the case of a learning agent communicating with an agent that uses fixed, classic schemes. In this setting, even random exploration would eventually find a modulation scheme that successfully interfaces with the fixed agent. By using feedback to guide exploration, we expect the \EPP{} protocol to perform much better than random guessing and quickly converge to a suitable modulation scheme.  We can think of such a fixed, friendly regenerative channel as a ``game'' that the learner plays where positive reward is achieved if what the channel echoes back can be decoded as what the learner encoded and sent in. Reinforcement learning is good at optimizing behaviors for simple games like this \cite{rl:Atari2013arXiv1312.5602M}. One of our main contributions is to show that the \EPP{} protocol works not only with fixed communication partners, but even in the case where two agents are learning simultaneously.

To verify the universality of the \EPP{} protocol and understand its performance relative to more structured or complex procedures, we run experiments with:
\begin{enumerate}
    \item Different learning protocols based on  varying amounts of information sharing as described in Section~\ref{sec:information_sharing}.
    \item Different levels of ``alienness''\footnote{``Alienness'' is a description of how different the models for the agents' modulators and demodulators are between the two agents. Factors that determine alienness include whether the agents are fixed or learning, the class of function approximators used by the learning agents and choice of hyperparameters and initializations.}  among agents as described in Section~\ref{sec:aliens}. 
    \item Different modulation order and levels of training SNR as described in Section~\ref{subsec:mod_order}.

\end{enumerate}



\section{Levels of Information Sharing}
\label{sec:information_sharing}

The \EPP{} protocol introduced in Section~\ref{subsec:procedure_metaprotocol} is designed to be minimalist in the sense that we share as little information as possible. However, using less information usually comes at the cost of performing worse. To quantify the value of shared information, in this section we describe the following protocols that allow an increasing amount of shared information:

\begin{enumerate}
    \item \textit{Echo with Shared Preamble} (ESP) protocol: Agents have access to shared preamble but can only get feedback via a round-trip during training;
    \item \textit{Loss Passing}  (LP) protocol: Agents have access to a shared preamble and share scalar loss values directly (without using the channel) during training; and 
    \item \textit{Gradient Passing} (GP) protocol: Agents have access to a shared preamble and share gradients directly (without using the channel) during training.
\end{enumerate}

\begin{figure*}[htb]
        \centering
        \includegraphics[width=0.8\linewidth]{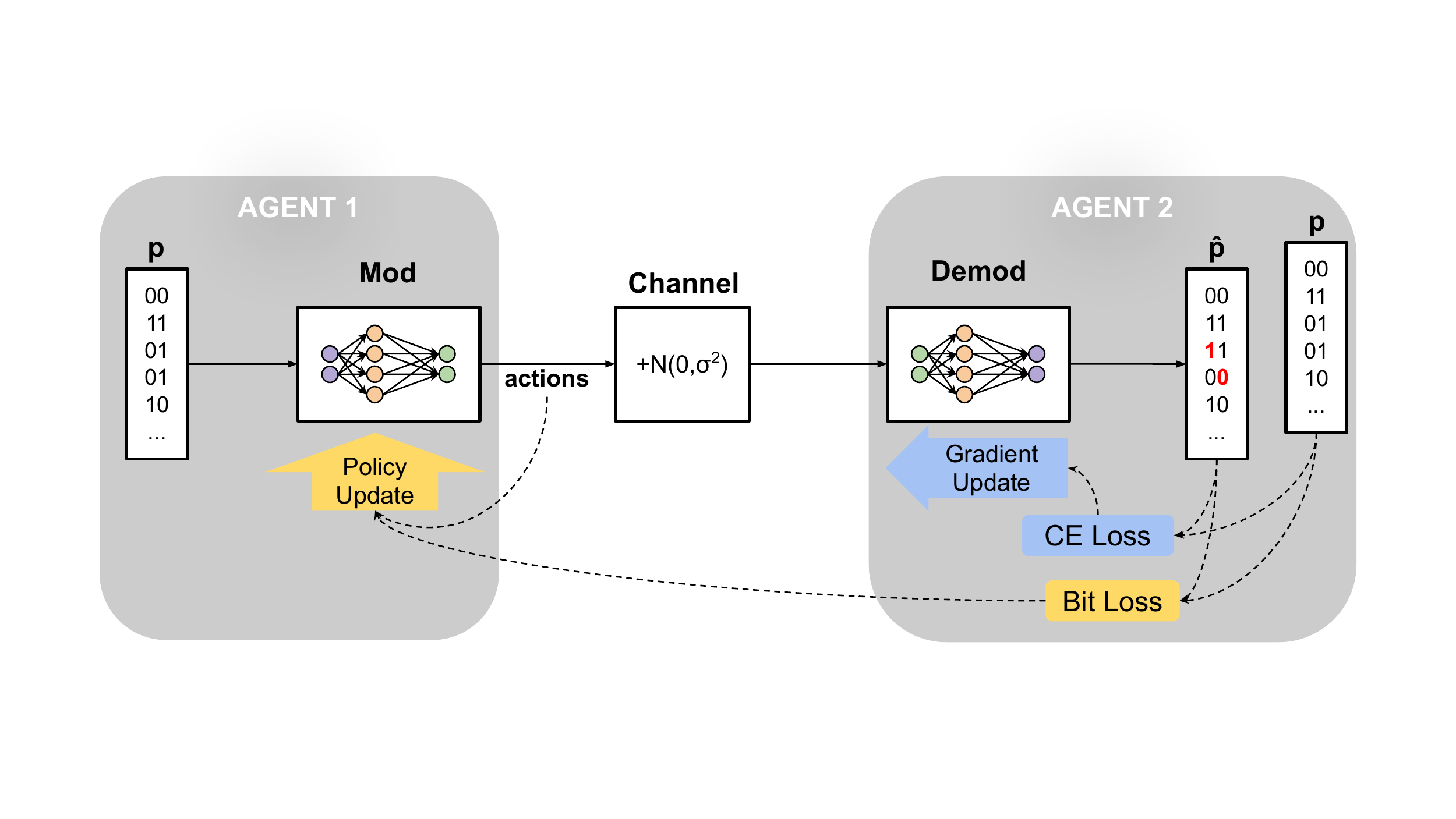}
        \caption{\textit{Loss Passing: Half-Trip}. In the \LP{} protocol, the preamble $p$ is modulated and sent from Agent 1 across the channel to Agent 2 where it is demodulated as $\hat{p}$. Using the shared preamble, Agent 2 performs a gradient update for its demodulator and shares a scalar bit loss value with agent 1. Agent 1 then uses this bit loss to perform a policy update of its modulator. Note that the loss is not passed through the channel. All implementations for the modulator currently use a Gaussian policy with mean and variance estimated by a function approximator as described in Section~\ref{subsec:learning}.}
        \label{fig:loss_passing}
\end{figure*}

\begin{figure*}[htb]
    \centering
    \includegraphics[width=0.8\linewidth]{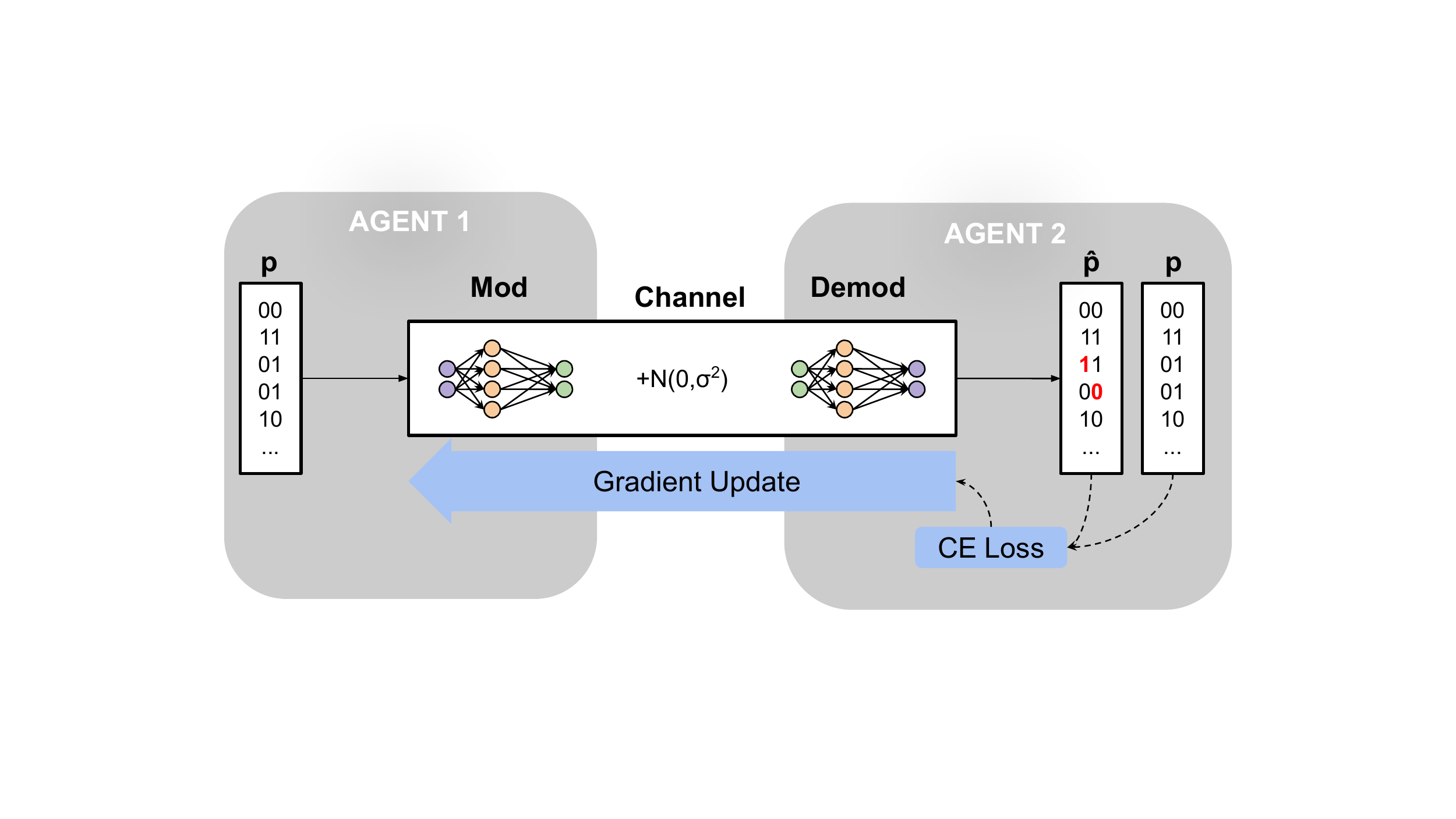}
    \caption{\textit{Gradient Passing: Half-Trip}.
    In the \GP{} protocol, the preamble $p$ is modulated and sent from Agent 1 through the channel to Agent 2 and is  demodulated as $\hat{p}$. Using the shared preamble, the modulator of Agent 1 and the demodulator of Agent 2 are updated using the cross entropy loss.} 
    \label{fig:gradient_passing}
\end{figure*}

Note that by sharing gradients or loss information directly across the channel, it is possible to truncate the learning process at step C in Fig.~\ref{fig:the_echo_protocol} and still update the modulator of the speaker. Examples of algorithms which stop at this step are shown later in Figs.~\ref{fig:loss_passing} and \ref{fig:gradient_passing}.  In fact, traditional autoencoder style training is like the gradient passing protocol described above. A reader who is already familiar with this concept may wish to read about the protocols in the reverse order from how we present them.

Our purpose in studying \LP{}, \GP{}, and \ESP{} protocols is primarily to understand the effect of shared information on learning since these are not new and have been studied independently in previous works such as \cite{cae:DBLP:journals/corr/OSheaH17}, \cite{wcm:DBLP:journals/corr/abs-1804-02276} and \cite{wcm:de2018cooperative}. The \LP{} and \GP{} protocols are not implementable in real world systems without a side channel to pass losses and gradients --- i.e.~they can be used in simulation at design-time, but not really used at run-time among distributed agents without depending on some existing communication infrastructure between them. \ESP{}, however, is practical and can be implemented by mandating that every agent use a common fixed preamble. The major difference is that \ESP{} requires agents to establish a shared preamble through some other mechanism before they can learn to communicate, whereas \EPP{} removes this requirement. Section~\ref{sec:results-info-sharing} reports the results of our experiments comparing the performance of these protocols and quantifying the value of shared information.

The following subsections describe the learning protocols for \EPP{}, \ESP{}, \LP{}, and \GP{} in detail, highlighting the important differences between them.

\begin{figure*}[htb]
        \centering
        \includegraphics[width=0.8\linewidth]{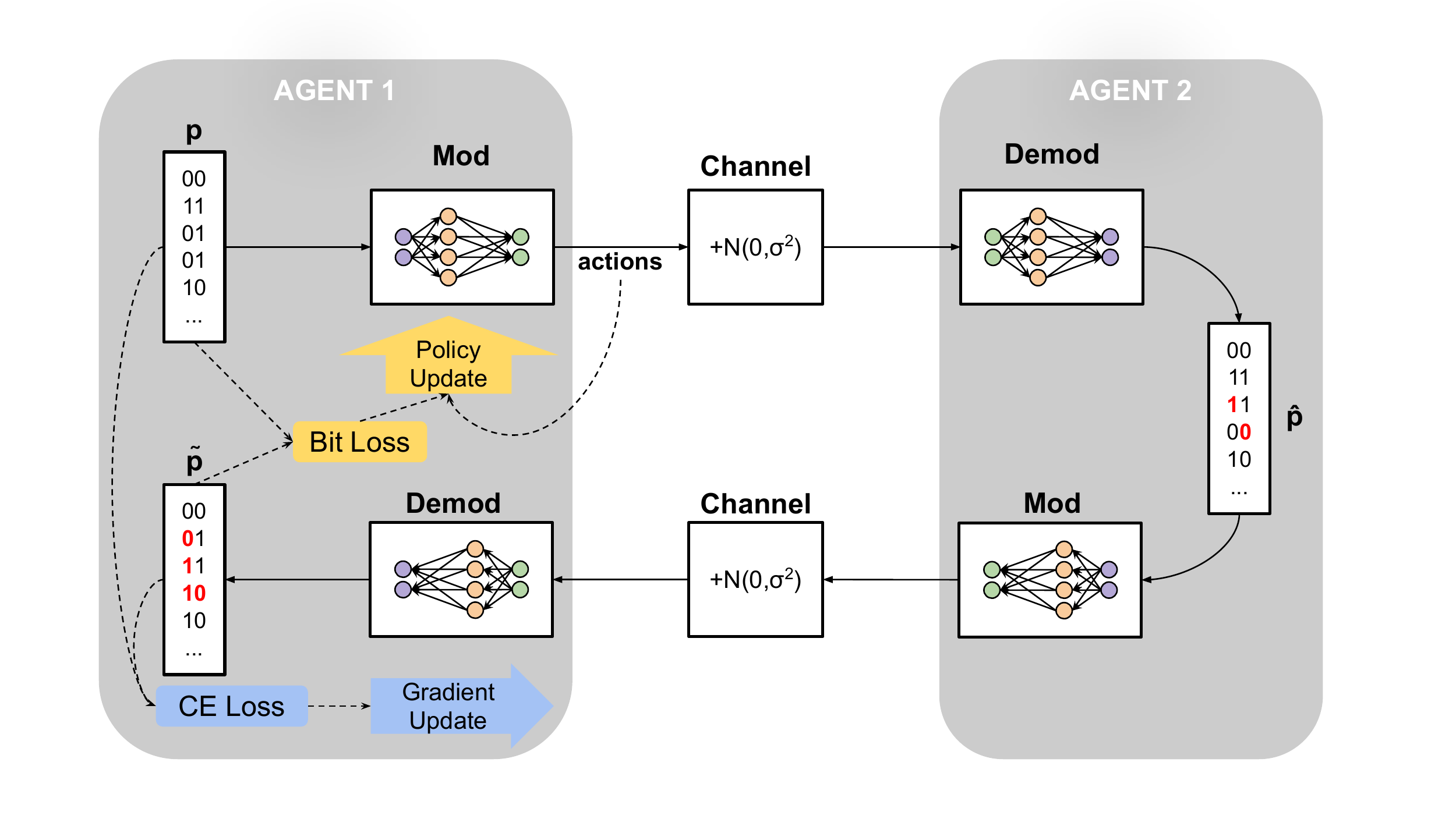}
        \caption{\textit{Echo with Private Preamble: Round-Trip.}
    In this diagram, the preamble $p$ is modulated and sent from Agent 1 through a channel to Agent 2 to be demodulated as $\hat{p}$. Agent 2 has no information about the message it received so it cannot update. It simply modulates and echos back the message it demodulated. Agent 1 then demodulates the echoed preamble as $\widetilde{p}$ and does a policy update of its modulator using a bit loss between the original preamble that Agent 1 sent, $p$, and echoed preamble that Agent 1 received, $\widetilde{p}$, as well as a supervised gradient update of its demodulator with cross entropy loss. Agent 1 and Agent 2 then switch roles so that now Agent 2 is the speaker and Agent 1 is the echoing agent. All implementations for the modulator currently use a Gaussian policy with mean and variance estimated by a function approximator as described in Section~\ref{subsec:learning}.}
        \label{fig:private_preamble}
\end{figure*}
    
\begin{algorithm*}[htb]
\begin{algorithmic}
\Procedure{EPP}{Agent 1, Agent 2}
\State Speaker $\gets$ Agent 1
\State Echoer $\gets$ Agent 2
\While{training}
    \State $p \gets n\mbox{ random bits}$\Comment{$p$ is known only to Speaker}\label{eppstart}
    \State $\mu, \sigma^2 \gets M(p; \theta_s)$\Comment{Speaker generates parameters for its Gaussian policy using $p$}
    \State $s \gets \mathcal{N}(\mu, \sigma^2 I)$\Comment{Speaker modulates by sampling from this distribution}
    \State $\hat{s} \gets f\textsubscript{channel}(s)$
    \State $\hat{p} \gets D(\hat{s}; \phi_e)$\Comment{Echoer demodulates received symbols}
    \State $\mu, \sigma^2 \gets M(\hat{p}; \theta_e)$\Comment{Echoer generates parameters for its Gaussian policy using $\hat{p}$}
    \State $\widetilde{s} \gets f\textsubscript{channel}\left(\mathcal{N}(\mu,\sigma^2 I)\right)$\Comment{Echoer modulates by sampling from this distribution}
    \State $\widetilde{p} \gets D(\widetilde{s}; \phi_s)$
    \State $\theta_s' \gets \theta_s + \Delta_{\theta_s}(s, \widetilde{p}, p)$\Comment{Policy gradient update for speaker's mod}
    \State $\phi_s' \gets \phi_s + \Delta_{\phi_s}(\widetilde{s}, \widetilde{p}, p)$\Comment{Cross-entropy loss gradient update for speaker's demod}
    \State Speaker $\longleftrightarrow$ Echoer\Comment{Agents switch Speaker and Echoer roles}
\EndWhile\Comment{Only the Speaker updates each round-trip}
\EndProcedure
\end{algorithmic}
\caption{Echo Protocol with Private Preamble}
\label{alg:epp}
\end{algorithm*}

\begin{figure*}[htb]
    \centering
    \subfloat[Agent 1 modulation scheme\label{fig:ambiguity-mod-1}]{
        \includegraphics[width=0.40\textwidth]{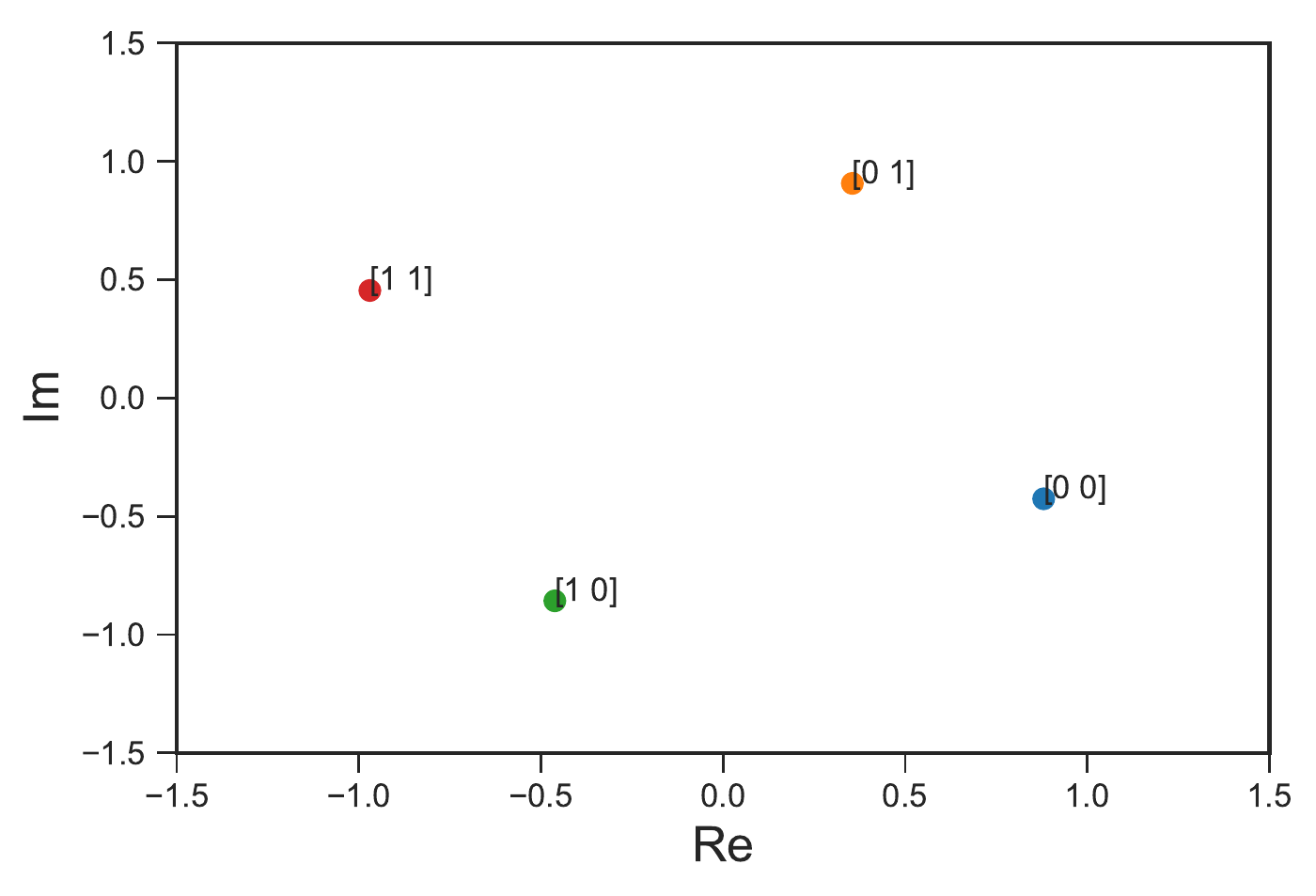}}
    ~
    \subfloat[Agent 2 demodulation scheme\label{fig:ambiguity-demod-2}]{
        \includegraphics[width=0.40\textwidth]{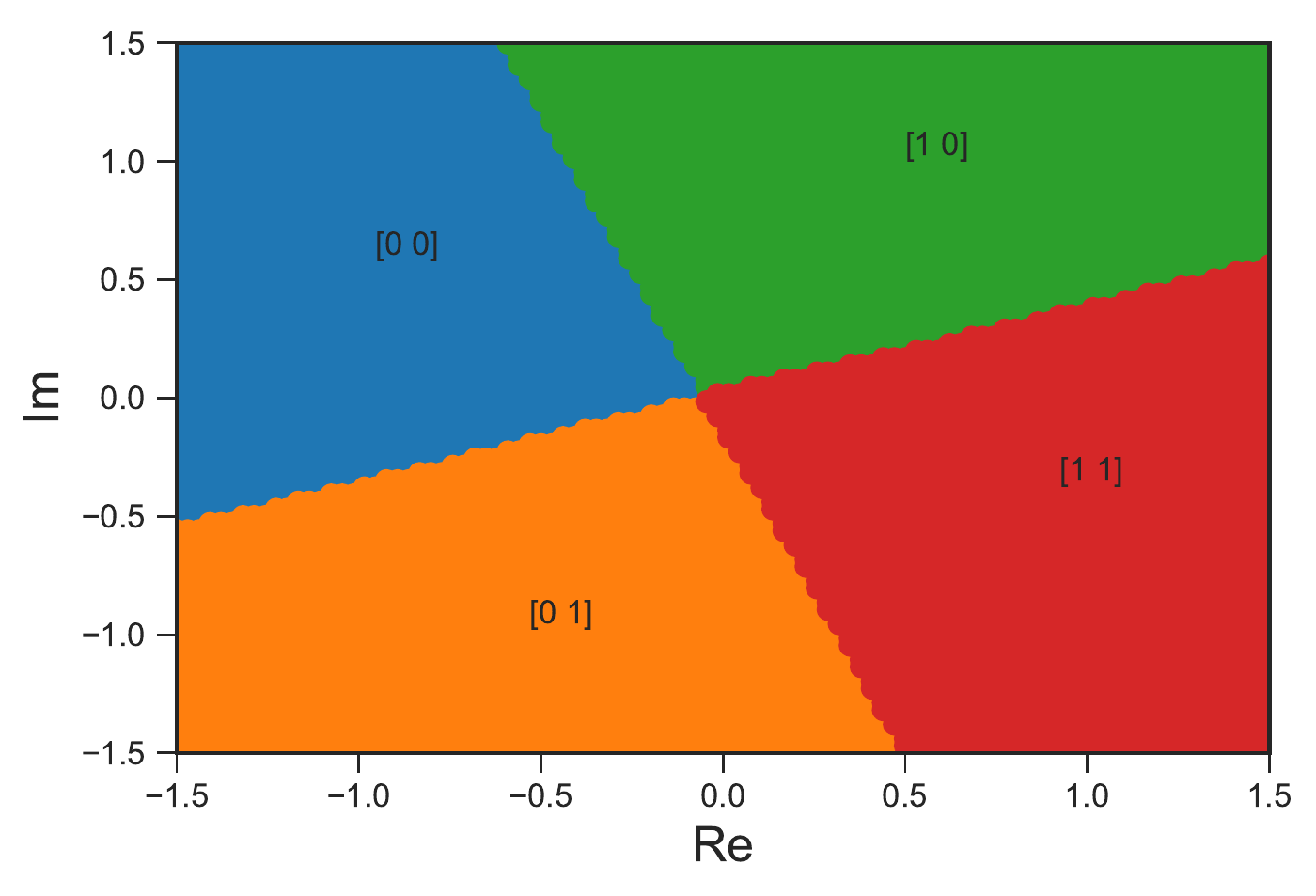}}

    \subfloat[Agent 2 modulation scheme\label{fig:ambiguity-mod-2}]{
        \includegraphics[width=0.40\textwidth]{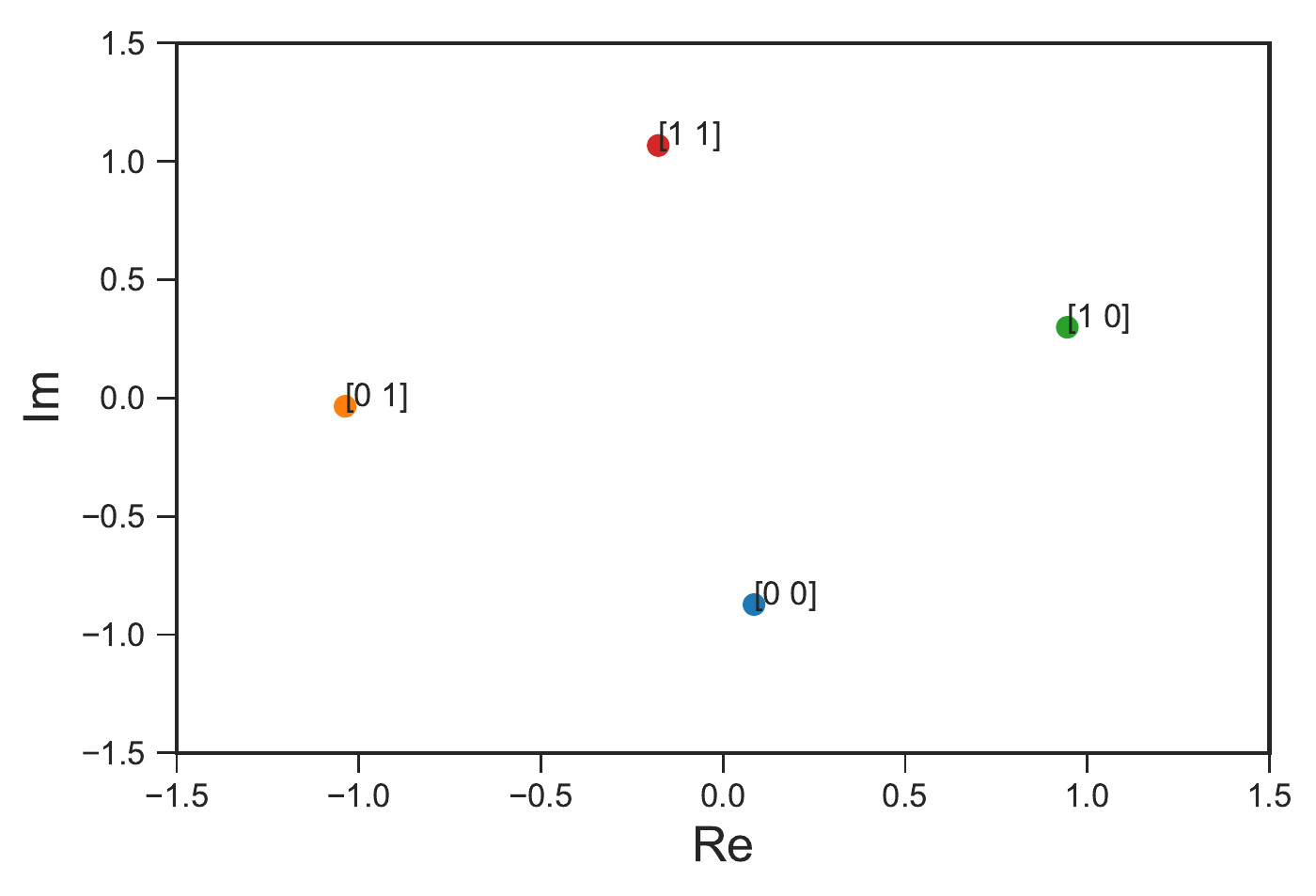}}
    ~
    \subfloat[Agent 1 modulation scheme\label{fig:ambiguity-demod-1}]{
        \includegraphics[width=0.40\textwidth]{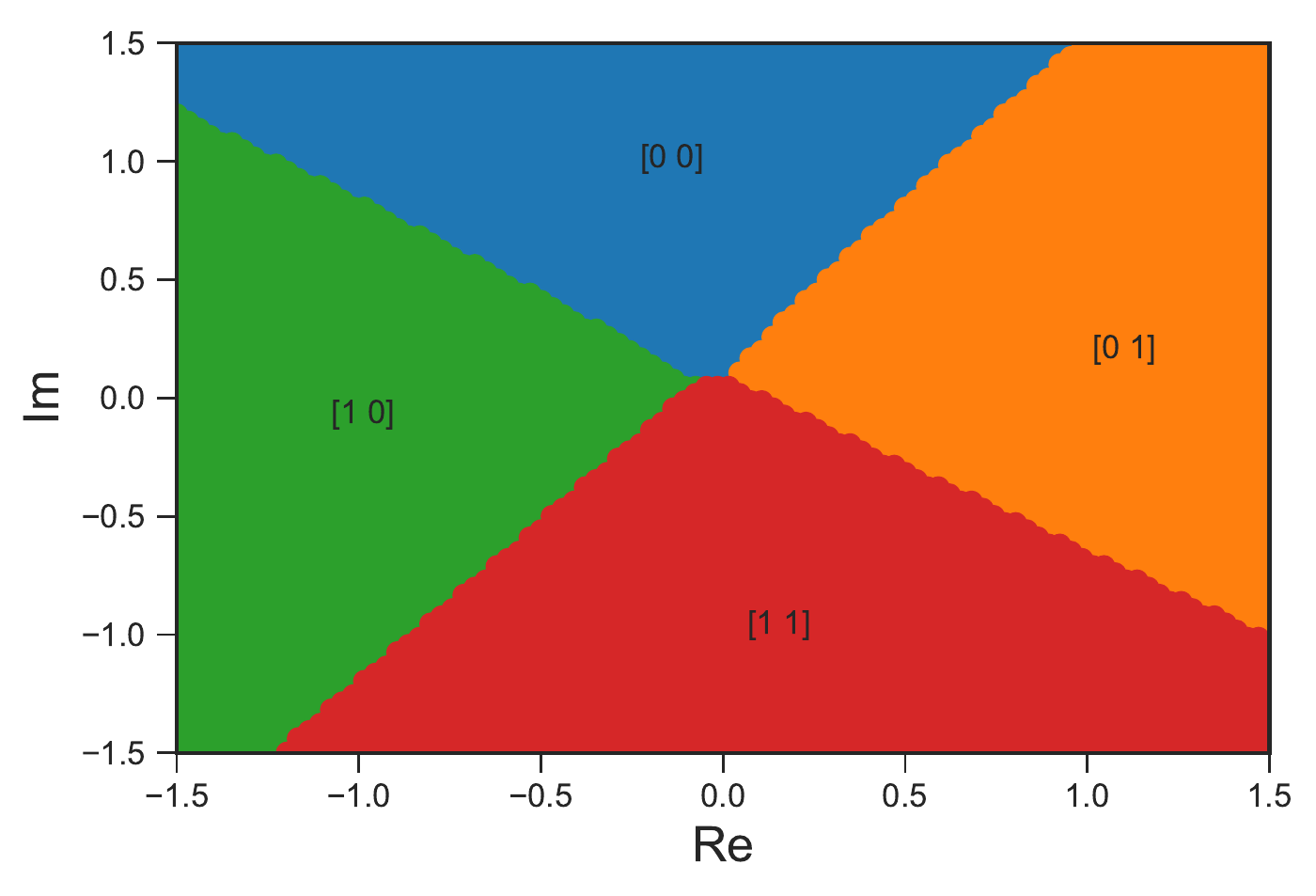}}
    \caption{An example modulation scheme learned by agents using the \EPP{} protocol to demonstrate ambiguity of communication after a half-trip, but coherence after a round-trip exchange.
    In this scheme Agent 1, maps the bit sequence `00' to the complex number $1 - 0.5j$, i.e $M_1($`00') = $1 - 0.5j$. Agent 2 demodulates this as the bit sequence `11', i.e $D_2(M_1($`00'$)) = $ `11' $ \neq $ `00'.
    However this mismatch is reversed when the round-trip is completed. Agent 2 modulates `$11$' as $0 - 1j$ and Agent 1 demodulates this as `$00$'. Thus $D_1(M_2(D_2(M_1($`00'$))) = $ `00'.
    }
    \label{fig:symbol-ambiguity}
\end{figure*}

\begin{figure*}[htb]
        \centering
        \includegraphics[width=0.8\linewidth]{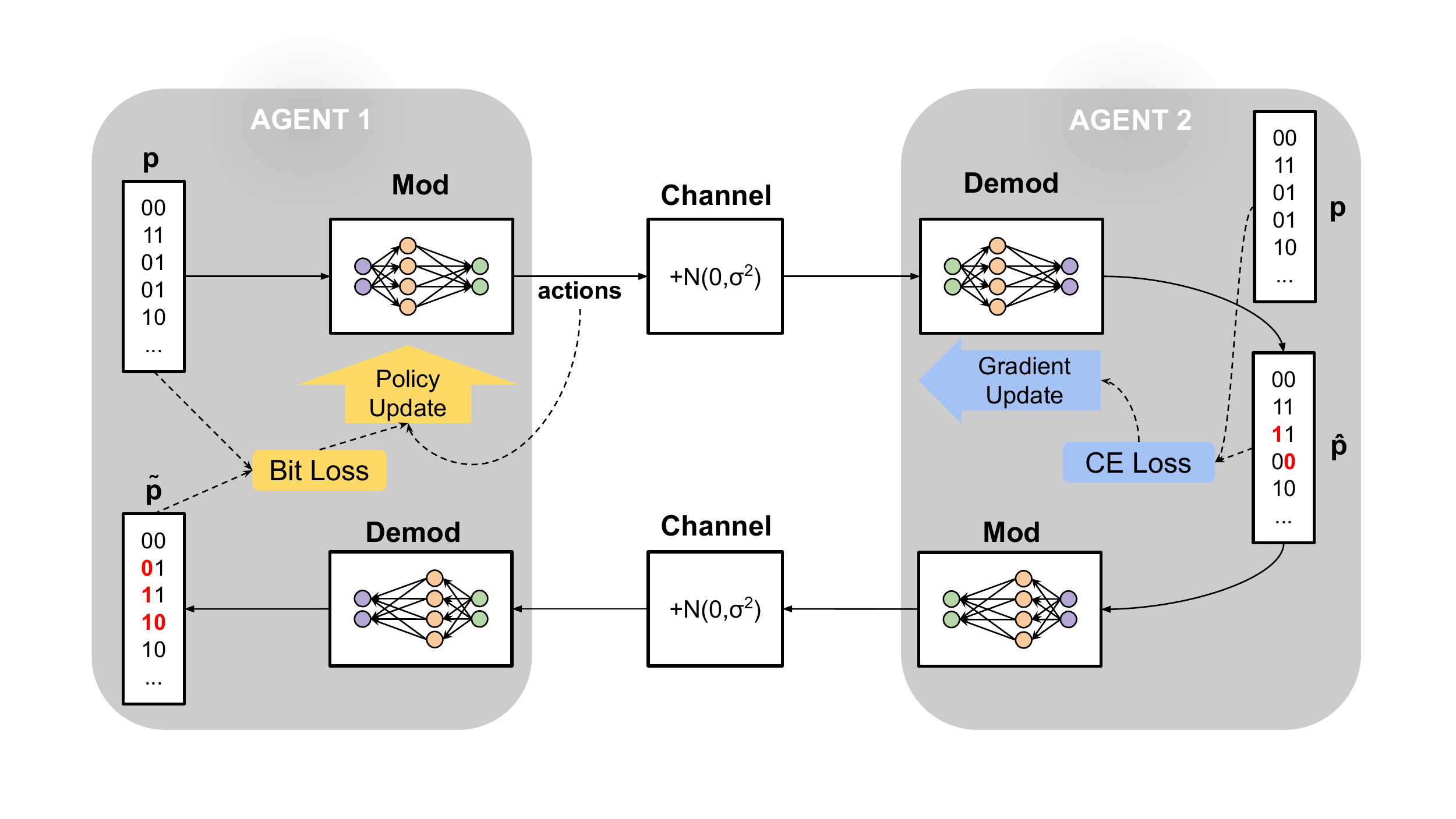}
        \caption{\textit{Echo with Shared Preamble: Round-Trip}.
    In the \ESP{} protocol, the preamble $p$ is modulated and sent from Agent 1 across the channel to Agent 2 and is demodulated as $\hat{p}$. Using the shared preamble, Agent 2 performs a gradient update on its demodulator and also modulates and sends back an echo, an estimate of the  preamble it received, $\hat{p}$, through the channel back to Agent 1. Agent 1 then demodulates the echo as $\widetilde{p}$ and does a policy update of its modulator using the bit loss between the original preamble $p$ and estimate of the echo  $\widetilde{p}$. Agent 1 and Agent 2 then switch roles and repeat the process.   All implementations for the modulator currently use a Gaussian policy with mean and variance estimated by a function approximator as described in Section~\ref{subsec:learning}.}
        \label{fig:shared_preamble}
\end{figure*}
 
\begin{algorithm*}[htb]
\begin{algorithmic}
\Procedure{ESP}{Agent 1, Agent 2}
\State Speaker $\gets$ Agent 1
\State Echoer $\gets$ Agent 2
\While{training}
    \State $p \gets n\mbox{ random bits}$\Comment{$p$ is known to Speaker and Echoer}\label{espstart}
    \State $\mu, \sigma^2 \gets M(p; \theta_s)$\Comment{Speaker generates parameters for its Gaussian policy using $p$}
    \State $s \gets \mathcal{N}(\mu, \sigma^2 I)$\Comment{Speaker modulates by sampling from this distribution}
    \State $\hat{s} \gets f\textsubscript{channel}(s)$
    \State $\hat{p} \gets D(\hat{s}; \phi_e)$\Comment{Echoer demodulates received symbols}
    \State $\phi_e' \gets \phi_e + \Delta_{\phi_e}(\hat{s}, \hat{p}, p)$\Comment{Cross-entropy loss gradient update for echoer's demod}
    \State $\mu, \sigma^2 \gets M(\hat{p}; \theta_e)$\Comment{Echoer generates parameters for its Gaussian policy using $\hat{p}$}
    \State $\widetilde{s} \gets f\textsubscript{channel}\left(\mathcal{N}(\mu,\sigma^2 I)\right)$\Comment{Echoer modulates by sampling from this distribution}
    \State $\widetilde{p} \gets D(\widetilde{s}; \phi_s)$
    \State $\theta_s' \gets \theta_s + \Delta_{\theta_s}(s, \widetilde{p}, p)$\Comment{Policy gradient update for speaker's mod}
    \State Speaker $\longleftrightarrow$ Echoer\Comment{Agents switch Speaker and Echoer roles}
\EndWhile\Comment{The Echoer's demodulator updates, not the Speaker's}
\EndProcedure
\end{algorithmic}
\caption{Echo Protocol with Shared Preamble}
\label{alg:esp}
\end{algorithm*}

\begin{algorithm*}[htb]
\begin{algorithmic}
\Procedure{LP}{Agent 1, Agent 2}
\State Speaker $\gets$ Agent 1
\State Echoer $\gets$ Agent 2
\While{training}
    \State $p \gets n\mbox{ random bits}$\Comment{$p$ is known to Speaker and Echoer}\label{lpstart}
    \State $\mu, \sigma^2 \gets M(p; \theta_s)$\Comment{Speaker generates parameters for its Gaussian policy using $p$}
    \State $s \gets \mathcal{N}(\mu, \sigma^2 I)$\Comment{Speaker modulates by sampling from this distribution}
    \State $\hat{s} \gets f\textsubscript{channel}(s)$
    \State $\hat{p} \gets D(\hat{s}; \phi_e)$\Comment{Echoer demodulates received symbols}
    \State $\phi_e' \gets \phi_e + \Delta_{\phi_e}(\hat{s}, \hat{p}, p)$\Comment{Cross-entropy loss gradient update}
    \State $L \gets \hat{p} \oplus p$
    \State $\theta_s' \gets \theta_s + \Delta_{\theta_s}(s, L, p)$\Comment{Policy gradient update}
    \State Speaker $\longleftrightarrow$ Echoer\Comment{Agents switch Speaker and Echoer roles}
\EndWhile\Comment{Only a half-trip is required for updates}
\EndProcedure
\end{algorithmic}
\caption{Loss Passing: Half-Trip}
\label{alg:lp}
\end{algorithm*}

\begin{algorithm*}[htb]
\begin{algorithmic}
\Procedure{GP}{Agent 1, Agent 2}
\State Speaker $\gets$ Agent 1
\State Echoer $\gets$ Agent 2
\While{training}
    \State $p \gets n\mbox{ random bits}$\Comment{$p$ is known to Speaker and Echoer}\label{gpstart}
    \State $s \gets M(p; \theta_s)$\Comment{Speaker modulates $p$ directly}
    \State $\hat{s} \gets f\textsubscript{channel}(s)$
    \State $\hat{p} \gets D(\hat{s}; \phi_e)$\Comment{Echoer demodulates received symbols}
    \State $\phi_e' \gets \phi_e + \Delta_{\phi_e}(\hat{s}, \hat{p}, p)$\Comment{Cross-entropy loss gradient update}
    \State $\theta_s' \gets \theta_s + \Delta_{\theta_s}(s, p, \Delta_{\phi_e})$\Comment{Gradient update}
    \State Speaker $\longleftrightarrow$ Echoer\Comment{Agents switch Speaker and Echoer roles}
\EndWhile\Comment{The Speaker performs a gradient-loss update}
\EndProcedure
\end{algorithmic}
\caption{Gradient Passing: Half-Trip}
\label{alg:gp}
\end{algorithm*}

\subsection{Echo Protocol With Private Preamble}
\label{subsec:relaxations-echo}

The \EPP{} protocol is the main contribution of our paper. It is described in detail in Alg.~\ref{alg:epp} and Fig.~\ref{fig:private_preamble}. The key details when comparing to \ESP{}, \LP{}, and \GP{} are the natures of the modulator and demodulator updates. For \EPP{}, the demodulator updates use supervised learning, but have to rely on noisy feedback because only $p$ is known, but the demodulator actually receives $\hat{p}$. The modulator updates use reinforcement learning based on the round-trip feedback. Because the preamble is known only to the speaker, only the speaker's modulator and demodulator can be updated during a round-trip. The choice of when to terminate training is arbitrary, but we choose to halt training after a fixed number of training iterations. Other implementations might halt training after a BER target is reached.

One important consideration that is unique to the \EPP{} protocol is that there is no way to ensure that the bit sequence sent by the modulator is interpreted as the same sequence after being demodulated. More formally, there is no way to ensure that
\begin{align}
    p = D_2\left(M_1(p; \theta_1); \phi_2\right).
\end{align} 
For example, Agent 1 might modulate the sequence $11$ as some symbol $c_1$, but Agent 2 might interpret $c_1$ as $00$. After a round-trip, however, any incorrect bit sequence to modulated symbol mappings will be reversed if the agents have trained properly. We can guarantee that
\begin{align} 
p = D_1\left(M_2\left(D_2\left(M_1(p; \theta_1); \phi_2\right); \theta_2\right); \phi_1\right).
\end{align}

Fig.~\ref{fig:symbol-ambiguity} demonstrates how this might happen. We address how we evaluate agents when this mapping ambiguity is present in Section~\ref{subsec:evaluation}. In general, it would require a protocol higher up the communication stack to disambiguate symbol mappings without access to a shared preamble --- some way of symmetry breaking is required, and this presumably requires knowing more about the context of communication.

\subsection{Echo With Shared Preamble}
\label{subsec:relaxations-shared-preamble}
The \ESP{} protocol is described in Fig.~\ref{fig:shared_preamble} and Alg.~\ref{alg:esp}. \ESP{} was first explored in \cite{wcm:de2018cooperative} where the modulator was neural network based and trained using policy gradients but where the demodulator used clustering methods\footnote{If the demodulator is using unsupervised clustering algorithms, acting cooperatively requires the clustering algorithm to be stable. If the label assigned to each cluster changes every iteration, the other agent will not be able to converge on a modulation scheme.} trained via supervised learning using the shared preamble. In our work, we use the \ESP{} protocol to train agents whose modulators and demodulators both use function approximators. (See Appendix \ref{app:agent-details} for more information). 

\ESP{} is similar to the \EPP{} protocol, but now both the speaker and echoer know the preamble $p$ that is transmitted. This allows the echoer to update its demodulator after the first half-trip since it knows exactly what it was supposed to have received. This demodulator update is typically of higher quality than the updates in \EPP{}, since those updates only have access to symbols based on the (possibly incorrect) estimate of the original preamble sent back by the echoer. The speaker agent does not bother to update its demodulator after the round-trip is complete, since it will receive higher quality feedback on the next training iteration after the speaker and echoer roles are switched. 

Importantly, the speaker's modulator still requires a full round-trip before it can receive feedback and be updated. In the next Sections~\ref{subsec:relaxations-shared-loss} and~\ref{subsec:relaxations-shared-gradient} this will no longer be the case. The consequence of round-trip feedback is that the speaker's modulator is actually optimizing for the performance of the \textit{speaker's} demodulator, since that is the only loss is has access to. Our presumption is that improving the round-trip performance of the speaker's demodulator will indirectly improve the half-trip performance of the echoer's demodulator, since the half-trip BER limits the round-trip BER. The consequences of this indirection are illustrated in Section~\ref{sec:results-info-sharing}.

\subsection{Loss Passing: Half-Trip}
\label{subsec:relaxations-shared-loss}

Now we remove the restriction that information can only be shared over the channel during training and allow the agents to magically pass losses back and forth. The loss passing protocol, as used in previous work such as \cite{wcm:DBLP:journals/corr/abs-1804-02276}, is detailed in Fig.~\ref{fig:loss_passing} and Alg.~\ref{alg:lp}. There is no longer a need for an echo from the second agent, since the speaker's modulator receives a loss value directly from the second agent's demodulator. This results in two major changes: a full training update can be completed after only a half-trip, and the speaker's modulator is optimizing for the echoer's demodulator performance directly. 

In the \EPP{} and \ESP{} protocols, the speaker's modulator has to optimize for the performance of the speaker's demodulator, only indirectly addressing the performance of the echoer's demodulator. The \LP{} protocol allows the speaker's modulator to directly optimize for the performance of the echoer's demodulator since the speaker has access to the relevant loss values. Although the speaker's modulator still has to use reinforcement learning rather than supervised learning to perform parameter updates, we expect the agents to be able to train much faster when using loss passing.

\subsection{Gradient Passing: Half-Trip}
\label{subsec:relaxations-shared-gradient}

    If we further allow the agents to share gradients during training, the system can naturally be treated as an end-to-end autoencoder\footnote{For classic modulators/demodulators, we were able to generate gradient updates by treating/approximating the modulator or demodulator as a differentiable function.} with channel noise introduced between the encoding and decoding sections. This method was employed successfully in \cite{cae:DBLP:journals/corr/OSheaKC16}. Our version of such an autoencoder-based training protocol, which we call the \GP{} protocol, is explained in detail in Fig.~\ref{fig:gradient_passing} and Alg.~\ref{alg:gp}.
    
    As in the \LP{} protocol, the speaker's modulator can be trained after only a half-trip because it has access to feedback from the echoer's demodulator. Instead of using reinforcement learning to train a Gaussian policy, however, the speaker in \GP{} trains its modulator to encode bits directly as complex numbers, and the gradients from the echoer's demodulator are used for supervised learning updates.




\section{Alienness of Agents}
\label{sec:aliens}

How can we determine if the \EPP{} is universal? We need to determine if it allows us to learn to communicate with strangers.
There are in principle three kinds of agents (strangers) that we might encounter with which we might wish to learn to communicate:

\begin{enumerate}
    \item A fixed agent that knows how to communicate;
    \item A learning agent that does not know how to communicate yet but is cooperative and willing to learn; or 
    \item An agent that does not know and will not learn how to communicate.
\end{enumerate}

The \textit{Classic} agent  uses a fixed modulation scheme  known to be optimal for AWGN channels for the given modulation order, for e.g. QPSK for 2 bits per symbol, 8PSK for 3 bits per symbol, and 16QAM for 4 bits per symbol \cite{qammod}. This is an example of an agent of the first kind. An example of an agent of the second kind is  one that uses a function approximator for its modulator and demodulator that can be trained. We consider \textit{Neural} agents, agents that use neural networks as function approximators, and \textit{Poly} agents, ones that use polynomials as function approximators.  We ignore the agents of the third kind since it is impossible to learn to communicate with such agents.

Note that there are several other examples of agents. A learning agent that has been pre-trained and frozen behaves like a fixed agent. We can in principle have learning agents with decision tree or nearest neighbor based function approximators.  However, in this paper, we restrict ourselves to \Classic{}, \Neural{}, and  \Poly{} agents. Details about these agents, including the hyperparameters used and training methods employed, are provided in Appendices~\ref{app:agent-details}~and~\ref{app:simulation-hyperparams}.

We perform experiments by pairing two agents with different levels of alienness, where alienness is determined by:

\begin{enumerate}
    \item Whether they are fixed agents or learning agents (e.g.~a Neural-and-Classic matchup)
    \item The class of function approximators used by the learning agents. We denote such agents as ``Aliens'' (e.g.~Neural-and-Poly).
    \item The random initialization and hyperparameters used by two learning agents using the same class of function approximators. We denote agents that use the same class of function approximators but different random initialization and hyperparameters  as ``Self-Aliens'' (e.g.~Neural-and-Self-Alien).
    \item The random initialization used by two learning agents using the same class of function approximators and the same hyperparameters. We denote agents that differ only in random initialization as ``Clones'' (e.g.~Neural-and-Clone).
\end{enumerate}
Results for these experiments that portray the effect of different levels of alienness on the performance of the \EPP{} protocol are provided in Section~\ref{sec:results-alieness}.



\newcommand{\dboff}{dB-off-optimal}

\section{Experiments}
\label{sec:experiments}
In addition to the effects of different levels of information sharing and alienness, modulation order, training SNR, and modulator constellation power constraints are other factors that affect the performance of our learning protocols. 

\subsection{Modulation Order and Training Signal to Noise Ratio}
\label{subsec:mod_order}

Modulation order, determined by the bits per symbol ($bps$) used, determines the number of unique symbols that can be sent and received. A $bps$ of $b$ corresponds to $2^b$ unique symbols. For instance for $bps=2$, we have $4$ unique symbols: `00', `01', `10', and `11'. We consider settings where bits per symbol is either 2, 3, or 4. For \Classic{} agents, $bps$ determines the fixed scheme, optimal for AWGN channels, used as a baseline. These are provided in Table~\ref{tab:ber-snr-mod} and visualized in Appendix~\ref{app:mod_order_mods_demods}.
For \Neural{} and \Poly{} agents, $bps$ determines the size of the inputs and outputs of the modulator and demodulator. Details about this are provided in Appendix \ref{app:agent-details}. 

Since higher modulation orders have higher bit error rates (BERs) at the same SNR, we must determine an appropriate SNR to use for training and testing to provide a fair comparison between different modulation orders. We do this by selecting the SNR based on the round-trip BER achieved when using the baseline (classic) schemes. For most experiments we use a training SNR corresponding to a BER of 1\% and for all experiments we test on SNRs corresponding to BERs ranging from 0.001\% to 10\% as described in  Table~\ref{tab:ber-snr-mod}. 
We explore the effect of modulation order and  training SNR on the performance of \EPP{} protocol in Appendix~\ref{app:addresults}.



\begin{table*}[ht]
\centering
\begin{tabularx}{0.59\linewidth}{@{}lrrrrr>{\columncolor{grey}}rr@{}}
\toprule
\multirow{2}{*}{\textbf{}} & \multirow{2}{*}{\makecell{Bits Per\\Symbol}} & \multirow{2}{*}{\makecell{\# Constell.\\Points}} &  \multicolumn{5}{c}{SNR Corresponding to Round-trip BER of} \\
\cmidrule{4-8}
& & & 0.001\% & 0.01\% & 0.1\% & 1\% & 10\% \\ \midrule
\textbf{QPSK} & 2 & 4 & 13~dB & 12~dB & 10.4~dB & 8.4~dB & 4.2~dB \\
\textbf{8PSK} & 3 & 8 & 18.2~dB & 17~dB & 15.4~dB & 13.2~dB & 8.4~dB\\
\textbf{16QAM} & 4 & 16 & 20~dB & 18.8~dB & 17.2~dB & 15.0~dB & 10.4~dB \\ \bottomrule
\end{tabularx}
\caption{SNRs corresponding to round-trip BER values for the modulation orders we investigate. The SNR-to-BER mappings are used to set test and train SNRs for performance measurements. The SNR corresponding to 1\% BER (shaded column) is the default training SNR for our experiments. }
\label{tab:ber-snr-mod}
\end{table*}

\subsection{Constellation Power Constraints}
\label{subsec:power-constraints}
As described in Section~\ref{subsec:problem-formulation}, the modulator maps symbols (bits) into complex numbers, i.e. points on the complex plane. Due to the presence of the AWGN channel, it is optimal to place these points as far away as possible to minimize the likelihood of an error. Thus to get non-degenerate solutions we must impose a constraint on how far these points can be from the origin. Note that this is similar to a real-world constraint on power used by a radio system. 


We introduce a hard power constraint
by requiring that the modulator outputs have an average power of less than 1. We experimented with other soft power constraints by including a penalty term in the loss function based on the power used while training, but chose not to use it in the end. For simulations, we observed that a hard power constraint was sufficient, and more importantly did not require tuning the hyperparameter corresponding to the weight of the power penalty.

\subsection{Training}
\label{subsec:learning}

For the \Neural{} and \Poly{} learning agents, the demodulator is trained using supervised learning with cross-entropy loss. In the \GP{} protocol, the modulator output is equal to the output of the underlying function approximator and its parameters are updated using supervised learning. In the \EPP{}, \ESP{}, and \LP{} protocols the modulator employs a Gaussian policy. The modulator output is sampled from a Gaussian distribution with mean and variance determined by the output of the underlying function approximator whose parameters are updated using vanilla policy gradients. More details about the  update procedure are provided in Appendix \ref{app:agent-details}.


We conduct multiple trials using different random seeds for each experiment to accurately estimate the performance of our protocols and agents. An experiment fixes the learning protocol, the agent types, training SNR, and modulation order. Each trial is run for a maximum number of training iterations that we determine empirically for each experiment. Easier learning tasks are run for fewer iterations to speed up the simulations. 
Note that instead of measuring training iterations we can also measure the number of preamble symbols transmitted. These two measurements are related via the preamble length, the number of symbols in the preamble. For all our experiments we set the preamble length to 256 symbols in order to allow fair comparison across experiments. This also reduces the relative cost of overheads in the implementation on real hardware radios. Certain protocols and modulation orders require fewer transmitted symbols to achieve good performance. Details about the maximum iterations (and thus maximum number of preamble symbols transmitted) can be found in Table~\ref{tab:exp_settings} in Appendix~\ref{app:simulation-settings}, and in the code itself at \cite{Repo}. 

\subsection{Evaluation}
\label{subsec:evaluation}
How do we determine the metrics that should be used to measure the performance of a learning protocol? These metrics should allow for a fair comparison across different protocols (\GP{}, \LP{}, \ESP{}, and \EPP{}) and must be informative in determining the effect of different levels of information sharing, alienness, and modulation order on the learning task. We are primarily interested in quantifying `efficiency', how long the protocol takes to learn a modulation scheme, and `robustness', how reliably the protocol learns this scheme.

First, we must decide on a metric to determine if the learned modulation scheme is `good'. 
Bit error rate (BER) is a natural choice in communication settings but since we have two agents we must determine whether to measure cross-agent BER (half-trip BER) or round-trip BER. 
In the \GP{}, \LP{}, and \ESP{} protocols both cross-agent and round-trip BER  are indicative of performance. In the \EPP{} protocol, since the two agents have no shared preamble, measuring cross agent BER is not a good indicator of performance since the two agents may have different bit interpretations of the same modulated symbol, as described in Section~\ref{subsec:relaxations-echo}. However, round-trip BER is a valid measure of performance in this case.
Consequently, we choose the round-trip BER to allow for a fair comparison between different protocols.
Note that when measuring the BER to evaluate performance of agent(s), instead of sampling from the Gaussian policy, the modulators deterministically use the mean of the Gaussian policy. This avoids introducing additional errors from ``exploration.''


Next, we must determine the SNR that we measure the round-trip BER at and whether the measured BER is indicative of good performance. As discussed in Section~\ref{subsec:mod_order}, we decide on the test SNRs based on the modulation order depending on the performance of the baseline. 

\begin{figure}[ht]
    \centering
    \includegraphics[width=0.45\textwidth]{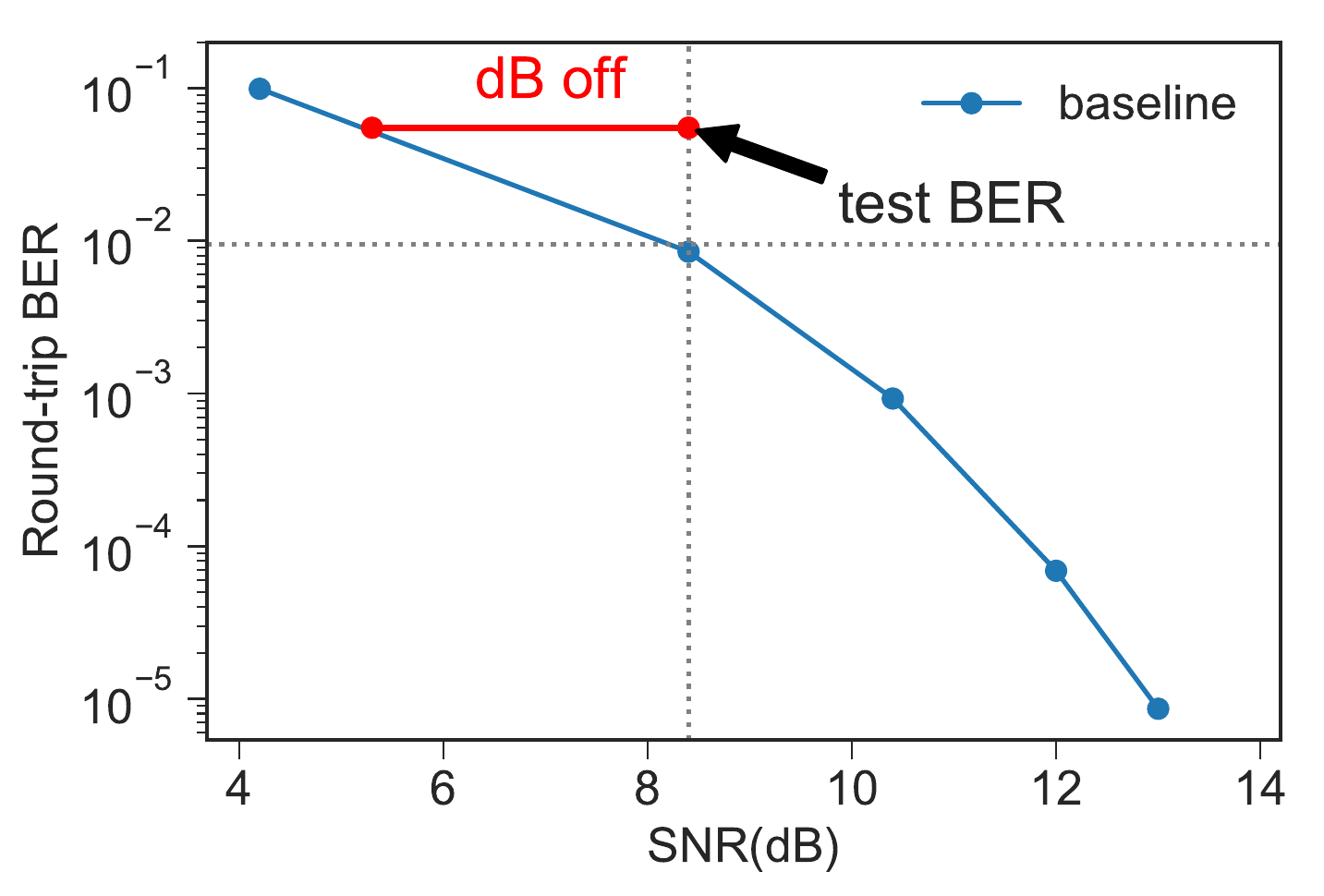}
    \caption{Example of \dboff{} calculation used to determine convergence for round-trip exchange.}
    \label{fig:db-off-calculation}
\end{figure}

To determine if the measured BER is indicative of good performance we measure the metric `dB off optimal', illustrated in Fig.~\ref{fig:db-off-calculation}.
To compute this metric we first measure the test BER achieved by our protocol at the SNR where the corresponding baseline scheme achieves 1\% BER. 
Then we compute the difference between this SNR(dB) and the 
minimum SNR(dB) required for the baseline scheme to achieve the measured BER. 
We measure this at different stages of the learning process corresponding to different numbers of preamble symbols transmitted. 
 
Using the round-trip BER and dB off optimal metrics we look at the following two graphs:
\begin{enumerate}
    \item \textbf{Round-trip BER vs SNR} \\
    Here we plot order statistics of the round-trip BER achieved by the learning protocol after it has converged or reached the maximum number of iterations allowed in our setup alongside that achieved by the baseline. This graph measures the limit of BER performance of our protocols subject to the maximum number of training iterations symbols we allow. Fig.~\ref{subfig:neurals-epp-ber} is a representative example.
    
    \item   \textbf{Fraction of trials that are 3~dB off optimal} \\
    Here we plot the fraction of trials that achieve \dboff{} values of less than 3 vs the number of preamble symbols transmitted. This tells us how robustly and how fast we are learning the modulation scheme. Fig.~\ref{subfig:neurals-epp-db3} is a representative example.
\end{enumerate}

Neither of these metrics is novel on its own; however, we are unaware of works which report both metrics. Our contribution is to combine these metrics to understand the performance of a learning protocol.



\section{Results}
\label{sec:results} 

In our experiments we consider the following agents: \Classic{}, \NF{}, \NS{}, \PF{}, and \PS{}.
The \NF{} and \NS{} agents (similarly \PF{} and \PS{}) are Self-Aliens; they share the same architecture but differ in learning rates and exploration parameters. They are named with -fast or -slow depending on their relative modulation scheme learning times using the \EPP{} protocol when paired with a clone.
To choose hyperparameters for our agents, we performed coarse hand-tuning for the \EPP{} protocol in the Agent-and-Clone setting and used the same hyperparameters for the \ESP{} and \LP{} protocols. The hyperparameters chosen this way were sufficient to obtain performance close to the baseline in terms of BER for all of these protocols in the Agent-and-Clone setting. However for the \GP{} protocol we observed that using the same hyperparameters as \EPP{} led to sub-optimal performance and thus we tuned parameters separately for the \GP{} case. While tuning hyperparameters, we further ensured that \PF{}-and-\PF{} had similar convergence times to \NS{}-and-\NS{}. It is known that hyperparameters matter, and finding good hyperparameters is hard (see for example \cite{DBLP:conf/iclr/NovakBAPS18}). However, we trust our conclusions about the relative performance of protocols in our experiments because we see order of magnitude differences across many trials. Hyperparameters for each agent are listed in
Appendix~\ref{app:simulation-hyperparams}.

We categorize the different pairings based on the different levels of alienness introduced in Section \ref{sec:aliens} as follows:
\begin{enumerate}
    \item Agent learning to communicate with its clone (Agent-and-Clone): \NF{}-and-\NF{}, \NS{}-and-\NS{}, \PF{}-and-\PF{}, \PS{}-and-\PS{}. 
    \item Agent learning to communicate with its self-alien (Agent-and-Self-Alien): \NF{}-and-\NS{}, \PF{}-and-\PS{}.
    \item Agent learning to communicate with an alien: \NF{}-and-\PS{}, \NF{}-and-\PF{}, \NS{}-and-\PS{}, \NS{}-and-\PF{}, and any agent with a \Classic{}.
\end{enumerate}

For the experiments in this section we use 2 bits per symbol and train at 8.4~dB SNR, corresponding to 1\% BER for the QPSK baseline. 
Tables~\ref{tab:convergence-info-gp-lp} and~\ref{tab:epp-esp-matchups} contain numerical results for our experiments on the effects of information sharing and alienness on the performance of modulation learning schemes. Sections~\ref{sec:results-info-sharing} and~\ref{sec:results-alieness} present additional figures and discuss the meaning of these results. 
Experiments on the effect of modulation order and training SNR on the performance of the \ESP{} and \EPP{} protocols can be found in Appendix~\ref{app:addresults}.

\begin{table*}[ht]
\begin{center}
\begin{tabular}{@{}lrrrr@{}}
\toprule
\textbf{} & 
\multicolumn{1}{l}{\textbf{\begin{tabular}[c]{@{}l@{}}Neural Mod, \\ \Classic{} Demod\end{tabular}}}  & 
\multicolumn{1}{l}{\textbf{\begin{tabular}[c]{@{}l@{}}Poly Mod, \\ \Classic{} Demod\end{tabular}}}  & 
\multicolumn{1}{l}{\textbf{\begin{tabular}[c]{@{}l@{}}Neural Mod, \\ Neural Demod\end{tabular}}} & 
\multicolumn{1}{l}{\textbf{\begin{tabular}[c]{@{}l@{}}Poly Mod,\\ Poly Demod\end{tabular}}} \\ \midrule
\textbf{Gradient Passing} & 2048 & 3328 & 2048 & 5632 \\
\textbf{Loss Passing} & 5888 & 9984 & 2816 & 20736  \\
\bottomrule
\end{tabular}
\end{center}
\caption{Number of symbols exchanged before $\geq 90\%$ of trials reached 3~dB off of optimal BER at 8.4~dB test SNR for the \GP{} and \LP{} protocols with various combinations of modulator and demodulator type and 2 bits per symbol. As expected, the results show learning using the \GP{} and \LP{} protocols to be fast (compared to \ESP{} and \EPP{} in Table~\ref{tab:epp-esp-matchups}). Furthermore, the \GP{} protocol converges faster than equivalent experiments using \LP{}. Gradients can carry more information than scalar loss values, so it makes sense for \GP{} to be a more effective learning protocol.}
\label{tab:convergence-info-gp-lp}
\end{table*}

\begin{table*}[ht]
\begin{center}
\begin{tabular}{lrrrrr}
\toprule
\hspace{0.5cm}Agent 2 & \multirow{2}{*}{\textbf{\Classic{}}} & \multirow{2}{*}{\textbf{\NF{}}} & \multirow{2}{*}{\textbf{\NS{}}} & \multirow{2}{*}{\textbf{\PF{}}} & \multirow{2}{*}{\textbf{\PS{}}} \\
 Agent 1 & & & & & \\
\midrule
\textbf{\ESP{}} & & & & \\
\textbf{\hspace{0.25cm}\NF{}} & 19456 & 25600 & - & 206336 & - \\
\textbf{\hspace{0.25cm}\PF{}} & 105472 & - & - & 347648 & - \\
\textbf{\EPP{}} & & & & \\
\textbf{\hspace{0.25cm}\NF{}} & 18432 & 115200 & 528384 & 404480 & 528384 \\
\textbf{\hspace{0.25cm}\NS{}} & 137728 & - & 528384 & 768000 & 690176 \\
\textbf{\hspace{0.25cm}\PF{}} & 107520 & - & - & 528384 & 690176 \\
\textbf{\hspace{0.25cm}\PS{}} & 176640 & - & - & - & 690176 \\
\bottomrule
\end{tabular}
\end{center}
\caption{Number of symbols exchanged before $\geq 90\%$ of trials reached 3~dB off of optimal BER at 8.4~dB test SNR for the \ESP{} and \EPP{} protocols with various agent types and 2 bits per symbol. An order of magnitude more symbols have to be exchanged before the learning agents converge compared to the \GP{} and \LP{} protocols in Table~\ref{tab:convergence-info-gp-lp}. Learning with a fixed agent (\Classic{}) is much easier than a clone learning agent, with convergence happening $1.3-3.3\times$ faster for \ESP{} and $3.8-6\times$ faster for \EPP{}. The extra shared information in \ESP{} seems to compensate for the increased difficulty of learning with another learning agent.}
\label{tab:epp-esp-matchups}
\end{table*}

\subsection{Effect of Information Sharing}
\label{sec:results-info-sharing}
In this first set of experiments, we explore the effect of information sharing on our learning protocols, seeking to quantify the value of shared information and understand the performance trade-off incurred by reducing shared information in the \ESP{} protocol. For this, we primarily consider the case of an agent learning to communicate with its clone. We choose this case because, in order to succeed at learning to communicate with others, one must first be able to communicate with (a copy of) oneself.

\begin{figure}[ht!]
     \subfloat[Round-trip median BER. The error bars reflect the \nth{10} to \nth{90} percentiles across 50 trials. All agents are evaluated at the same
SNR but error bars have been dithered for readability. ]{
 \includegraphics[width=1.0\linewidth]{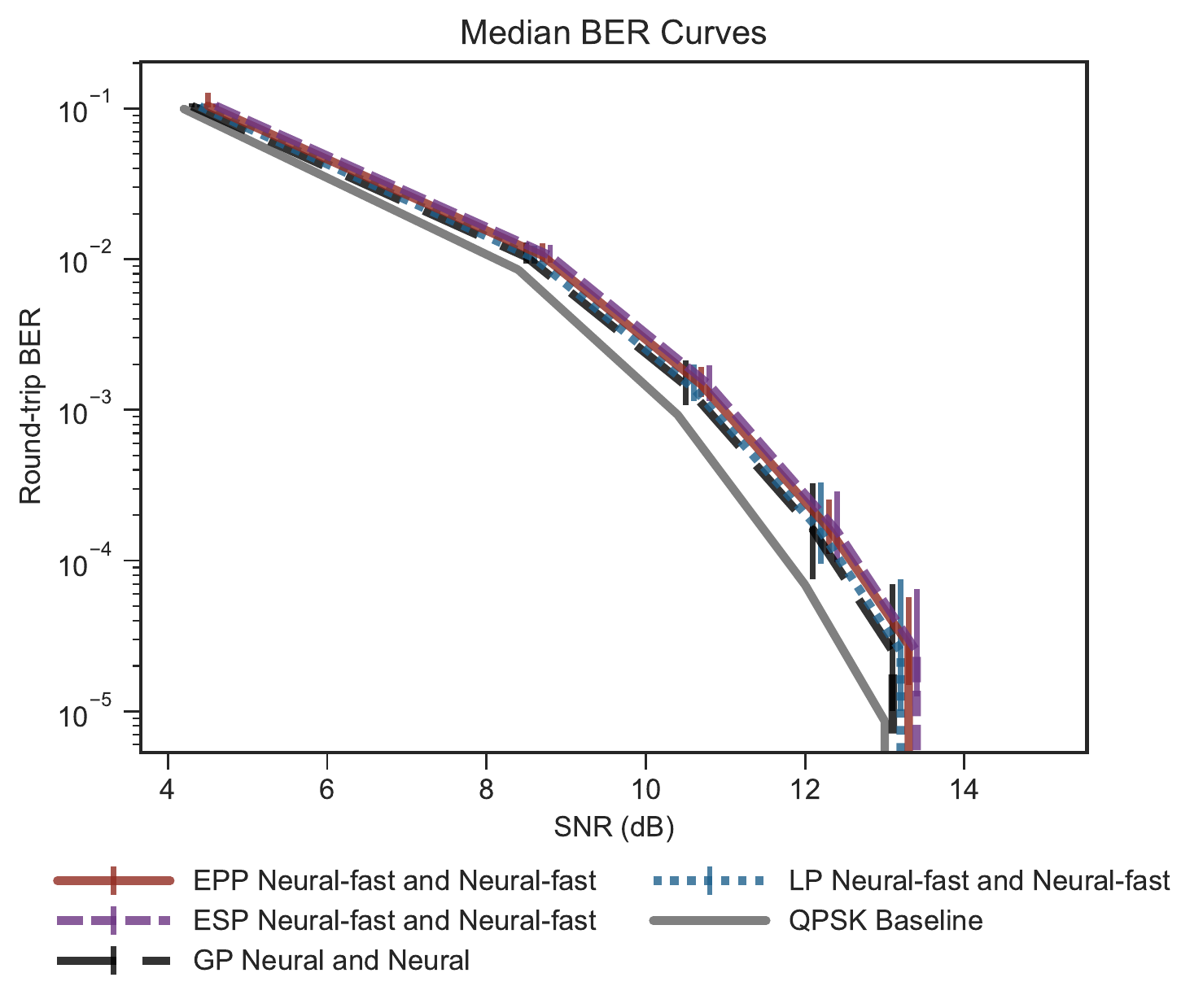}
 \label{subfig:neural-fast-neural-fast-allproto-ber}
 }\\
    \subfloat[Convergence of 50  trials  to be within 3~dB at testing SNR 8.4~dB.]{ \includegraphics[width=1.0\linewidth]{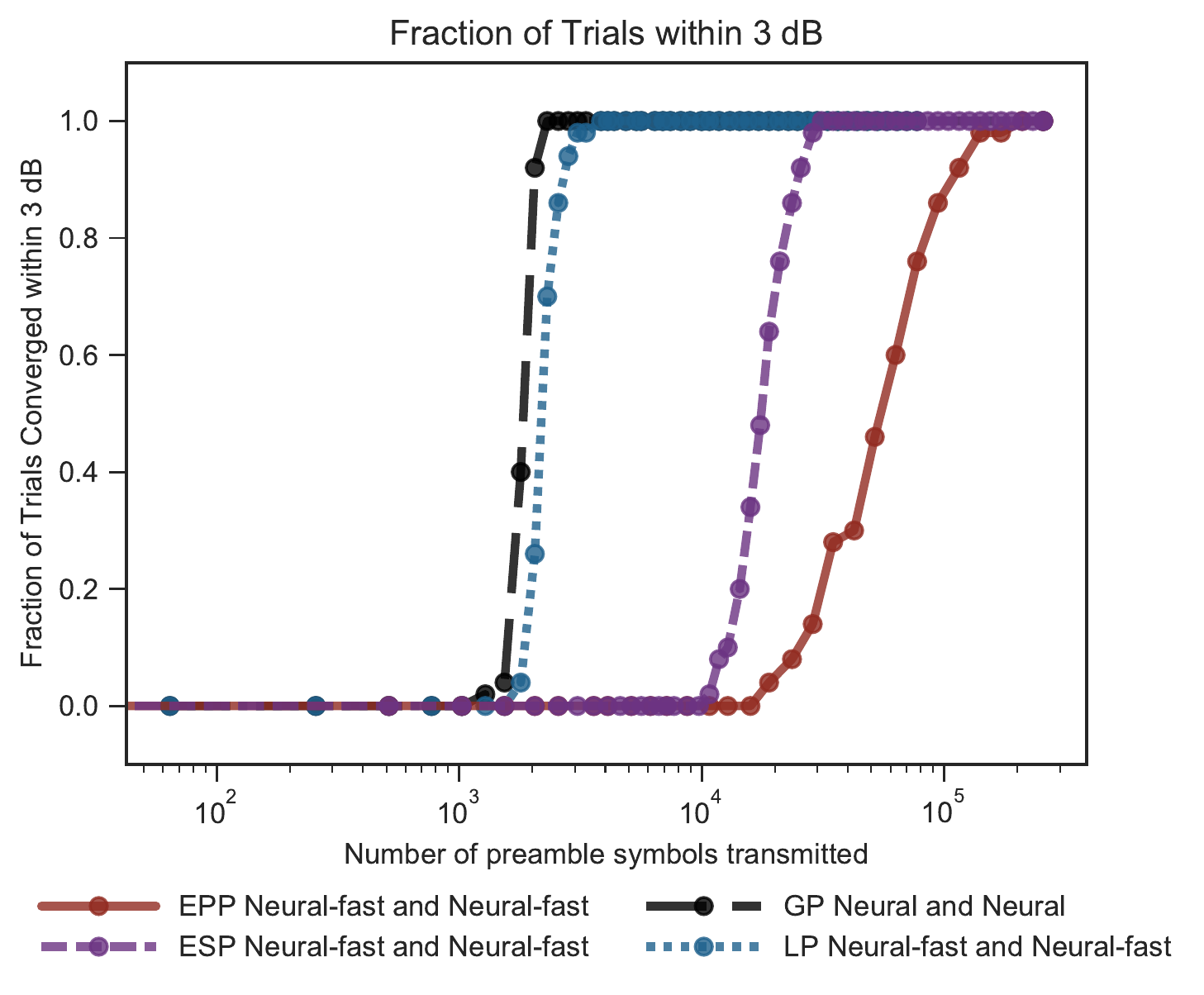}
     \label{subfig:neural-fast-neural-fast-allproto-db3}
    }
    \caption{Effect of Information Sharing, \NF-and-\NF: The BER plot (a) shows that all protocols achieve BER close to that of QPSK baseline. From the convergence plot (b), we observe that \EPP{} is much slower than \ESP{}, which in turn is an order of magnitude slower than \LP{} and \GP{}. Protocols with greater information sharing lead to faster convergence.}
    \label{fig:neural-fast-neural-fast-allproto}
\end{figure}

Fig.~\ref{fig:neural-fast-neural-fast-allproto} compares the performance of the \GP{}, \LP{}, \ESP{} and \EPP{} protocols for a \Neural{} agent learning to communicate with its clone. From the round-trip BER curves in Fig.~\ref{subfig:neural-fast-neural-fast-allproto-ber}, we observe that all protocols achieve similar values for the median BER and upper and lower percentiles. Furthermore, the median BER is close to the QPSK baseline. This is one of the main results of our work. \textbf{ \EPP{} can perform as well as \ESP{}, \LP{} and \GP{} and achieve performance close to an optimal baseline.}  From Fig.~\ref{subfig:neural-fast-neural-fast-allproto-db3} we observe that all protocols are robust, with the fraction of trials that converge going to 1 after sufficient preamble symbols are exchanged. The \EPP{} protocol needs the most preamble symbols to converge, followed by the \ESP{} protocol, and both these protocols take a much larger number of preamble symbols to converge than the \GP{} and \LP{} protocols. Thus, we conclude that with decreasing amount of information sharing it takes longer to learn to communicate, highlighting the value of shared information.

Tables~\ref{tab:convergence-info-gp-lp}~and~\ref{tab:epp-esp-matchups} tabulate the number of preamble symbols that have to be exchanged for more than 90\% of trials to converge within 3 dB-off-optimal for the different protocols. From these tables we see that there is an order of magnitude or more difference in the number of preamble symbols required between the protocols that use a side channel (\GP{} and \LP{}) and ones that don't (\ESP{} and \EPP{}).
We  performed similar experiments using \Poly{} agents and observed the same behavior, as shown in Fig.~\ref{fig:poly-fast-poly-fast-allproto}. These results are included in Appendix~\ref{app:addresults}.
    
In the rest of the experiments, we determine the effect of alienness on the \EPP{} protocol to address the universality of the \echo{} protocol.

\subsection{Effect of Alienness}
\label{sec:results-alieness}

We explore the effects of alienness with the following cases:
\begin{enumerate}
    \item Learning with fixed: An agent learning to communicate with a fixed agent. (\NF{}, \NS{}, \PF{}, \PS{} and \Classic{}.)
    \item Learning with clone: An agent learning to communicate with its clone. (\NF{} and \NF{}, \NS{} and \NS{}, \PF{} and \PF{}, \PS{} and \PS{}.)
    \item Learning with self-alien: An agent learning to communicate with its self-alien. (\NF{} and \NS{}, \PF{} and \PS{}.) 
    \item Learning with alien. An agent learning to communicate with an alien (\NS{} and \PF{}, \NF{} and \PS{}.)
\end{enumerate}
 We are primarily interested in answers to the following questions:
\begin{enumerate}
    \item Is it possible to learn to communicate with self-aliens and alien agents using the \EPP{} protocol? 
    \item Is it intrinsically more difficult to learn to communicate  with aliens or self-aliens than learning with clones?
    \item Can we say something about the performance of the \EPP{} protocol with alien agents based on the individual performances when trained with clones? (e.g Can we say something about the performance of \NS{}-and-\PF{} by looking at the performances of \NS{}-and-\NS{} and \PF{}-and-\PF{}?)
\end{enumerate}

\subsubsection{Learning with Fixed Agents}
\label{sssec:results-fixed-alien}
We first address the question of whether it is possible for learning agents to learn to communicate with fixed agents. This is important, since we are likely to encounter agents that use fixed modulation schemes in the real world and our learning agent must be compatible with them. 
To do this, we run experiments with a \Neural{} agent learning to communicate with a fixed \Classic{} agent, \NF-and-\Classic{}. After confirming that learning agents can work with fixed \Classic{} agents, we compare this to the case when a \Neural{} agent trains with another learning agent. In particular, we examine whether learning to communicate with a clone is harder than learning to communicate with a fixed agent, and whether increasing alienness (self-alien and completely alien) further increases the difficulty of the task.

\begin{figure}[ht!]
     \subfloat[Round-trip median BER. The error bars reflect the \nth{10} to \nth{90} percentiles across 50 trials. All agents are evaluated at the same
SNR but error bars have been dithered for readability. ]{
       \includegraphics[width=1.0\linewidth]{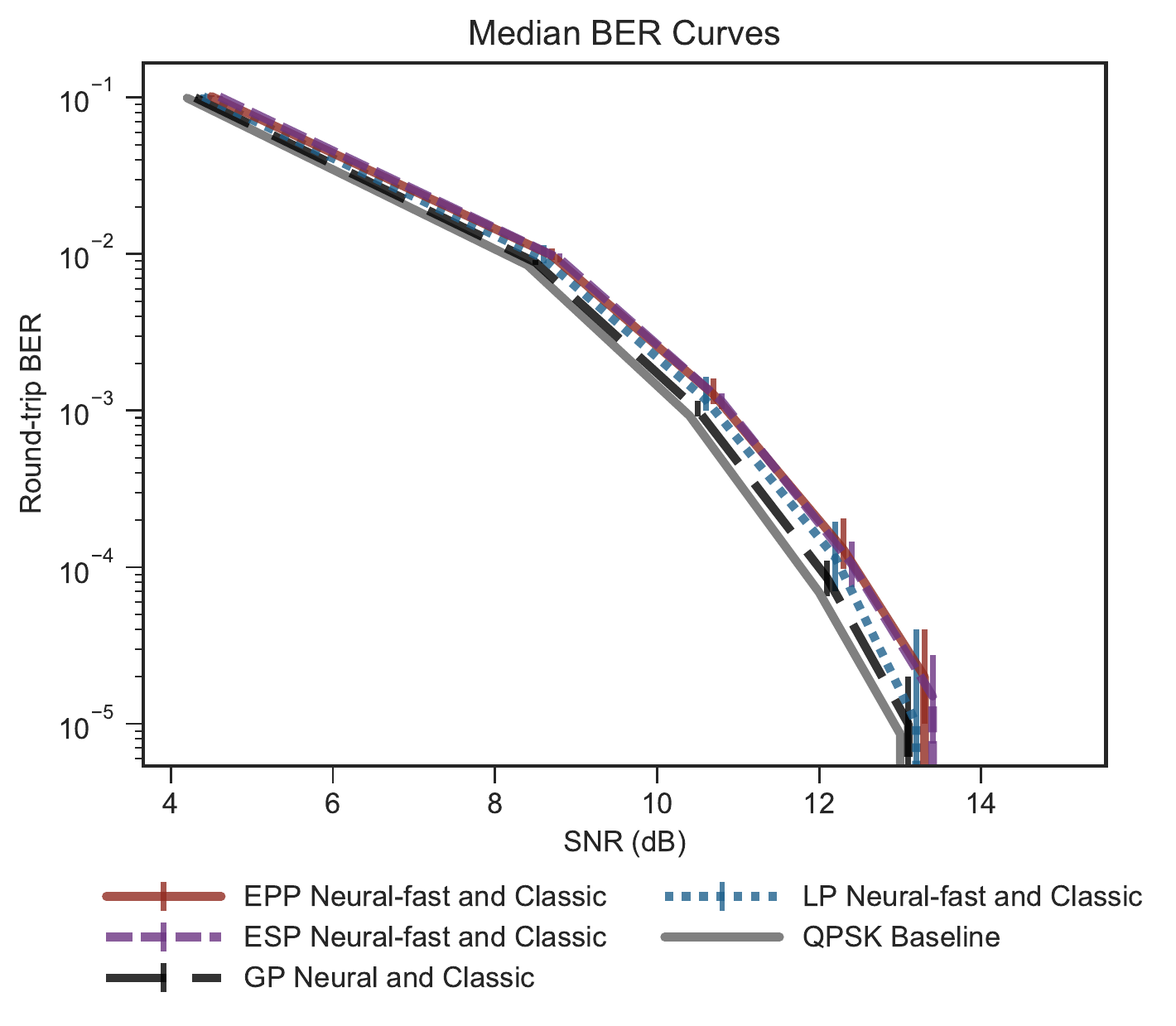}
     \label{subfig:neural-fast-classic-allproto-ber}
     }\\
        \subfloat[Convergence of 50  trials to be within 3~dB at testing SNR 8.4~dB.]{ \includegraphics[width=1.0\linewidth]{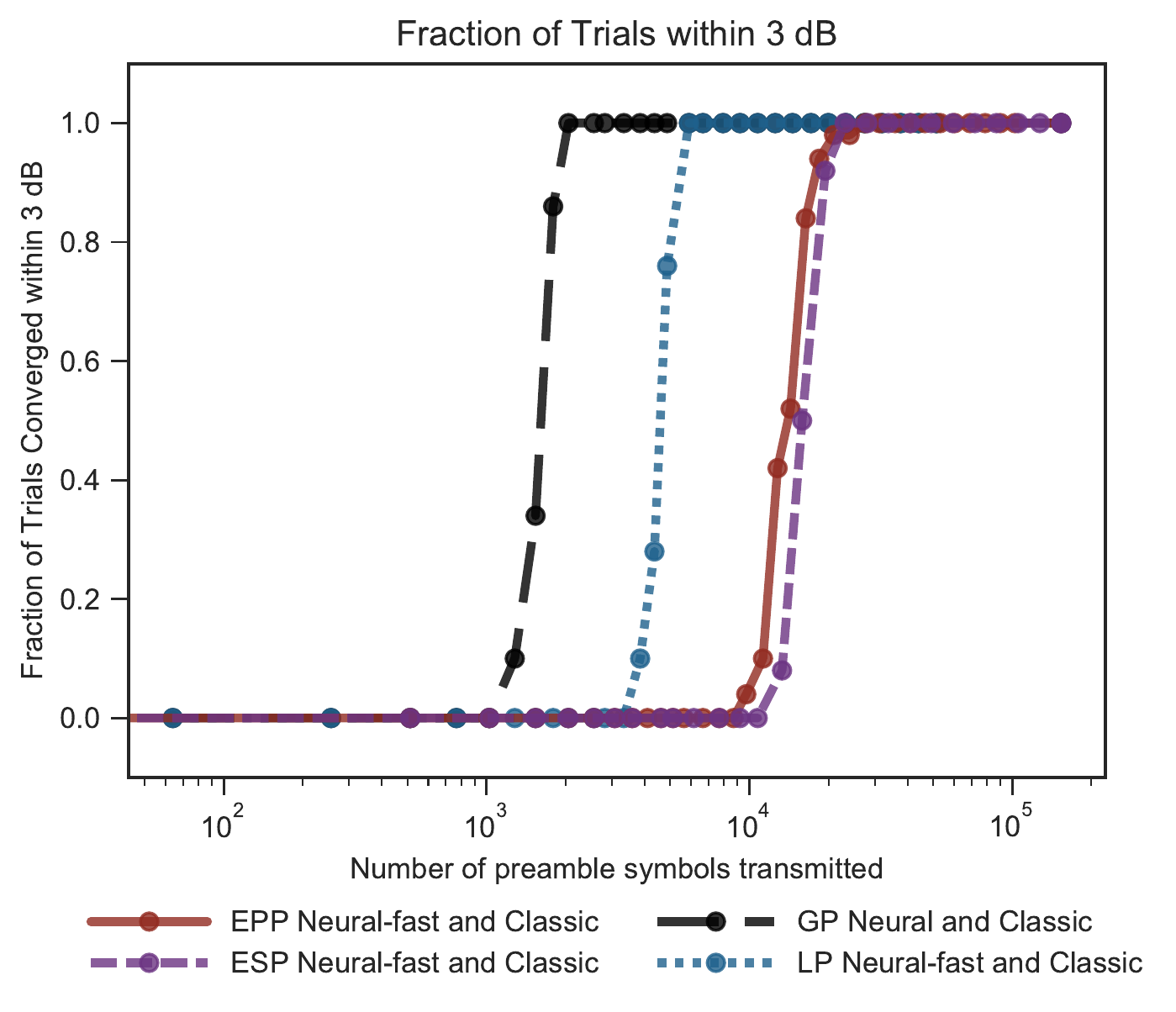}
         \label{subfig:neural-fast-classic-allproto-db3}
        }
    \caption{Learning to communicate with a fixed agent, \NF-and-\Classic{}: The BER plot (a) shows that all protocols achieve BER close to that of QPSK baseline. From the convergence plot (b), we observe that \EPP{} and \ESP{} have similar convergence behavior and are an order of magnitude slower than \LP{} and \GP{}. For the \ESP{} and \EPP{} protocol convergence is much faster than when learning with a clone (Fig.~\ref{fig:neural-fast-neural-fast-allproto})  }
    \label{fig:neural-fast-classic-allproto}
\end{figure}

Fig.~\ref{fig:neural-fast-classic-allproto} compares the performance of the \GP{}, \LP{}, \ESP{} and \EPP{} protocols for a \Neural{} agent learning to communicate with a \Classic{} agent. From the round-trip BER curves in Fig.~\ref{subfig:neural-fast-classic-allproto-ber}  we observe that all protocols achieve round-trip BER close to the QPSK baseline. Fig.~\ref{subfig:neural-fast-classic-allproto-db3} shows the \EPP{} and \ESP{} protocol have similar convergence behavior but are an order of magnitude slower than the \GP{} and \LP{} protocols. All protocols lead to robust convergence. Furthermore, comparing against Fig.~\ref{subfig:neural-fast-neural-fast-allproto-db3}, we see that learning to communicate with a \Classic{} agent is much easier than learning to communicate with a clone learning agent. Table~\ref{tab:epp-esp-matchups} shows a difference in convergence speed of up to $5.5\times$ between these two cases when using the \EPP{} and \ESP{} protocols. This matches what we expect intuitively, since when both agents are learning each agent is trying to improve its own behavior and \textit{simultaneously} track the behavior of the other agent. When one agent is fixed, the learning agent only has to match a static behavior. Graphical results for a \Poly{} agent learning to communicate with a \Classic{} agent can be found in Fig.~\ref{fig:poly-fast-classic-allproto} in Appendix~\ref{app:addresults}.

\subsubsection{Learning with Self-aliens}
\label{sssec:results-self-alien}

\begin{figure}[ht!]
     \subfloat[Round-trip median BER. The error bars reflect the \nth{10} to \nth{90} percentiles across 50 trials. All agents are evaluated at the same
SNR but error bars have been dithered for readability. ]{
         \includegraphics[width=1.0\linewidth]{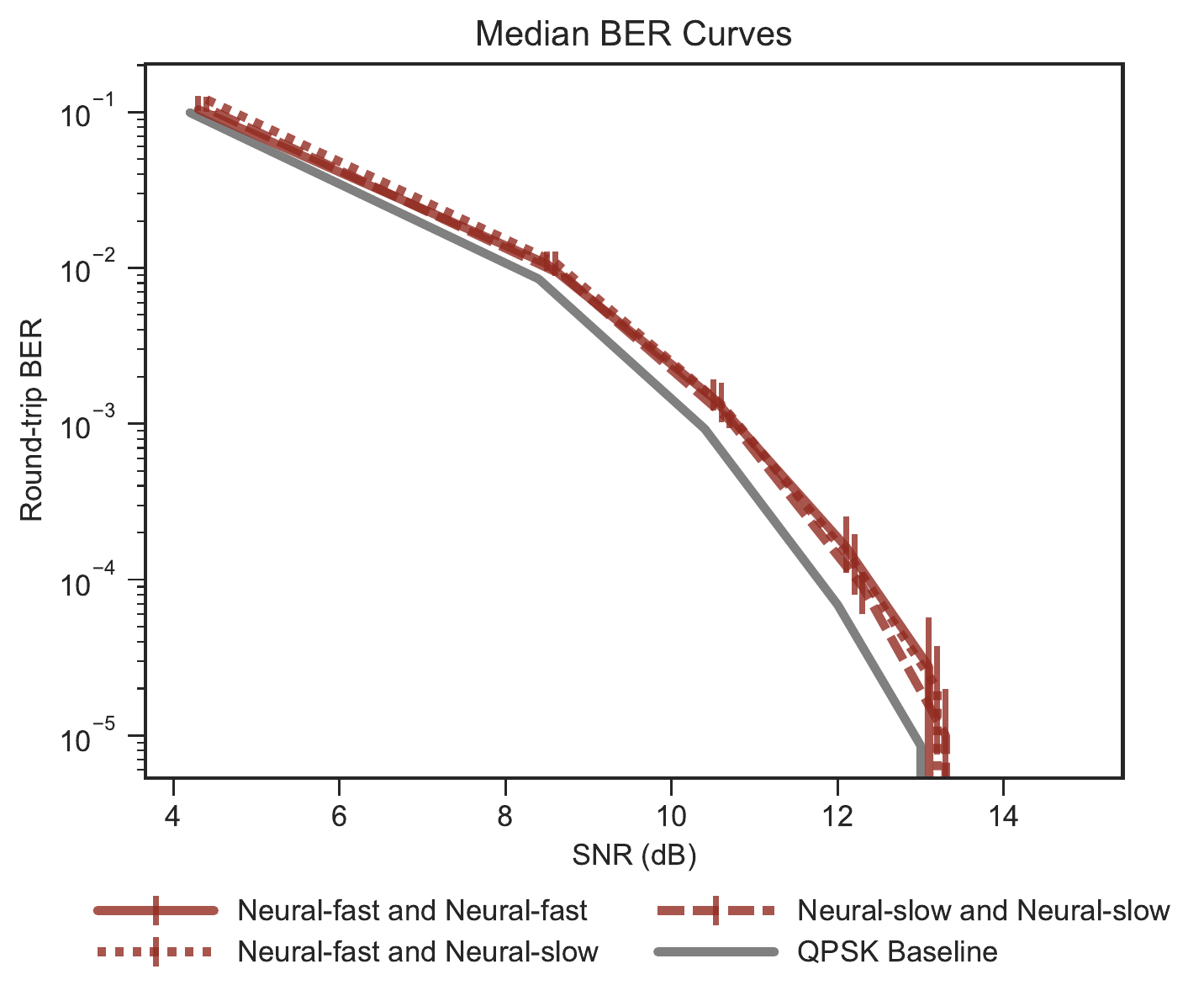}
         \label{subfig:neurals-epp-ber}
     }\\
    \subfloat[Convergence of 50  trials to be within 3~dB at testing SNR 8.4~dB.]{                \includegraphics[width=1.0\linewidth]{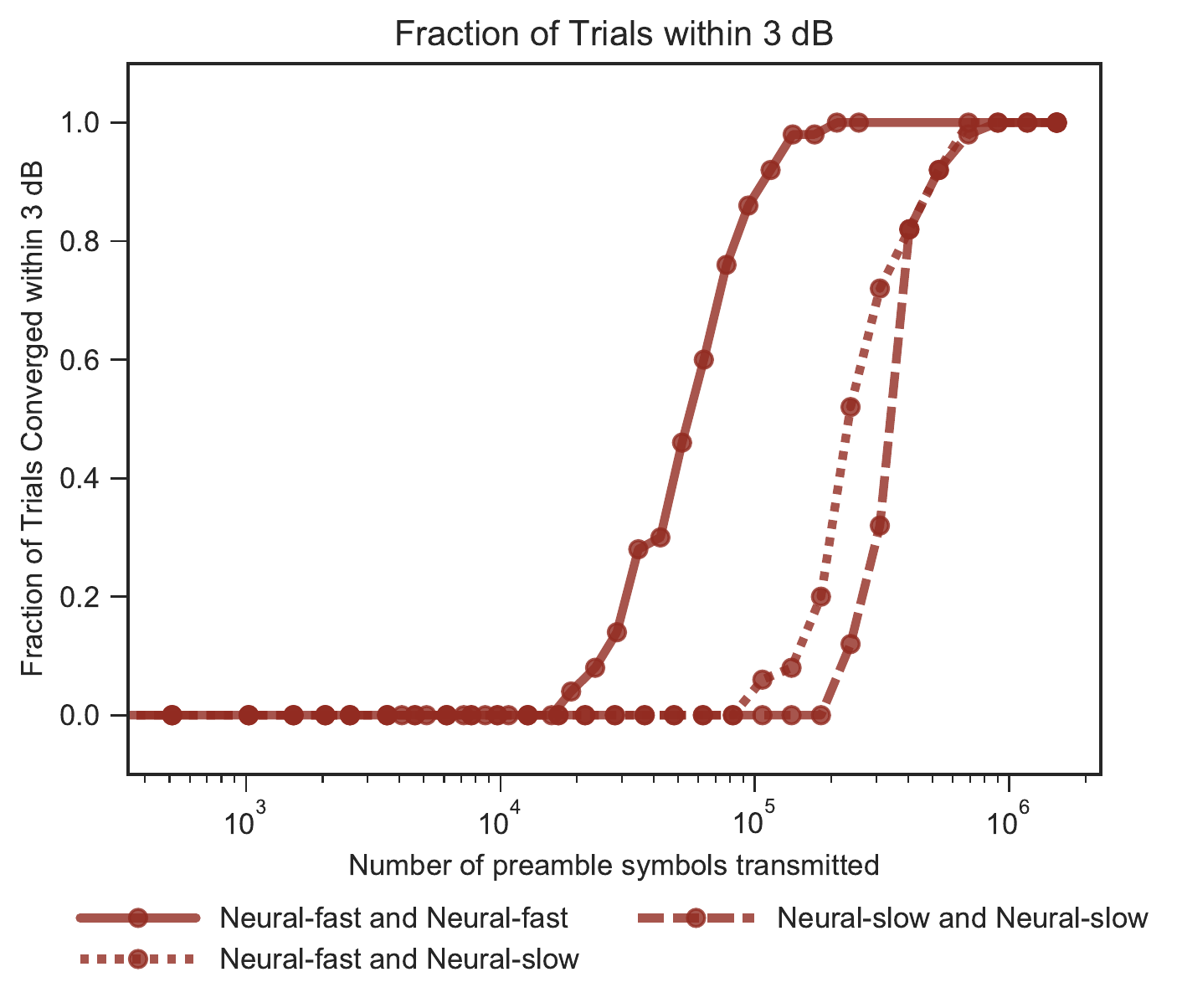}
        \label{subfig:neurals-epp-db3}
    }
    \caption{Learning with clones (\NF-and-\NF, \NS-and-\NS) compared to learning with self-alien (\NF-and-\NS) using the \EPP{} protocol. The BER plot (a) shows that the round-trip BER for \NS-and-\NS{} is slightly lower than the others but all BERs are close to the QPSK baseline. From the convergence plot (b), we observe that \NF-and-\NF{} is much faster than \NS-and-\NS. However, pairing \NF{} with \NS{} helps the latter learn faster, resulting in a convergence time for \NF-and-\NS{} between those of the clone pairings.}
    \label{fig:neurals-epp}
\end{figure}

\begin{figure}[ht!]
     \subfloat[Round-trip median BER. The error bars reflect the \nth{10} to \nth{90} percentiles across 50 trials. All agents are evaluated at the same
SNR but error bars have been dithered for readability. ]{
 \includegraphics[width=1.0\linewidth]{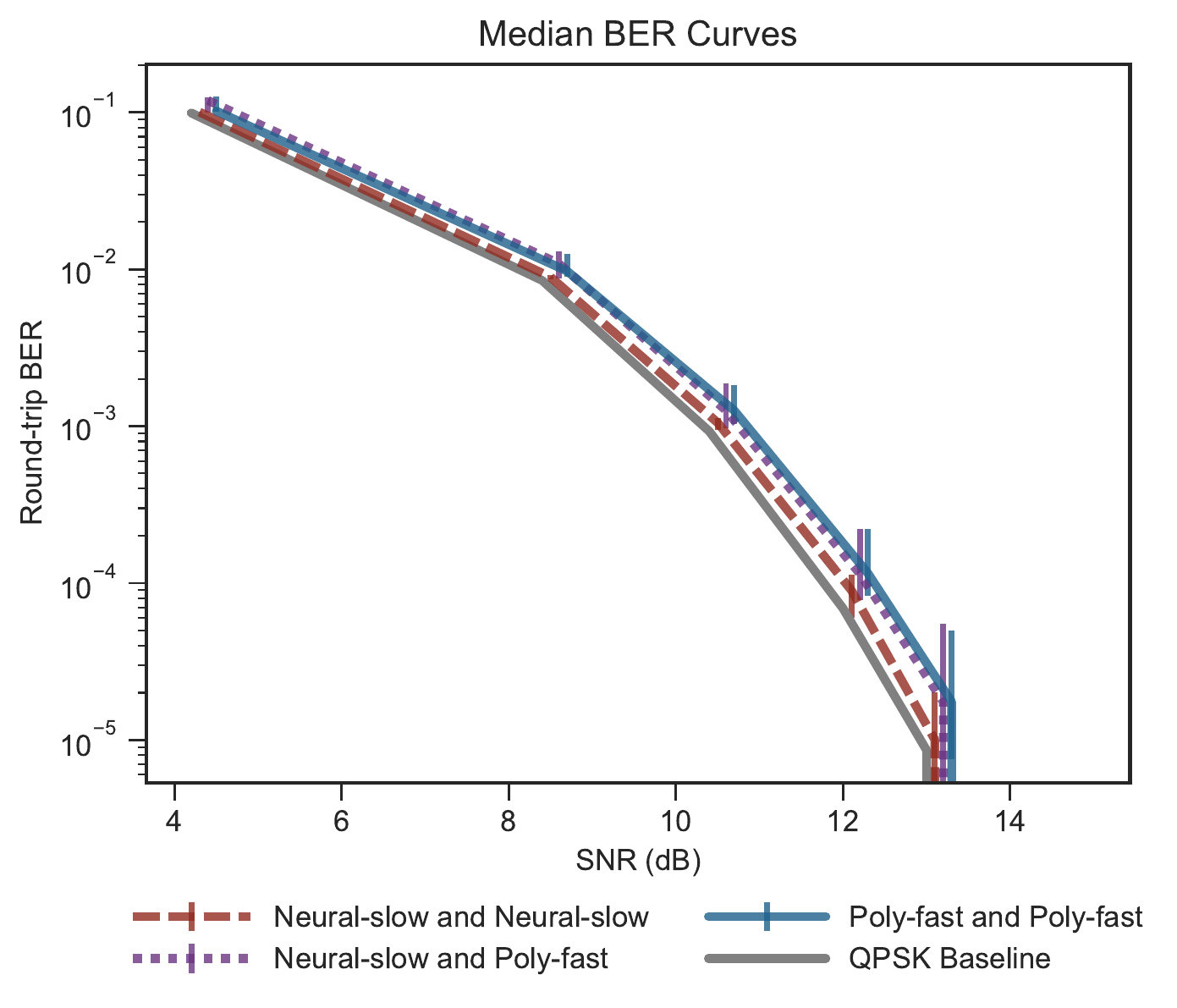}
 \label{subfig:neural-poly-cross-epp-ber}
 }\\
    \subfloat[Convergence of 50  trials to be within 3~dB at testing SNR 8.4~dB.]{ \includegraphics[width=1.0\linewidth]{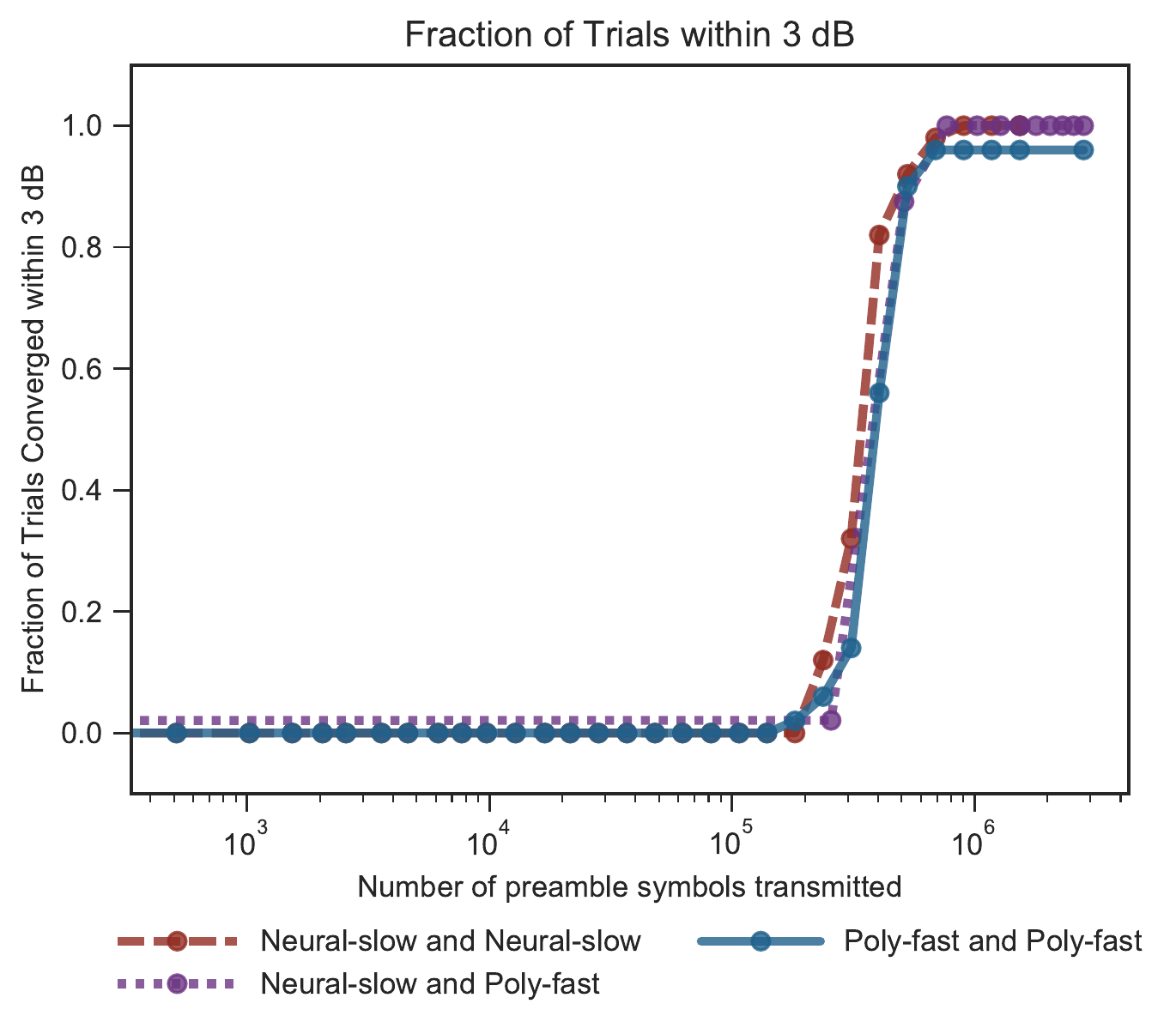}
     \label{subfig:neural-poly-cross-epp-db3}
    }
    \caption{Learning with clones (\NS-and-\NS, \PF-and-\PF) compared to learning with an alien (\NS-and-\PF) using the \EPP{} protocol. Here the \NS{} and \PF{} agents have similar convergence behavior when paired with a clone. The BER plot (a) shows that in all cases, the BER achieved is close to the QPSK baseline. From the convergence plot (b), we observe that when both clone parings show similar convergence behavior, the alien pairing also has the same behavior. Learning to communicate with an alien is not intrinsically more difficult than learning to communicate with a clone.}
    \label{fig:neural-poly-cross-epp}
\end{figure}

\begin{figure}[ht!]
     \subfloat[Round-trip median BER. The error bars reflect the \nth{10} to \nth{90} percentiles across 50 trials. All agents are evaluated at the same
SNR but error bars have been dithered for readability.]{
        \includegraphics[width=1.0\linewidth]{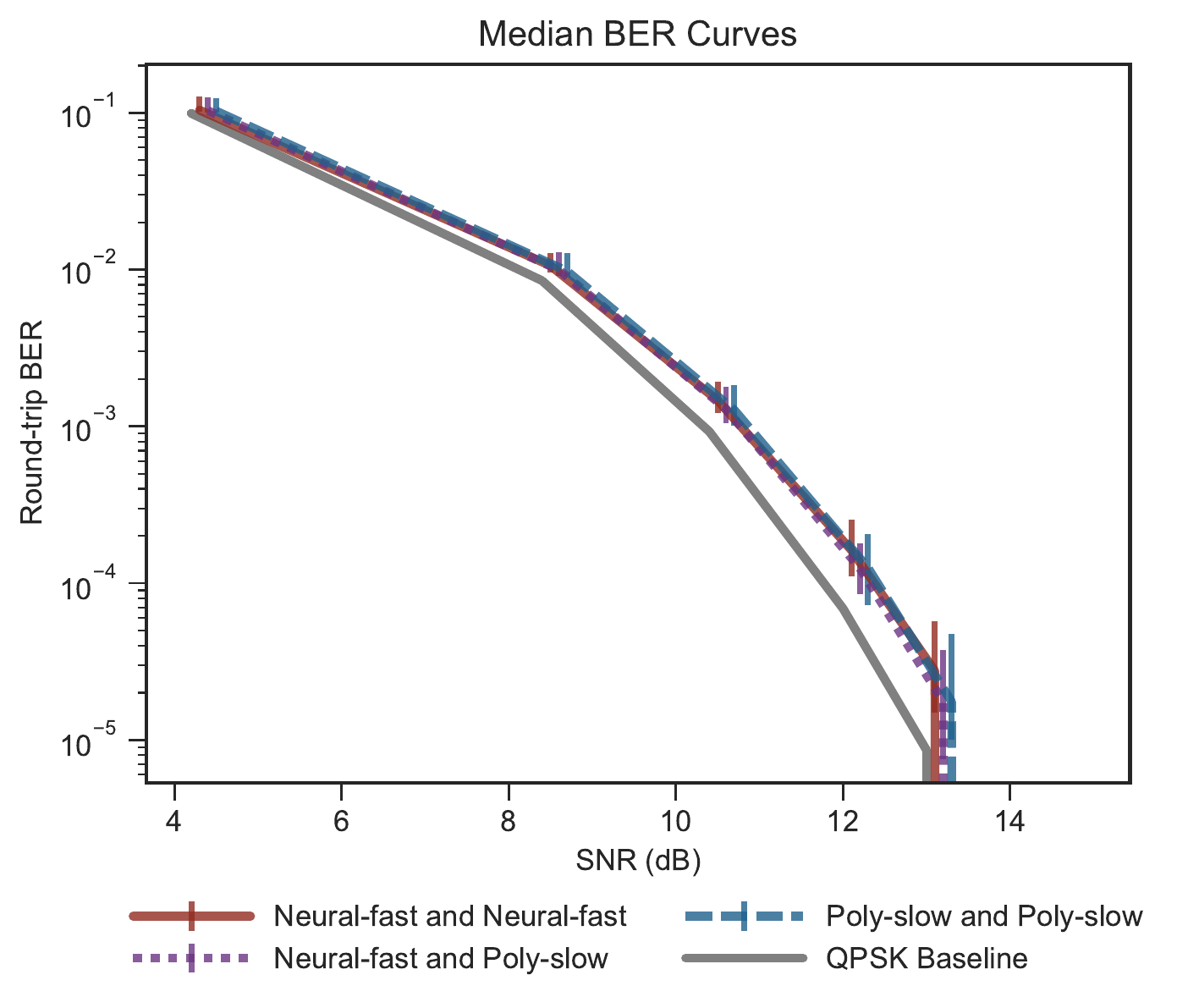}
         \label{subfig:neural-poly-cross-gap-epp-ber}
    }\\
    \subfloat[Convergence of 50 trials to be within 3~dB at testing SNR 8.4~dB.]{ 
        \includegraphics[width=1.0\linewidth]{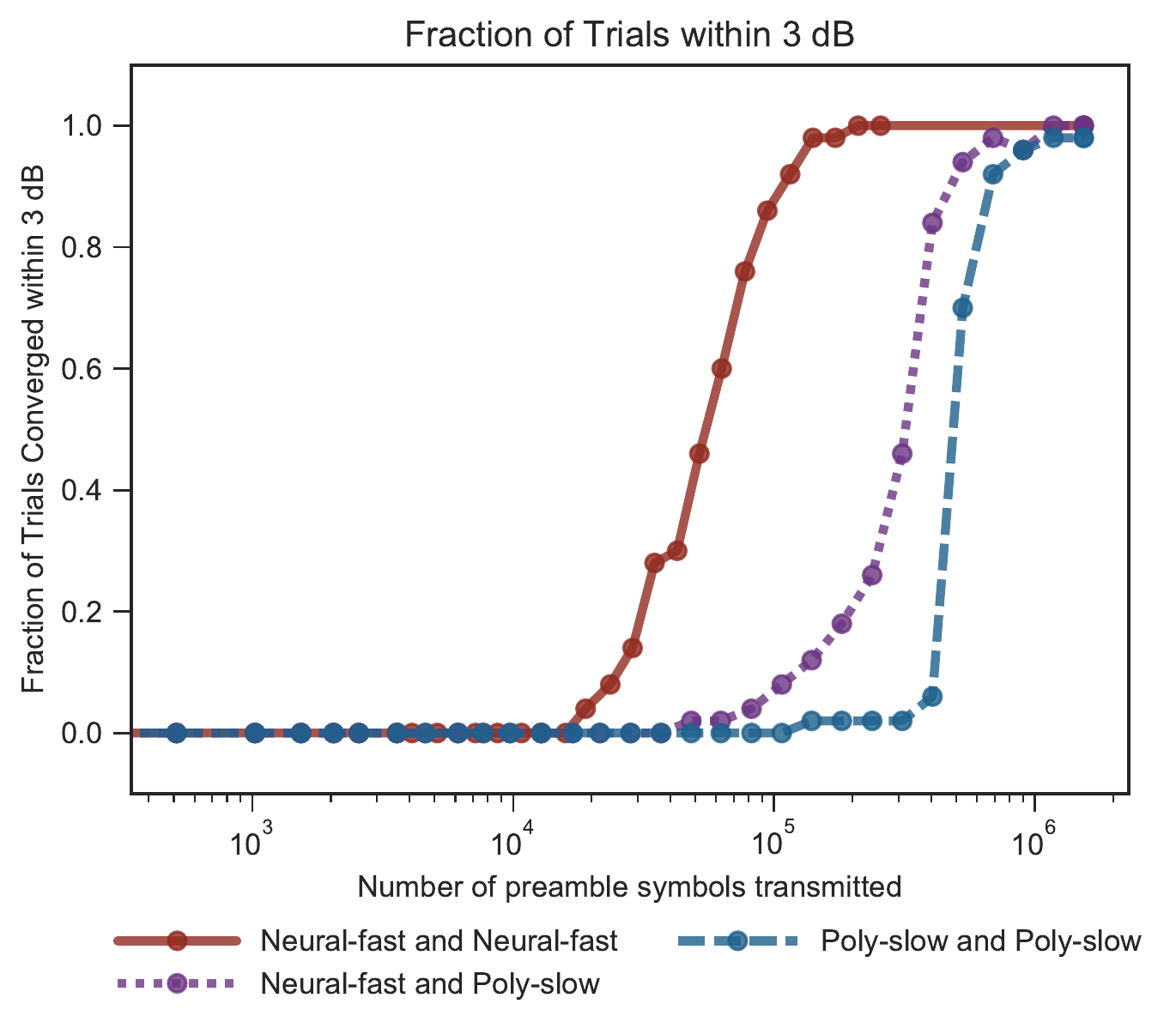}
        \label{subfig:neural-poly-cross-gap-epp-db3}
    }
    \caption{Learning with clones (\NF-and-\NF, \PS-and-\PS) compared to learning with an alien (\NF-and-\PS) using the \EPP{} protocol. Here the \NF{} and \PS{} agents have vastly different convergence behavior when paired with a clone. The BER plot (a) shows that in all cases, the BER achieved is close to the QPSK baseline. From the convergence plot (b), we observe that when the clone pairings have vastly different convergence behavior, the alien pairing shows convergence behavior somewhere in between. All alien pairings converged to within 3~dB off optimal, even though the \PS{} with clone failed at least once. The robustness of the \NF{} learner may have helped the \PS{} agent converge more reliably.}
    \label{fig:neural-poly-cross-gap-epp}
\end{figure}

Fig.~\ref{fig:neurals-epp} compares learning with clones to learning with a self-alien for \Neural{} agents using the \EPP{} protocol. Here we have one ``fast'' agent and one ``slow'' agent, defined based on speed of convergence when paired with a clone. From the BER curves in Fig.~\ref{subfig:neurals-epp-ber} we see that all cases achieve round trip accuracy close to the QPSK baseline. From the convergence plot in Fig.~\ref{subfig:neurals-epp-db3} we make an interesting observation. The fast agent helps the slower agent to learn more quickly, resulting in convergence times for the self-alien pairing in between those of the clone pairings. This is very encouraging since it suggests that not only is learning with self-aliens possible, it can also be faster than learning with clones for a slow agent.  We repeated this experiment using \Poly{} agents and found similar behavior, shown in Fig.~\ref{fig:polys-epp} in Appendix~\ref{app:addresults}.

\subsubsection{Learning with Aliens}
\label{sssec:results-full-aliens}

Next we compare learning with aliens to learning with clones. Fig.~\ref{fig:neural-poly-cross-epp} depicts the results for the case where a \Neural{} and \Poly{} agent that show similar convergence behavior when learning with a clone are paired with each other. From the BER curve in Fig.~\ref{subfig:neural-poly-cross-epp-ber} we see that the \Neural{} agent when paired with a clone has slightly lower BER than the \Poly{} agent paired with clone, but more importantly the BER for the alien pairing is very similar to the others and close to the QPSK baseline. From the convergence plot in Fig.~\ref{subfig:neural-poly-cross-epp-db3} we see that when the two clone pairings show similar convergence behavior the alien pairing does not deviate. This is another main result of our work. \textbf{Learning to communicate with an alien agent using \EPP{} is not intrinsically more difficult than learning with a clone}. 

In the next experiment we investigate whether it is possible for an agent to learn to communicate with an alien agent when the two agent types have vastly different convergence behaviors while learning with clones. From the BER curve in Fig.~\ref{subfig:neural-poly-cross-gap-epp-ber} we see that the two clone pairings have round trip error rates close to the QPSK baseline. Interestingly,  Fig.~\ref{subfig:neural-poly-cross-gap-epp-db3} shows that the alien pairing always learned a good modulation scheme even though the \PS{}-and-clone pairing occasionally failed. The alien pairing convergence speed lies in between the two clone pairings. These results suggest that a fast, reliable agent can help a slow agent, alien or not, learn more quickly and more robustly. This phenomenon is not unique to the \EPP{} protocol, as can be seen from results with the \ESP{} protocol in Fig.~\ref{fig:esp-combos} in Appendix~\ref{app:addresults}. Another interesting result from this experiment found in Table~\ref{tab:epp-esp-matchups} is that for \ESP{} the difference between convergence times for  \Neural{}-and-\Classic{} and \Neural{}-and-Clone (and similarly for \Poly{}-and-\Classic{} and \Poly{}-and-Clone) is smaller than for \EPP{}. This could be partially because the hyperparameters were tuned for the \EPP{} agent-and-clone setting, but also suggests that as we increase information sharing the performance gap between learning with fixed agents and other learning agents shrinks.

We can now provide answers to the questions we raised at the start of this subsection. It is possible to learn to communicate with self-aliens and aliens using the \EPP{} protocol. In fact, neither of these tasks is intrinsically more difficult than learning with a clone. Furthermore, a self-alien/alien pairing shows convergence behavior in between the two clone pairings. A fast agent paired with a slower agent can help the slow agent learn faster. An interesting experiment would be to map out the range of conditions where these observations continue to hold. Can learning become impossible if the difference in convergence behavior of the two agents is large enough? Can two agents have convergence behaviors that are similar, but differ so fundamentally in the way they learn that they fail to learn in the alien setting? We leave this as an area for future research.



\section{Implementation in Software Defined Radios}
\label{sec:implementation-real-radios}

In order to corroborate our simulation results, we implement the \ESP{} (\ref{sec:information_sharing}) and \EPP{} (\ref{subsec:procedure_metaprotocol}) protocols on Ettus USRP software defined radios using GNU Radio \cite{gnr:gnu-radio}. The goal of this implementation is not to provide a real-time implementation of the \echo{} protocol, since in general the real-time components of radio communications are implemented in ASICs, and even software components are run in special real-time operating systems to achieve deterministic or bounded latencies. The focus of our work is to learn modulation schemes, so the primary goal of the GNU Radio implementation  is to  demonstrate that the learning protocols work not only in simulations but also when trained in real, physical systems. Other work such as  \cite{sync:2019arXiv190510468S} and \cite{cae:DLCommOverAir18} have also demonstrated that end-to-end learning of communication schemes is possible over the air in real radio systems. 
We plan to address other components of radio communications such as channel equalization and error correction coding in future works. Only after all of these processing components have been addressed will it be necessary to have real-time hardware implementations of components such as the modulation learning.

\subsection{Additional Processing}
\label{subsec:gnuradio-wrapper}
The GNU Radio implementation attempts to abstract away the details of packet transmission, reception, and non-AWGN channels in order to provide as close an approximation as possible to the training environment of the previous sections. The implementation corrects for carrier frequency offset (CFO), multitap channels, and arbitrary packet arrival times using several algorithms implemented with NumPy \cite{gnr:numpy}. We detect packets using correlation against a fixed prefix and constant false alarm rate detection \cite{Barrett1987}. CFO and channel effects are corrected using the same prefix. We perform coarse sample timing synchronization by upsampling to two samples per symbol for transmission, then downsampling after the start of the packet has been detected.

The additional processing adds significant overhead to each round-trip training cycle. The results from a typical run with a 50-unit single hidden layer modulator and demodulator and 256 symbols per preamble are shown in Table~\ref{tab:gnr-overhead}. As shown in the table, the packet wrapper comprises more than one third of the execution time during a run. In addition to the computation time, the GNU Radio implementation introduces latency by sending data between packet processing blocks and modulator or demodulator blocks. 
\begin{table}[ht]
\begin{center}
\begin{tabular}{@{}lrrr@{}}
\toprule
\textbf{} & \thead[l]{Neural \\ Modulator} & \thead[l]{Neural \\ Demodulator} & \thead[l]{Packet \\ Processing} \\ \midrule
\textbf{\begin{tabular}[c]{@{}l@{}}Median Processing \\ Time (ms)\end{tabular}} & 13.4 & 26.1 & 20.6 \\
\textbf{\begin{tabular}[c]{@{}l@{}}Percent of Execution \\ Time\end{tabular}} & 22.3 & 43.3 & 35.8 \\ \bottomrule
\end{tabular}
\end{center}
\caption{Average execution times for \Neural{} agent training update and GNU Radio wrapper processing. The additional processing required for transmission over USRP radios is about $1/3$ of the total execution time. These times do not account for the additional latency of moving data between components of the GNU Radio processing chain.}
\label{tab:gnr-overhead}
\end{table}

\subsection{Training Procedure Modifications}
\label{subsec:gnuradio-modifications}
Constraints introduced by running on physical radios required several changes to the Neural agent training procedure before we could successfully train these agents. The constraints and the modifications necessary to overcome them are detailed in Sections~\ref{subsubsec:gnr-maximum-tx-amp} and \ref{subsubsec:gnr-dc-offset}.

\subsubsection{Maximum Transmit Amplitude}
\label{subsubsec:gnr-maximum-tx-amp}

Signals sent through USRP radios cannot exceed a maximum amplitude, and any signals sent to the radio which exceed this amplitude are silently clipped to the maximum amplitude. However, the \EPP{} implementation in our simulations only restricts the average energy of a constellation. This means that any individual constellation point can have almost arbitrarily large amplitude, and exploration can drive the amplitude of a transmitted symbol even higher. It turns out that clipping a significant number of transmitted symbols breaks the training process for neural modulators, and they never converge to a reasonable constellation. In order to prevent clipping, we restrict the average power of a constellation during training to significantly less than the radio's cap, and rely on the vast majority of symbols which are not clipped to produce good training feedback. 

Because we control the environment for our tests, we can ensure that we train and test at the desired SNRs for any given constellation. However, in the real world a system may need to use all of its transmit power to achieve a usable SNR. In such a case, restricting the average power of a constellation to less than the maximum would prevent learning from taking place. We hope to address the problem of exploring out to a bounding box, without exceeding it, while maintaining training performance in future work.

\subsubsection{DC Offset Correction}
\label{subsubsec:gnr-dc-offset}

USRP radios use an adaptive DC offset canceler in the receive chain which causes the IQ that the demodulator eventually receives to be centered around the origin, regardless of the originally transmitted constellation. However, the base \echo{} implementation does not place any restriction on the mean of a constellation. The most energy efficient constellation possible is always centered at the origin, so the constellations achieved after training are approximately centered at the origin as well. Unfortunately, the constellation center commonly moves far from the origin during the training process before being forced back as the constellation is optimized. This causes a significant DC offset in the transmitted signal. The receive chain DC offset corrections change the round-trip feedback that a modulator receives significantly enough that neural modulators fail to train. 

The adaptive DC offset cancellation can be disabled, but this would require a calibration period at the start of each run, or even after each received packet, to measure the true DC offset and set the DC offset canceler manually. Instead, we explored methods of forcing the constellations to be approximately centered while training. We settled on a loss term for the squared magnitude of the constellation center --- this was done individually at each agent and so did not violate the spirit of the problem. See Appendix~\ref{app:constellation-centering} for more details.

\subsection{Experiments}
\label{subsec:gnuradio-experiments}
\begin{figure}[ht]
    \centering
    \includegraphics[width=\linewidth]{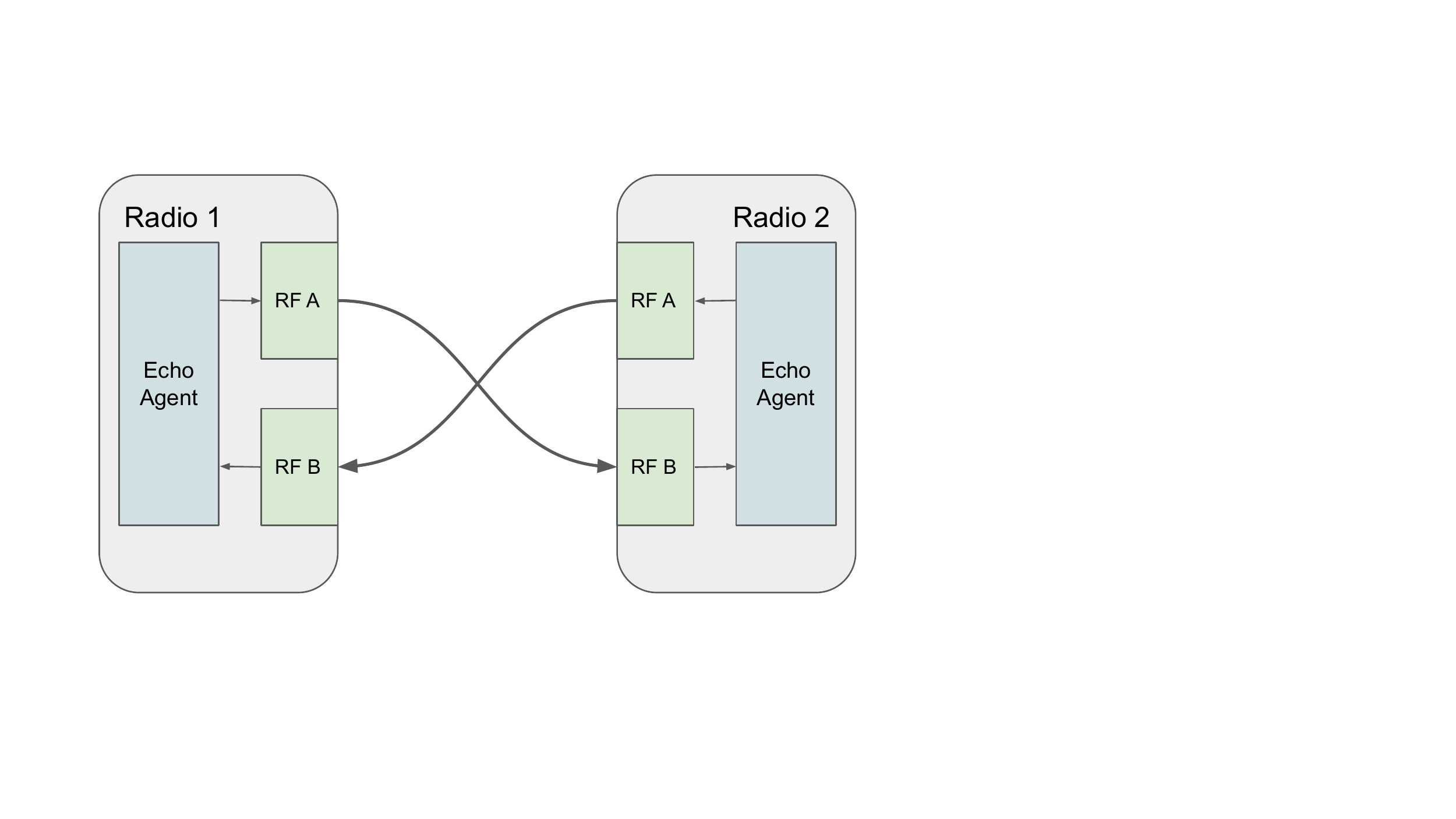}
    \caption{USRP radio connections.}
    \label{fig:usrp-setup}
\end{figure}

The radio experiments were conducted using two Ettus USRP X310 software defined radios (SDRs) connected to each other with SMA cables as shown in Fig.~\ref{fig:usrp-setup}. 75~dB of attenuation was added between the radios both to simulate path loss and to allow us to achieve desired SNRs with the available internal transmit and receive gains.

We tuned hyperparameters for the radio experiments separately from the main simulation hyperparameters because of the extra hyperparameter introduced for DC offset correction. We use these same hyperparameters in simulations when comparing with the radio experiment results. After coarse hand-tuning we achieved performance similar to \NS{}-and-clone. The hyperparameters are listed in Appendix~\ref{app:gnr-hyperparams}. Each experiment was run 20 times with random seeds at an SNR which resulted in 1\% round-trip BER for two classic agents (Section~\ref{subsec:alien-classic}).

\subsubsection{Echo with Shared Preamble Comparison}
\label{subsubsec:gnr-shared-preamble-comparison}

\begin{figure}[ht]
    \centering
    \includegraphics[width=\linewidth]{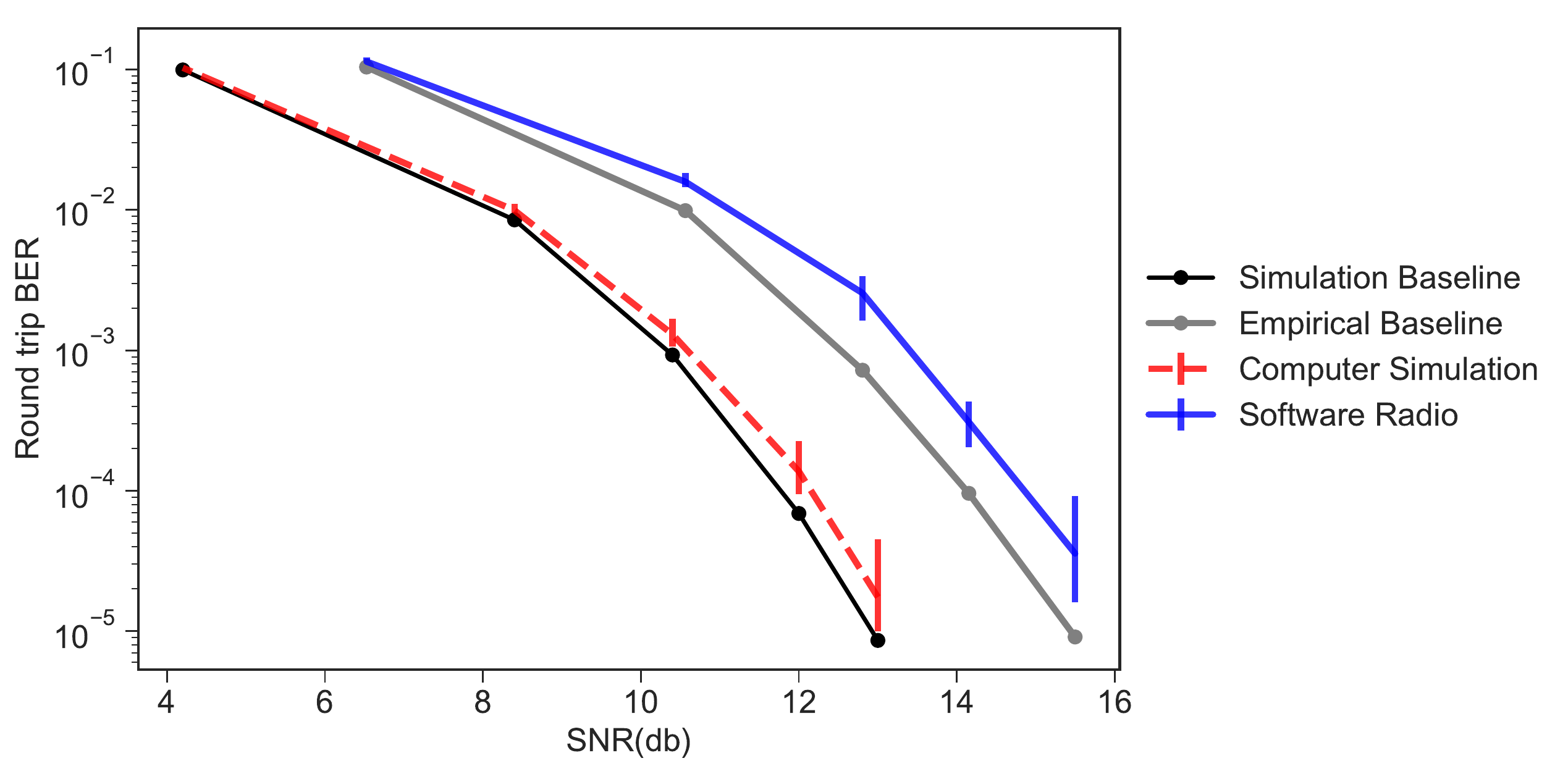}
    \caption{ Round-trip median bit error curves for \Neural{}-and-Clone python simulation and GNU Radio agents learning QPSK under the \ESP{} protocol at training SNRs corresponding to 1\% BER. Alongside the bit error curves of the learned modulation schemes is the baseline. In all cases, modulation constellations are constrained as detailed in Section~\ref{subsec:gnuradio-modifications}.  Although 2~dB SNR extra is required to achieve the same baseline performance due to processing losses in the GNU Radio implementation, the trained agents show only slightly greater loss in performance against the baseline than the pure simulation agents.}
    \label{fig:gnr-shared-nn-snr_ber}
\end{figure}

\begin{figure}[ht]
    \centering
    \includegraphics[width=\linewidth]{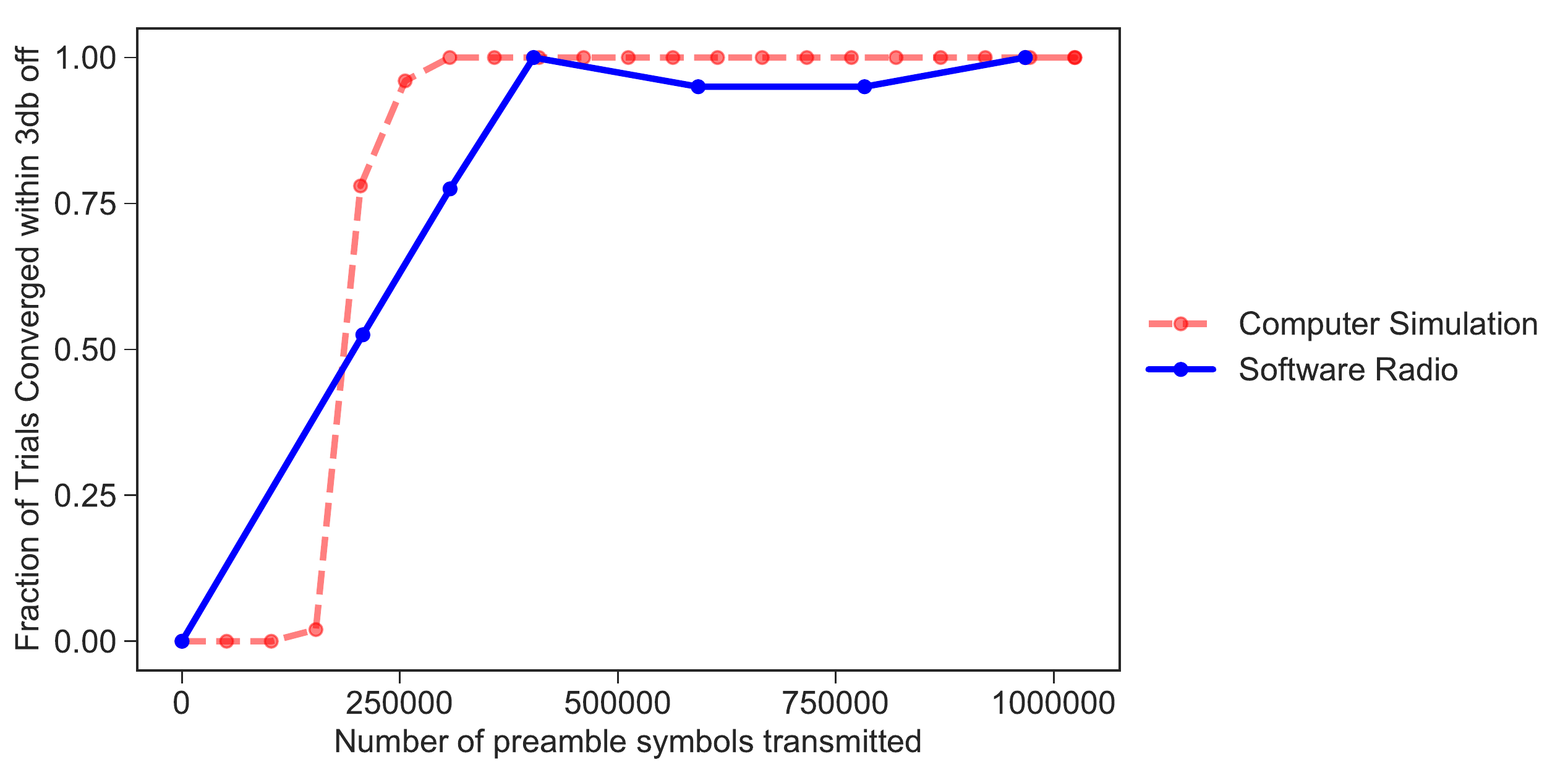}
    \caption{Convergence of 50 simulation and 20 GNU Radio trials to be within 3~dB (at testing SNR corresponding to 1\% BER) of the corresponding baseline for \ESP{} trials of neural agent and clone at training SNR corresponding to 1\% BER for QPSK modulation. The GNU Radio agents were only trained at 1\% BER SNR, equivalent to SNR\_dB=8 among the simulation curves. Unlike the simulation curve which is sampled over time for one batch of agents, the GNU Radio data points come from separate batches with different seeds. The dip in performance around 600000 symbols is a result of variance in how many seeds converge, not agents losing performance after they've initially reached the performance threshold.}
    \label{fig:gnr-shared-nn-train_convergence}
\end{figure}

Figs.~\ref{fig:gnr-shared-nn-snr_ber} and \ref{fig:gnr-shared-nn-train_convergence} compare the performance of the GNU Radio implementation to our simulations for \ESP{} neural-clone training. Fig.~\ref{fig:gnr-shared-nn-snr_ber} shows that the additional processing required to handle channel equalization and CFO correction requires 2~dB additional empirical SNR to achieve the same baseline BER performance for classic agents. In Fig.~\ref{fig:gnr-shared-nn-snr_ber}, the agents trained on SDRs perform slightly worse relative to the baseline than agents trained in simulation. Fig.~\ref{fig:gnr-shared-nn-train_convergence} shows that learning agents train at approximately the same rate on SDRs as in simulation. Although the simulation curve comes from sampling one set of seeds over time as they train, each data point on the software radio curve comes from a separate set of seeds trained for a given amount of time. There is some variance in how many seeds eventually converge which causes the droop in the curve around 600000 symbols transmitted. For the \ESP{} case with neural agents, the simulated performance is similar to that obtained while using SDRs. This is evidence in support of \echo{} style protocols being practically implementable procedures for learning to communicate.

\subsubsection{Echo with Private Preamble Comparison}
\label{subsubsec:gnr-private-preamble-comparison}

\begin{figure}[ht]
    \centering
    \includegraphics[width=\linewidth]{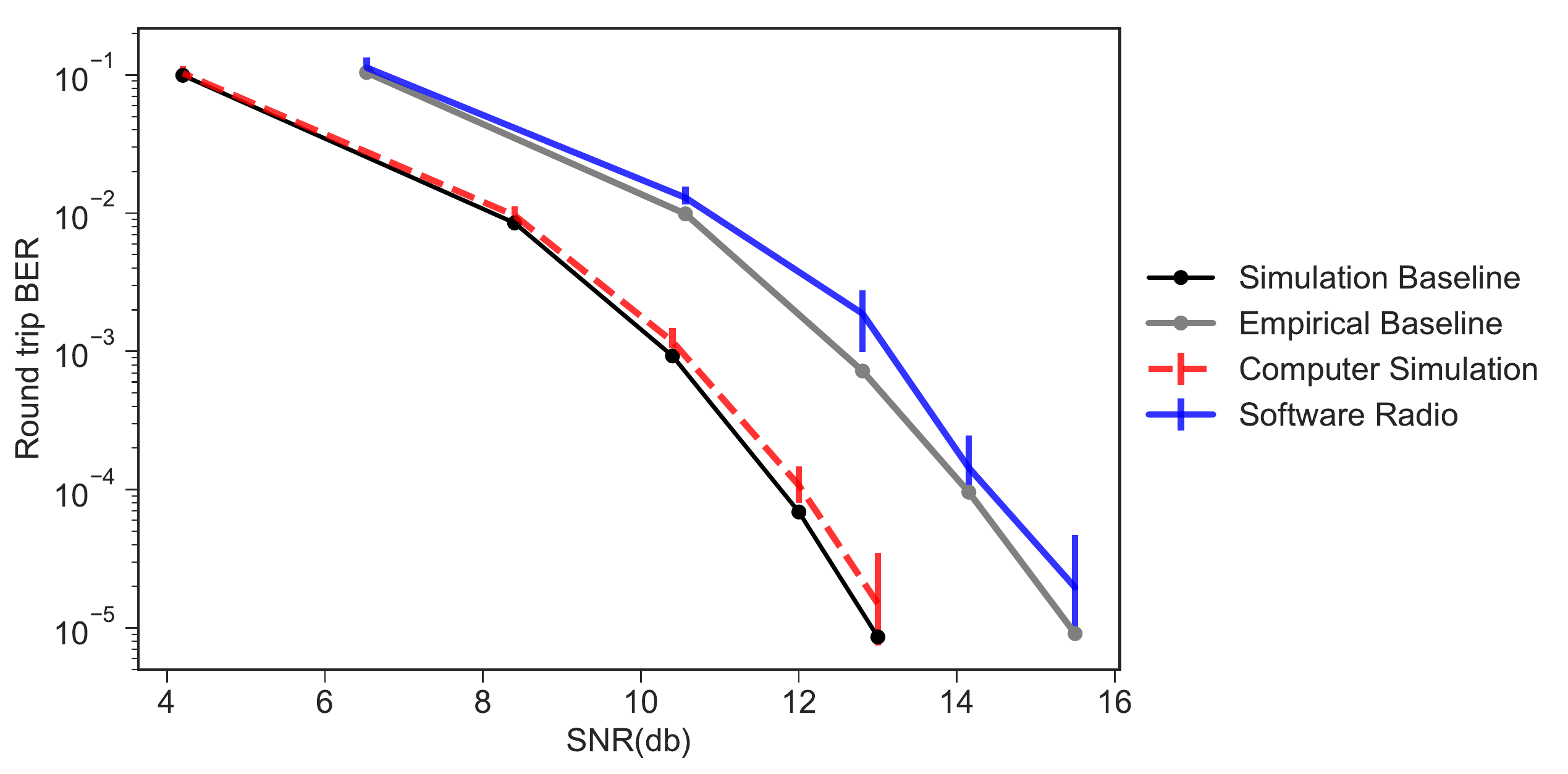}
    \caption{ Round-trip median bit error curves for \Neural{}-and-Clone python simulation and GNU Radio agents learning QPSK under the \EPP{} protocol at training SNRs corresponding to 1\% BER. Alongside the bit error curves of the learned modulation schemes are the baselines. In all cases, modulation constellations are constrained as detailed in Section~\ref{subsec:gnuradio-modifications} to constrain the average signal power.  Although 2~dB SNR extra is required to achieve the same baseline performance due to processing losses in the GNU Radio implementation, the trained agents show similar loss in BER performance compared to the baseline.}
    \label{fig:gnr-private-nn-snr_ber}
\end{figure}

\begin{figure}[ht]
    \centering
    \includegraphics[width=\linewidth]{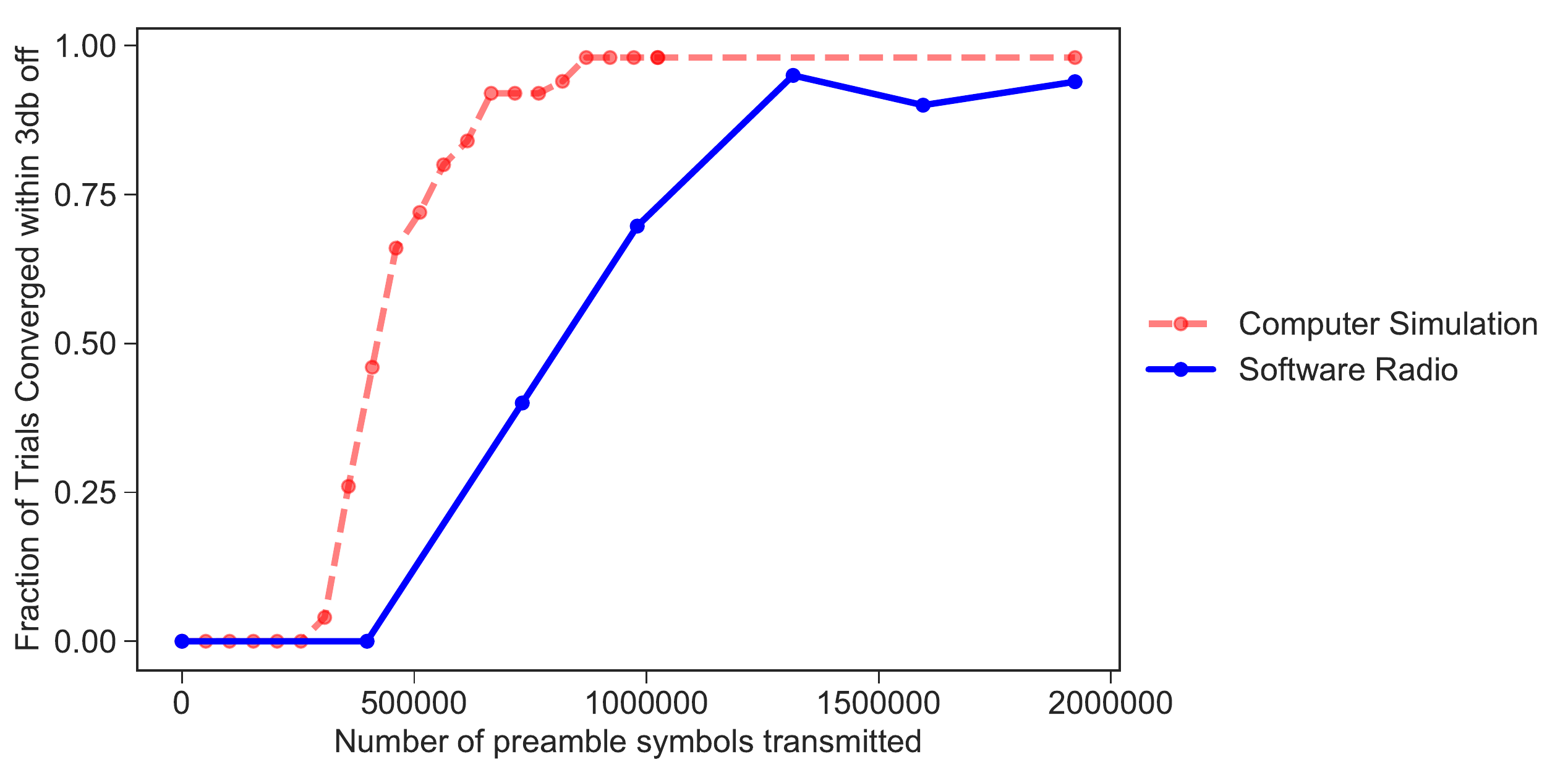}
    \caption{Convergence of 50 simulation and 20 GNU Radio trials to be within 3~dB (at testing SNR corresponding to 1\% BER) of the corresponding baseline for \EPP{} trials of \Neural{} agent vs clone at training SNR corresponding to 1\% BER for QPSK modulation. The GNU Radio agents were only trained at 1\% BER SNR, equivalent to SNR\_dB=8 among the simulation curves. Note that the GNU Radio agents take longer to converge to within 3~dB of optimal, but after sufficient time a similar proportion of trials converge.}
    \label{fig:gnr-private-nn-train_convergence}
\end{figure}

Figs.~\ref{fig:gnr-private-nn-snr_ber}~and~\ref{fig:gnr-private-nn-train_convergence} compare the performance of the GNU Radio implementation to our simulations for \EPP{} neural-clone training. Apart from the additional SNR required to achieve the same baseline performance, the trained neural agents show a similar spread in final BER performance across SNRs. This is another main result of our work, \textbf{\EPP{} is successful at learning modulation schemes over the wire while using software defined radios.}

Fig.~\ref{fig:gnr-private-nn-snr_ber} compares the convergence rate for many trials with training time for the GNU Radio implementation to simulation. Clearly it takes longer for the GNU Radio agents to converge to 3~dB off of optimal BER than the simulation agents, but the final proportion of successful trials is similar. We speculate that there may be more noise in the feedback given to agents during the GNU Radio training process than in the simulation training. This could slow down convergence by reducing the consistency of feedback without reducing its average quality, i.e. some very good feedback mixed with poor feedback. Over time the good feedback would prevail, since it will be self-consistent, whereas the poor feedback will not be consistent and will eventually be averaged away. We will address this discrepancy further in future work. 

\section{Conclusion}

In this work we studied whether the \echo{} protocol enables two agents to learn modulation schemes with minimal information sharing. We proposed a variation of the generic \echo{} protocol, denoted \EPP{} (\echo{} with private preamble), that assumes no shared knowledge apart from knowledge of the echo protocol and the ability to perform turn taking. To evaluate the cost of minimal information sharing, we explored a range of protocols varying in the amount of information shared. We observed that reduced information sharing comes at the cost of slower convergence, meaning more symbols need to be exchanged before a good modulation scheme is learned.
A learning agent when paired with a clone can robustly learn a two bits per symbol modulation scheme in~$2\times10^3$ symbols if we allow gradient passing and in~$3\times10^3$ symbols if we allow loss passing. If we restrict information sharing further, the number of symbols required to learn a scheme robustly goes up exponentially. Allowing only sharing of preambles takes~$2.5\times10^4$ symbols while the case without shared preambles takes~$10^5$ symbols.

Despite the increase in sample complexity, we showed that even under these minimal assumptions, agents can learn to communicate. The \EPP{} protocol is universal, in that it allows agents of diverse types to learn to communicate with each other, and also works when one of the agents uses a fixed communication scheme. 

Our results suggest that learning with ``alien'' agents is not intrinsically more difficult than learning with agents of the same type. For instance, with the learning agents Neural-slow and Poly-fast we observed that the clone pairings (Neural-slow and Neural-slow, Poly-fast and Poly-fast) as well as the alien pairing (Neural-slow and Poly-fast) required a very similar number of training symbols of around $7\times10^5$ to robustly learn a modulation scheme. However, learning to communicate with an agent that uses a fixed modulation scheme is much easier with Neural-fast and Classic requiring only~$10^4$ symbols before a good scheme is learned. 

In Appendix~\ref{app:results-mod-order-snr} we investigated performance of the learning protocols for higher modulation orders and noticed that the difficulty of the learning task increases substantially with modulation order, and the number of preamble symbols that must be transmitted before a good scheme is learned increases exponentially. Confirming the results of others (\cite{cae:DBLP:journals/corr/OSheaKC16}, \cite{cae:2019arXiv190510468S}), we observed that moderate levels of noise have a regularizing effect and facilitate learning but too much noise can be detrimental to the learning process. 

Overall, learning modulation schemes has a high up-front cost in complexity and some cost in loss of optimality for AWGN channels, relative to a designed optimal scheme. For simple known channels it is possible to design a scheme which is provably optimal and has no cost in time spent converging to a common method. However, our goal is to extend this learning protocol to more complex channels for which optimal schemes are not known. Only if it is able to achieve near-optimal performance for a simple channel can we hope that it will also perform well on a harder channel.

This work raises some intriguing questions and opens up several exciting new avenues of research. On the universality of the learning process, one might wonder if it is always possible for two alien agents to learn to communicate with each other when each has the ability to learn to communicate with a clone. What happens when these two agents have vastly different convergence behavior in terms of how fast they learn, measured in terms of number of preamble symbols transmitted? Is learning still possible? Or is there something fundamental to the learning process that determines whether two agents can learn when paired together and two agents with seemingly similar convergence behavior can fail to learn to communicate with each other because their inherent learning behavior is different? What would optimal blind learning look like?

Meta-learning techniques have shown promise for decreasing the sample complexity of learning tasks and have enabled few shot learning in several applications \cite{DBLP:MAML}, \cite{ecc:mind:2019arXiv190302268J}. Can we apply meta-learning techniques to initialize our learning agents in a favorable state that allows them to learn to communicate with others much faster than they would when initialized at a random state? 

We also hope to relax some of the most restrictive assumptions of this paper. Although the \EPP{} protocol aims to share as little information as possible, currently we assume a fixed and known number of bits per symbol. Removing this assumption would be a further step towards a complete learning protocol. Currently a single pair of agents take turns in perfect order, but in real-world environments there are likely to be many agents with imperfect turn-taking. We would like to explore how the \echo{} protocol works with multiple agents, and when agents do not always echo the most recent message or even echo at all.

Can other parts of the communication pipeline, such as equalization and error correcting codes, be integrated into the learning process? Can all these processing stages be learned end-to-end, and does that provide a benefit in terms of training time or communication performance? End-to-end training might allow us to discover new communication strategies for certain types of channels that beat the current best known strategies for the channel. All these research avenues are aimed at bringing us closer to a world where a machine learning-based communication ``standard'' can become a reality. Such a standard would be a minimal set of guidelines which, if followed by agents, would enable them to \textit{learn} how best to communicate with each other based on the current channel conditions.



\begin{appendices}

\section{Code}
\label{app:code}
Our code for the \echo{} protocol, simulation environment, and experiment runs can be found at \url{https://github.com/ml4wireless/echo} in the \textit{ieee-paper} branch. Code for the GNU Radio implementation of the \echo{} protocol can be found at \url{https://github.com/ml4wireless/gr-echo} \cite{Repo}. 

\section{Detailed Agent Descriptions}
\label{app:agent-details}

\subsection{Classic}
\label{subsec:alien-classic}
    \begin{description}
        \item[\textit{Modulator} --] The modulator uses a  fixed strategy known to be optimal for AWGN channels (e.g., Gray coded QPSK for 2 bits per symbol, 8PSK for 4 bits per symbol, and 16QAM for 4 bits per symbol) \cite{qammod}.
        \item[\textit{Demodulator} --] 
        The demodulator uses the 1 nearest-neighbor method to return the closest neighbor from the constellation of the corresponding optimal modulator. Essentially, the demodulator partitions the complex plane into different regions and demodulates based on which region the input to the demodulator lies in.  When using classic demodulation schemes for the \GP{} protocol we require that the output be differentiable, and here we output probabilities for each symbol by taking a softmax of the squared distance of the point to each symbol from the optimal constellation. 
    \end{description}

\subsection{Neural}
\label{subsec:alien-neural}
    First we describe parameter settings that are common to both modulators and demodulators:
    \begin{description}
        \item     \textit{Network architecture}:  We use one layer networks  with fully connected layers with the `tanh' activation. Input and output sizes are different for the modulator and demodulator, as described below. 
        \item     \textit{Initialization}: 
    The weights for each layer are initialized by sampling from the distribution $U[\frac{-1}{\sqrt{n}}, \frac{1}{\sqrt{n}}]$, where $n$ is the number of input units to the layer; the biases are initialized as $0.01$.
    \item \textit{Optimizer}:  We use the Adam optimizer \cite{kingma2014adam}.
     \end{description}
     Next we describe the modulator and demodulator specific parameters and details about their update methods.
     For the rest of the section let $b$ denote bits per symbol (equivalently the modulation order). 
     
     \subsubsection{Modulator}
     \begin{description}
     \item \textit{Input width}: $b$. We take in input in bit format (but treat these 0-1 values as floats).
     \item \textit{Output width}: 2. The output width is fixed since it represents a complex number to be sent over the channel.
     \item \textit{Parameters}: In addition to the network weights and biases, $\theta$, we also include a separate learned parameter $\sigma$, a scalar denoting the standard deviation of the Gaussian distribution we sample from for our policy. 
     \item \textit{Modulation procedure}: The neural net outputs $\mu$. Here $\mu$ is the output of the neural network and is the mean of the Gaussian distribution that we sample from. Note that if the input is of size $[N, b]$, $\mu$ will have size $[N,2]$ (first dimension corresponding to the real part of a complex number and the other corresponding to the imaginary part of the complex number).
     While training, the modulator outputs symbols $s$ sampled from a Gaussian distribution with mean $\mu$ and standard deviation $\sigma$ ($\sigma$ is bounded by minimum and maximum values.), i.e. $s \sim \mathcal N(\mu,\sigma^2I)$.
     \item \textit{Update procedure}: Suppose for our given actions $s$ we receive the reward $r$, the negative of the number of incorrect bits (comparing the original bit sequence to the received echo). The log probability for each action is given by,
     \begin{align}
         \log p_i = -C\frac{(s_i - \mu_i)^2}{2\sigma^2},
     \end{align}
     for some constant $C$.
     The loss function we minimize is given by,
     \begin{align}
         l_{pg} = -\left(\sum\limits_{i} \log p_i * r_i\right),
     \end{align}
     In some settings we modify the reward $r$ to include penalty terms such as one for distance of average output from origin as detailed in Appendix \ref{app:constellation-centering}.
     We update our parameters as,
     \begin{align}
         \theta &\leftarrow \theta + \operatorname{Adam~update}(\theta, \eta_\mu, l_{pg}) \\
         \sigma &\leftarrow \sigma + \operatorname{Adam~update}(\sigma, \eta_\sigma, l_{pg}),
     \end{align}
     where $\eta_\mu$ and $\eta_\sigma$ denote the separate learning rate parameters for the network parameters and the standard deviation $\sigma$.
     \end{description}
     
    \subsubsection{Demodulator}
     \begin{description}
     \item \textit{Input width}: $2$
     \item \textit{Output width}: $2^b$. The demodulator is a classifier which outputs logits for each class that, on application of the softmax layer, correspond to the probabilities of the classes. The classes are the set of possible bit sequences for the modulation order.
     \item \textit{Parameters}: The network weights and biases denoted as $\phi$.
     \item \textit{Demodulation procedure}: Given input of size $[N, 2]$ , the neural net outputs logits $(\mathsf{logits})$ of shape $[N,2^b]$. On applying the softmax operation these correspond to a probability distribution over classes. The demodulated symbols, $\hat p$, are computed by choosing the class with the highest probability,
     \begin{align}
         \hat{p} = \arg\max(\mathsf{softmax(logits)})).
     \end{align}
     
     \item \textit{Update procedure}:
     Suppose after applying the softmax layer we have probability $q_{i,c}$ corresponding to the true class label of symbol $i,\,i=1\ldots N$. We compute the cross-entropy loss as
    \begin{align}
        l_{CE} = -\sum_i{ \log\left(\frac{\exp{q_{i,c}}}{ \sum_{j=1}^{2^b} \exp q_{i,j} }\right)}
    \end{align}
    We update our parameters as
    \begin{align}
        \phi \leftarrow \phi - \operatorname{Adam~update}(\phi, \eta_\phi, l_{CE})
    \end{align}
    where $\eta_\phi$ is the learning rate parameter for the demodulator updates.
    
     \end{description}

\subsection{Polynomial}
\label{subsec:alien-polynomial}
    First we describe parameter settings that are common to both modulators and demodulators:
    \begin{description}
        \item \textit{Network architecture}: 
        The inputs to the network are used to form a polynomial of degree $d$. We use a single fully connected linear layer to connect the polynomial terms to the output. Input and output sizes are different for the modulator and demodulator, as described below.
        \item \textit{Initialization}: 
        The weights for each layer are initialized by sampling from the distribution $U[\frac{-1}{\sqrt{n}}, \frac{1}{\sqrt{n}}]$, where $n$ is the number of input units to the layer; we do not use biases for polynomial agents.
        \item \textit{Optimizer}:
        We use the Adam optimizer \cite{kingma2014adam}.
    \end{description}
    Next we describe the modulator and demodulator specific parameters and details about their update methods. For the rest of the section let $b$ denote bits per symbol and $d$ the degree of the polynomial.
    
    \subsubsection{Modulator}
    \begin{description}
        \item \textit{Input width}: $b$. We take in input in bit format (but treat these 0-1 values as floats).
        \item \textit{Output width}: 2. The output width is fixed since it represents a complex number to be sent over the channel.
        \item \textit{Parameters}:
        Internally, the input bits are used to calculate all unique polynomial terms of order $d$. Since the bits $b_i$ are in $\{0,1\}$, terms including $b_i^2,b_i^3,\ldots$ are redundant and omitted from our calculations, thus allowing us to determine a unique maximum-degree polynomial. The polynomial terms are fed into the single fully connected layer with parameters $\theta$. We also include a separate parameter $\sigma$, a scalar denoting the standard deviation of the Gaussian distribution we sample from for our policy.
        \item \textit{Modulation procedure}:
        The polynomial network outputs $\mu$. Here $\mu$ is the mean of the Gaussian distribution that we sample from. Note that if the input is of size $[N, b]$, $\mu$ will have size $[N,2]$ (first dimension corresponding to the real part of a complex number and the other corresponding to the imaginary part of the complex number).
        While training, the modulator outputs symbols $s$ sampled from a Gaussian distribution with mean $\mu$ and standard deviation $\sigma$, i.e. $s \sim \mathcal N(\mu,\sigma^2I)$.
        \item \textit{Update procedure}:
        The update procedure for polynomial modulators is identical to the procedure for neural modulators.
    \end{description}
    
    \subsubsection{Demodulator}
    \begin{description}
        \item \textit{Input width}: 2
        \item \textit{Output width}: $2^b$. The demodulator is a classifier which outputs logits for each class that, on application of the softmax layer, correspond to the probabilities of the classes. The classes are the set of possible bit sequences for the modulation order.
        \item \textit{Parameters}:
        Internally, the input symbols is used to calculate all unique polynomial terms of order $d$ containing the real part and imaginary part of the symbol. For example,
        \begin{align}
            P(s, 2) = & \left[ \real{s},\real{s}^2,\imag{s}, \right. \\
            & \,\, \left. \imag{s}^2,\real{s}\imag{s}\right]^\top
        \end{align} 
        The polynomial terms are fed into the single fully connected layer with parameters $\phi$.
        \item \textit{Demodulation procedure}: The demodulation procedure is the same as the neural agent.
        \item \textit{Update procedure}: The update procedure is the same as the neural agent, except for an $L1$ penalty added to the demodulator's loss term. 
    \end{description}

\section{Additional Results}
\label{app:addresults}

This appendix contains additional experimental results which, although not required to support our primary conclusions, we believe are of interest to anyone who wants to replicate or build upon our work. Appendix~\ref{app:results-mod-order-snr} shows the effects of modulation order and training SNR on the performance of the \ESP{} and \EPP{} protocols with clone agents. Since any learning communications system in the wild will be exposed to multiple SNR conditions and desired signalling rates, understanding performance variation across SNR and modulation order will be crucial. Our results indicate that moderately high training SNR leads to the best performance confirming observations by others (\cite{cae:DBLP:journals/corr/OSheaKC16}, \cite{cae:2019arXiv190510468S}). Appendix~\ref{app:results-polys-replicate-neural} presents experiments with \Poly{} clone and self-alien agents demonstrating similar behavior to \Neural{} clone and self-alien agents.

\subsection{Effect of Modulation Order and Training Signal to Noise Ratio}
\label{app:results-mod-order-snr}

\begin{table*}[ht]
\begin{center}
\begin{tabular}{@{}lrrr@{}}
\toprule
\textbf{Training SNR (BER)} & \textbf{QPSK} & \textbf{8PSK} & \textbf{16QAM} \\
\textbf{} & (2 BPS) & (3 BPS)  & (4 BPS)  \\\midrule
\textbf{\ESP{}} & & & \\
\textbf{\hspace{0.25cm}8.4~dB (1\%)} & 25600 & 39936 & 152064 \\
\textbf{\EPP{}} & & & \\
\textbf{\hspace{0.25cm}4.2~dB (10\%)} & 901120$^*$ & - & - \\
\textbf{\hspace{0.25cm}8.4~dB (1\%)} & 115200 (404480$^*$) & 238080 & 2734592 \\
\textbf{\hspace{0.25cm}13.0~dB (0.001\%)} & 309760$^*$ & - & - \\
\bottomrule
\end{tabular}
\end{center}
\caption{Number of symbols exchanged before $\geq 90\%$ of trials reached 3~dB off of optimal BER for the \ESP{} and \EPP{} protocols with \Neural{} agents and varying bits per symbol (BPS) and SNR. The results show the increased difficulty of learning at higher modulation orders. The \EPP{} protocol is impacted much more than \ESP{} by high modulation order, requiring $24\times$ more symbols for 16QAM than QPSK compared to $6\times$ for \ESP{}. *The experiments comparing performance with training SNR were conducted using a different set of hyperparameters, found in Table~\ref{tab:echo-neural-snr-qpsk-hyperparams} in Appendix~\ref{app:simulation-hyperparams}. Lower training SNRs require longer to converge.}
\label{tab:convergence-mod-snr}
\end{table*}

\begin{figure}[ht!]
     \subfloat[Round-trip median BER curves for \Neural{} agent learning with clone with mod orders (bits per symbol) 2, 3, 4 using the \EPP{} protocol at training SNRs corresponding to 1\% BER. Alongside the BER curves of the learned modulation schemes is the baseline QPSK (order 2), 8PSK(order 3) and 16QAM(order 4). In all cases, modulation constellations are normalized to constrain the average signal power.  ]{
 \includegraphics[width=1.0\linewidth]{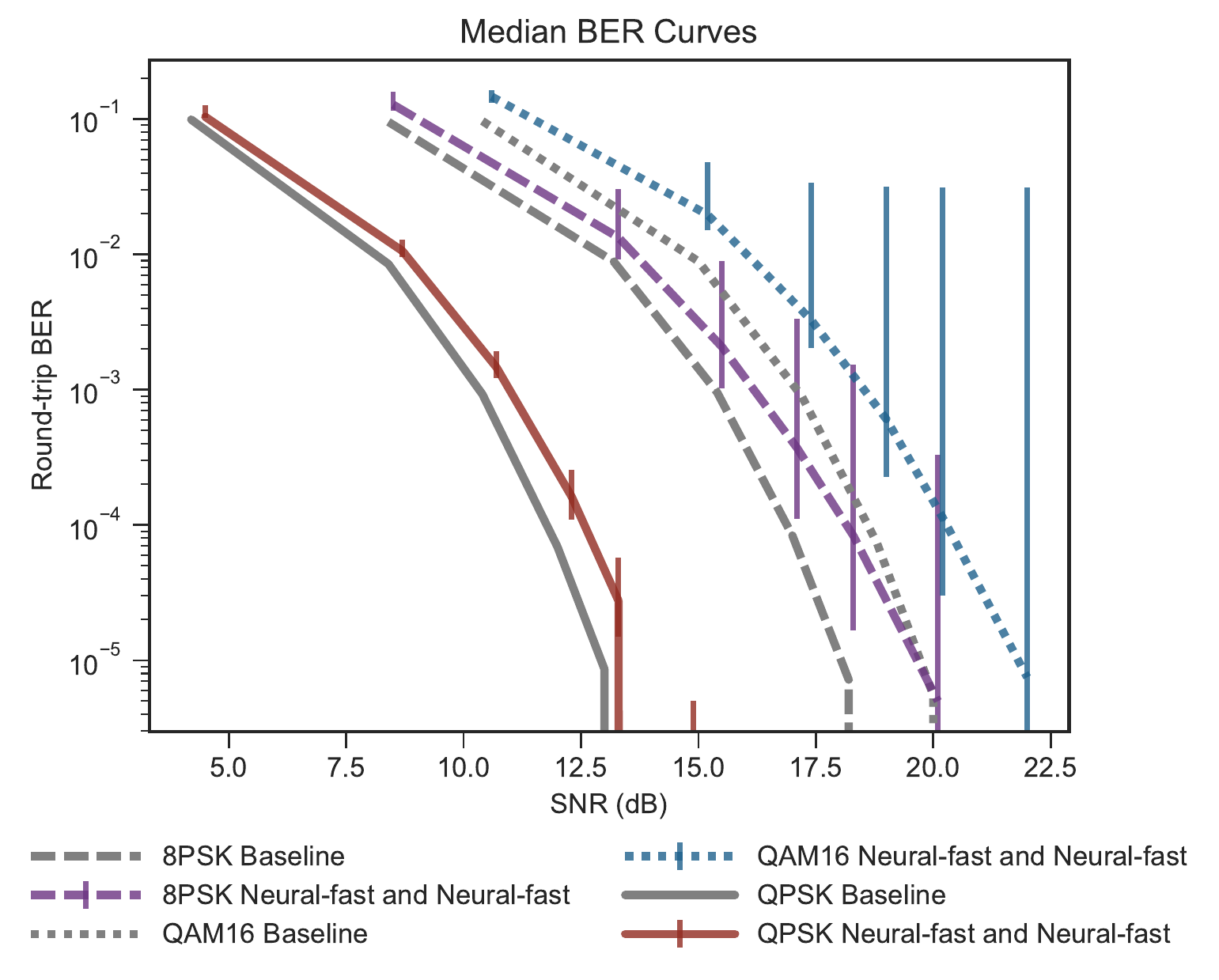}
 \label{subfig:mod-orders-epp-ber}
 }\\
    \subfloat[Convergence of 50 trials to be within 3~dB (at testing SNR corresponding to 1\% BER) of the corresponding baseline for \EPP{} trials of at training SNR corresponding to 1\% BER for increasing modulation order. 16QAM, with the highest modulation order 4, takes much longer to converge than QPSK (order 2) and 8PSK (order 3).]{ \includegraphics[width=1.0\linewidth]{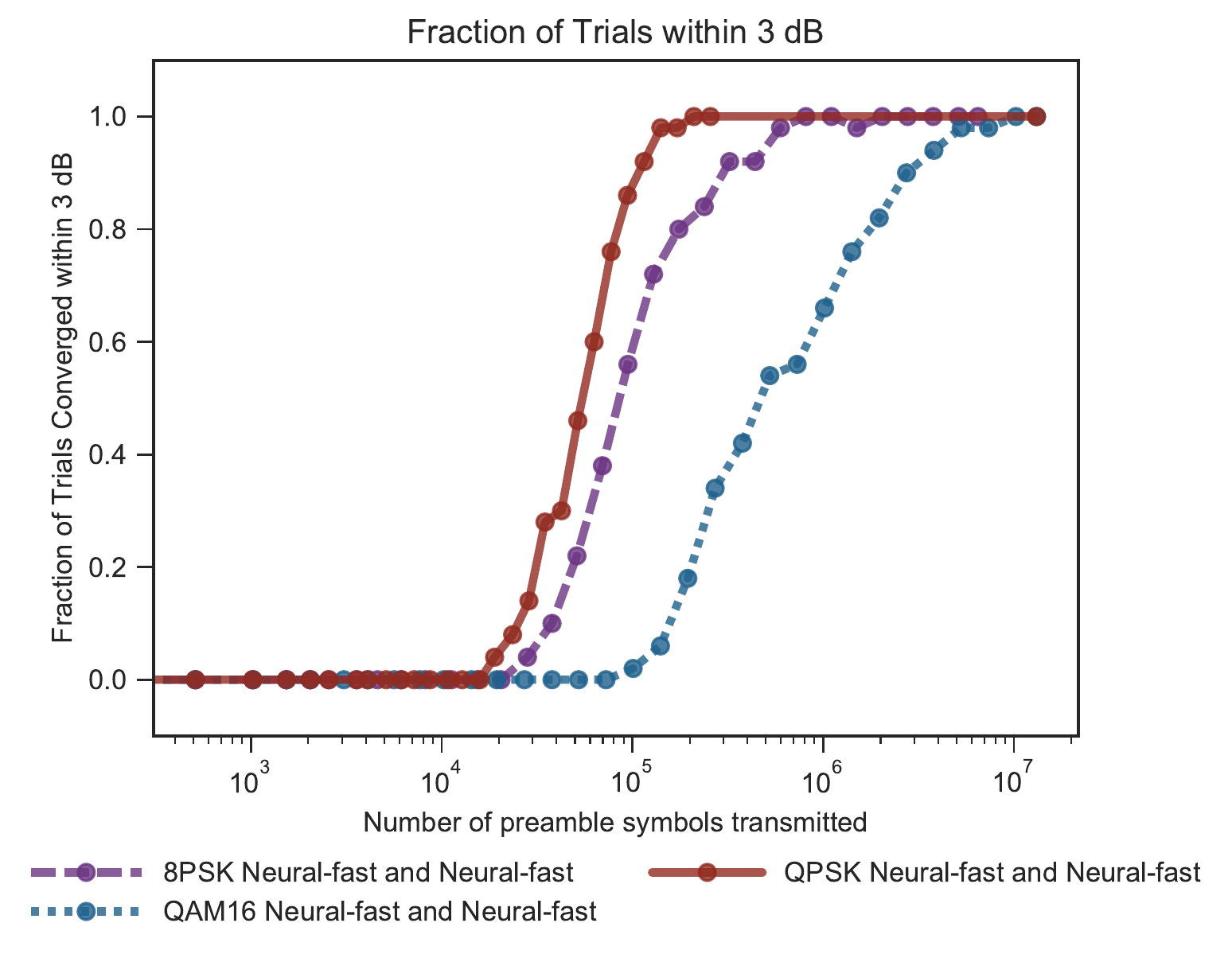}
     \label{subfig:mod-orders-epp-db3}
    }
        \caption{\Neural{} agents learning with a clone using the \EPP{} protocol for different modulation orders. The BER plot (a) shows that the gap between the median BER of the learned scheme and the corresponding baseline increases for higher mod orders. From the convergence plot (b), we observe that higher mod orders take exponentially longer to converge to a good strategy.  }
    \label{fig:mod-orders-epp}
\end{figure}

In the experiments detailed in Sec.~\ref{sec:results} we learned to modulate with 2 bits per symbol. Here we explore whether the learning protocols continue to work for higher modulation orders, i.e. more bits per symbol. We conduct experiments using the \EPP{} protocol for a \Neural{} agent learning to communicate with a clone for 3 and 4 bits per symbol. We compare these cases, and the 2 bits per symbol case, in Fig.~\ref{fig:mod-orders-epp}. From Fig.~\ref{subfig:mod-orders-epp-ber} we observe that, at higher modulation orders, there is a larger gap between the BER curves of the learned agents and the corresponding baselines. Although some agents continue to approach the baseline BERs, as evidenced by the error bars, the median agent no longer achieves near-optimal performance at high SNRs.
Fig.~\ref{subfig:mod-orders-epp-db3} shows that, for higher modulation orders, fewer trials learn a good modulation scheme and it takes longer to learn good schemes. From Table~\ref{tab:convergence-mod-snr} we see that the increase in convergence times is exponential, with \EPP{} requiring $2\times$ and $24\times$ more symbols for convergence for 8PSK and 16QAM, respectively. \ESP{} requires $1.5\times$ and $6\times$ more symbols for convergence. Still, even for the highest modulation order examined (16QAM), 96\% of trials eventually converge to a good scheme. This phenomenon of performance degradation with increasing modulation order is expected since the modulation functions for higher order modulation schemes are more complex.

\begin{figure}[ht!]
    \subfloat[Round-trip median BER curves for a \Neural{} agent learning with a clone  using     the \EPP{} protocol at training SNRs  13.0, 8.4, and 4.2~dB corresponding to 0.001\%,     0.1\%, and 10\% BERs for the baseline. The error bars reflect the \nth{10} to \nth{90}     percentiles across 50 trials. All agents are evaluated at the same SNR but error bars     have been dithered for readability.]{
         \includegraphics[width=0.85\linewidth]{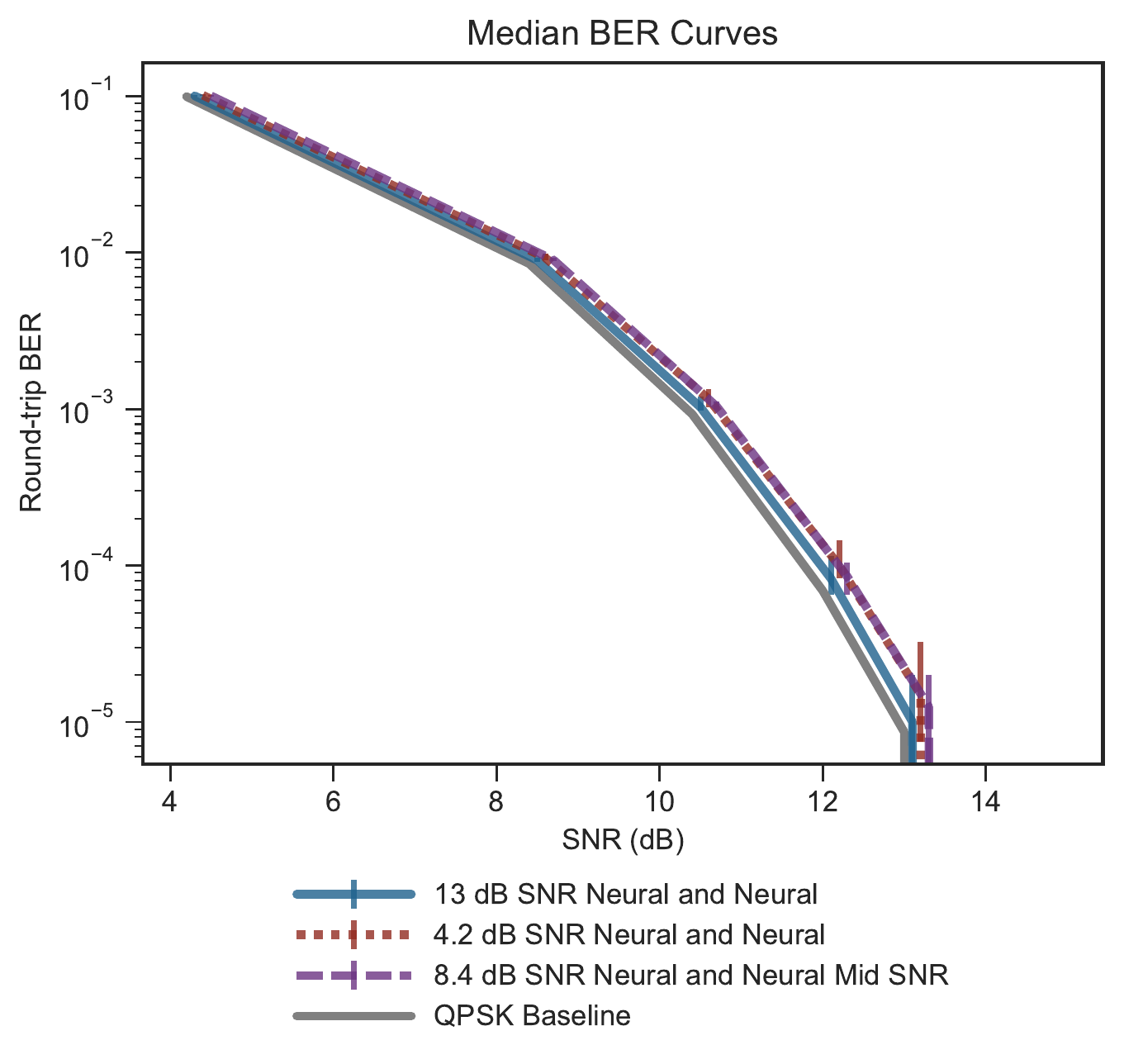}
         \label{subfig:train-snrs-epp-ber}
     }\\
    \subfloat[Convergence of  50  trials to be within 3~dB at testing SNR 8.4~dB  at training     SNRs 13.0, 8.4, and 4.2~dB. Training at higher SNR reduces the number of symbols required for most trials to converge.]{
        \includegraphics[width=0.85\linewidth]{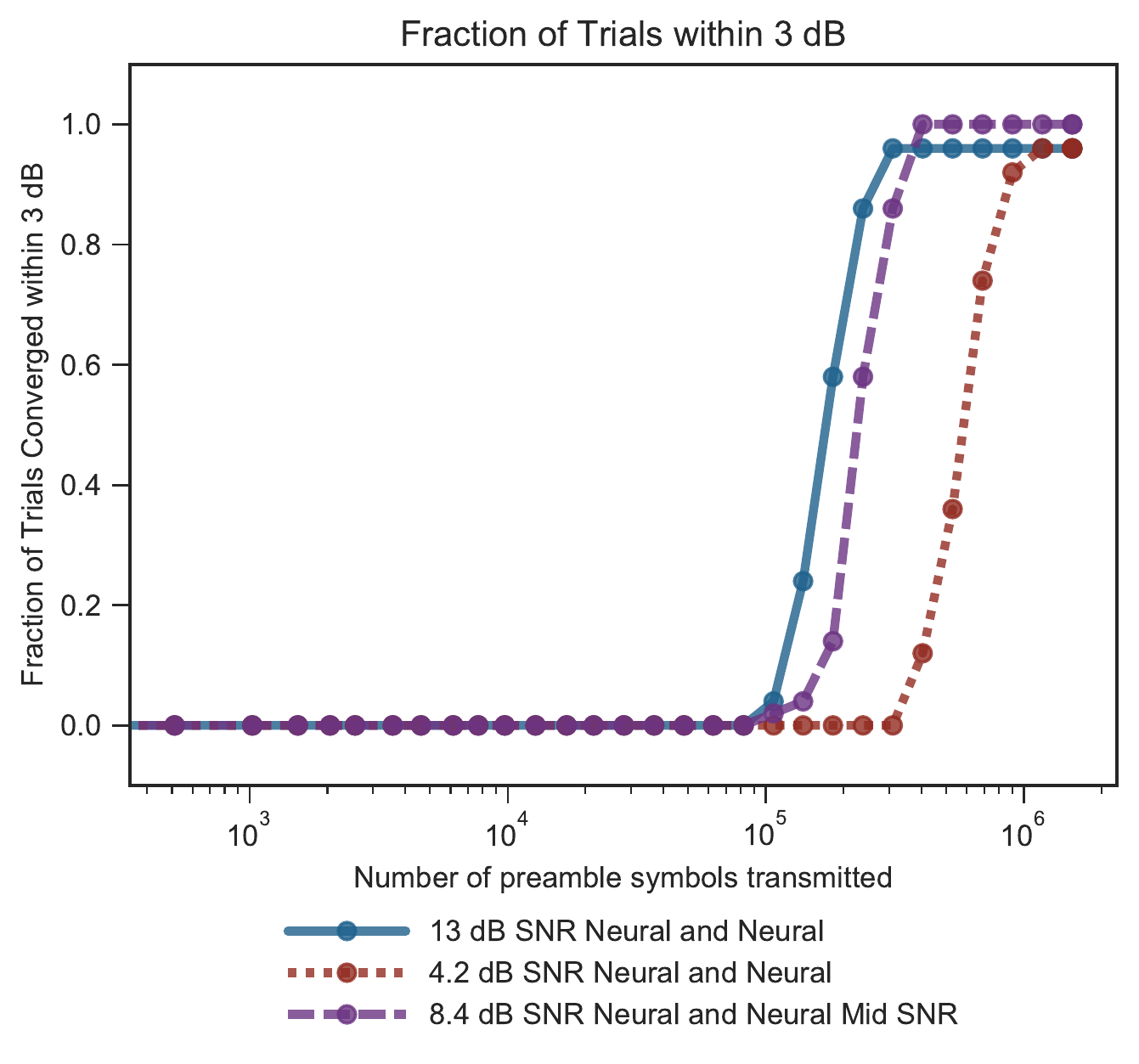}
        \label{subfig:train-snrs-epp-db3}
    }
    \caption{\Neural{} agents learning with a clone using \EPP{} for different training signal to noise ratios. The BER plot (a) shows that training at higher SNR leads to lower BERs across all SNRs. From the convergence plot (b), we observe that limited noise plays a regularizing effect, helping more trials to converge. Too much noise, however, has a detrimental effect and slows down convergence. Training at higher SNRs helps agents to converge more quickly, although not every trial converges.}
    \label{fig:train-snrs-epp}
\end{figure}

Next, we investigate the effect of training SNR. In all other experiments we trained our agents at the SNR corresponding to 1\% BER for the baseline scheme of the given modulation order. Is this the optimal SNR to train at? Does the learning protocol work at lower SNRs? We explore answers to these questions by conducting experiments using the \EPP{} protocol for the setting where a \Neural{} agent learns to communicate with a clone at various training SNRs corresponding to 0.001\%, 0.1\%, and 10\% BERs for the baseline modulation scheme. From our results in Fig.~\ref{fig:train-snrs-epp} we observe that in all 3 settings we achieve a BER close to the QPSK baseline. It takes longer to learn a modulation scheme at lower SNRs. However, not every trial converges when trained at high SNR. This can be explained by a regularization role that noise seems to have on our learning task. At very low SNRs, some trials fail to converge but those that do converge achieve similar BERs to agents trained at higher SNRs. It is possible that the agents are taking gradient steps which are too large and being forced into local minima by steps in poor directions caused by noisy feedback. This suggests the question of whether there is a ``speed limit'' to how fast agents can reliably (i.e. having all trials converge to within 3~dB off optimal) learn at a given SNR. We hope to answer this question in future work.


\subsection{Polynomial Agent Experiments}
\label{app:results-polys-replicate-neural}

\begin{figure}[ht!]
     \subfloat[Round-trip median BER. The error bars reflect the \nth{10} to \nth{90} percentiles across 50 trials. All agents are evaluated at the same
SNR but error bars have been dithered for readability. ]{
 \includegraphics[width=1.0\linewidth]{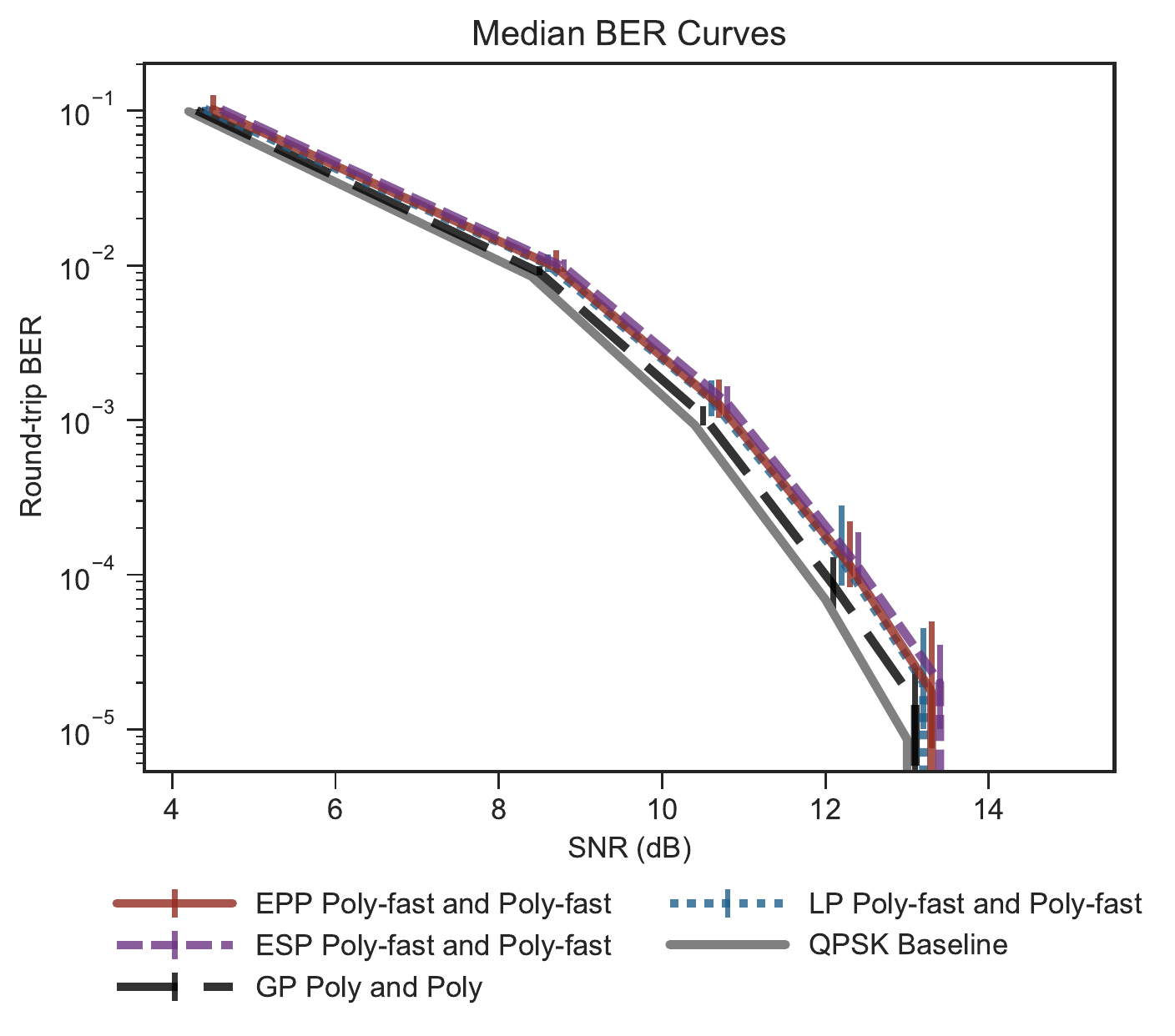}
 \label{subfig:poly-fast-poly-fast-allproto-ber}
 }\\
    \subfloat[Convergence of 50  trials to be within \dboff at 8.4~dB training SNR.]{ \includegraphics[width=1.0\linewidth]{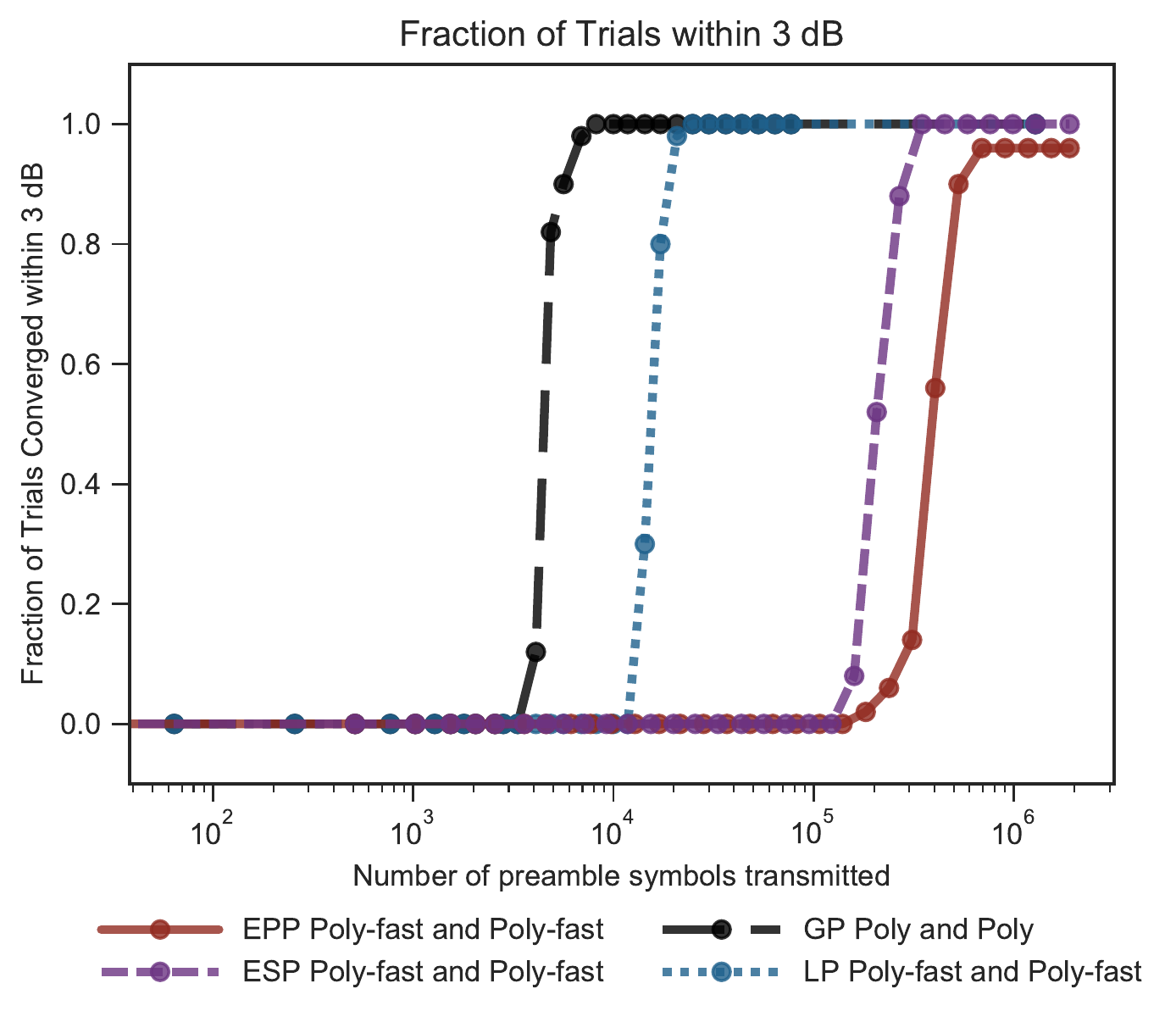}
     \label{subfig:poly-fast-poly-fast-allproto-db3}
    }
        \caption{Effect of Information Sharing while learning to communicate with a clone, \PF-and-\PF: The BER plot (a) shows that all protocols achieve BER close to that of QPSK baseline. From the convergence plot (b), we observe that \EPP{} is much slower than \ESP{} which is an order of magnitude slower than \LP{} and \GP{}. \GP{} converges the fastest of all the protocols.}
    \label{fig:poly-fast-poly-fast-allproto}
\end{figure}

\begin{figure}[h!]
     \subfloat[Round-trip median BER. The error bars reflect the \nth{10} to \nth{90} percentiles across 50 trials. All agents are evaluated at the same
SNR but error bars have been dithered for readability. ]{
 \includegraphics[width=1.0\linewidth]{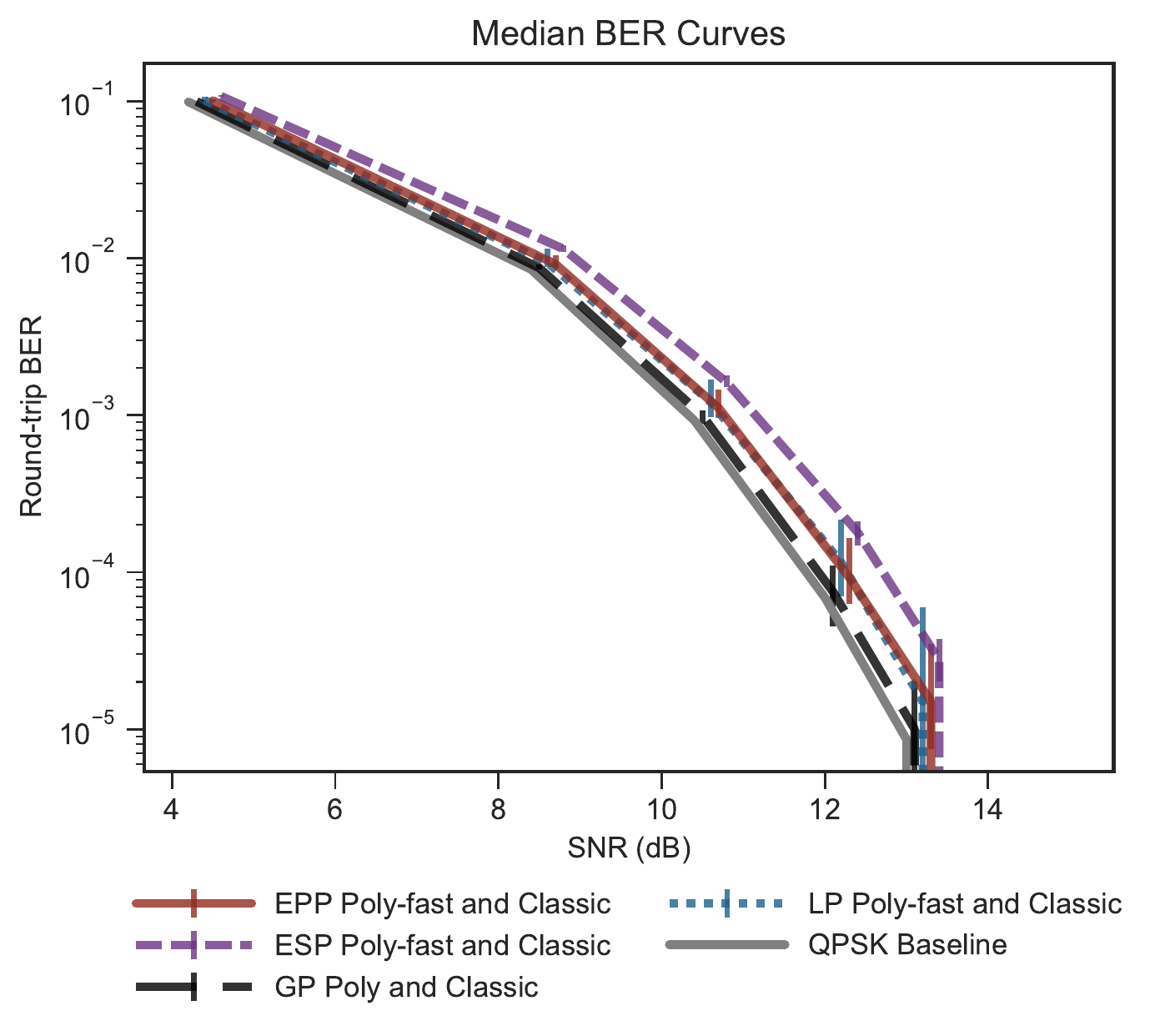}
 \label{subfig:poly-fast-classic-allproto-ber}
 }\\
    \subfloat[Convergence of 50  trials to be within 3~dB at testing SNR 8.4~dB.]{ \includegraphics[width=1.0\linewidth]{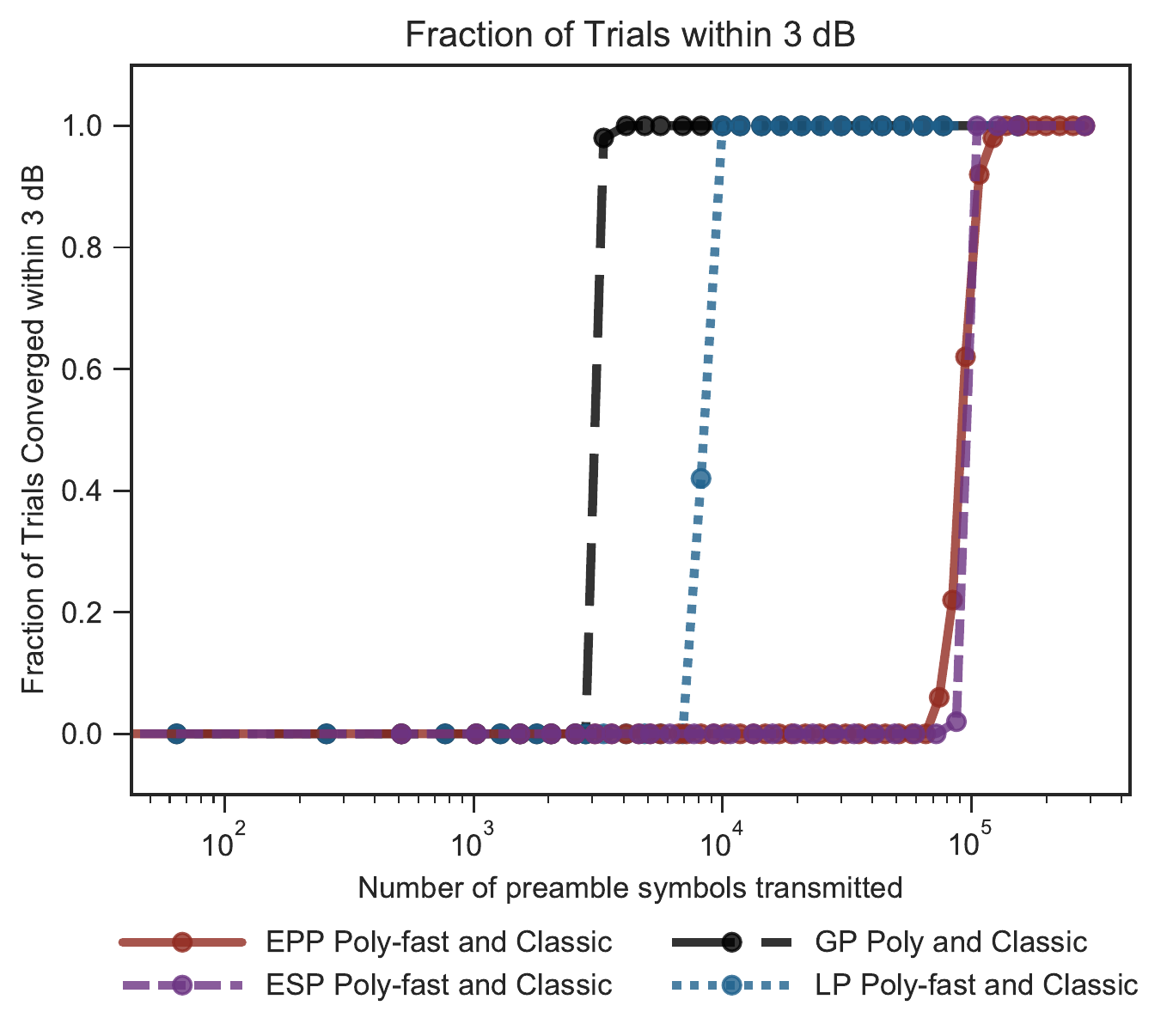}
     \label{subfig:poly-fast-classic-allproto-db3}
    }
        \caption{Learning to communicate with fixed agent,  \PF-and-\Classic: The BER plot(a) shows that all protocols achieve BER close to that of QPSK baseline. From the convergence plot (b), we observe that \EPP{} and \ESP{} have similar convergence behavior and are an order of magnitude slower than \LP{} and \GP{}. \GP{} converges the fastest of all the protocols. Across all protocols, convergence is faster than learning with a clone (Figure: \ref{fig:poly-fast-poly-fast-allproto})  }
    \label{fig:poly-fast-classic-allproto}
\end{figure}

\begin{figure}[h!]
     \subfloat[Round-trip median BER. The error bars reflect the \nth{10} to \nth{90} percentiles across 50 trials. All agents are evaluated at the same
SNR but error bars have been dithered for readability. ]{
        \includegraphics[width=1.0\linewidth]{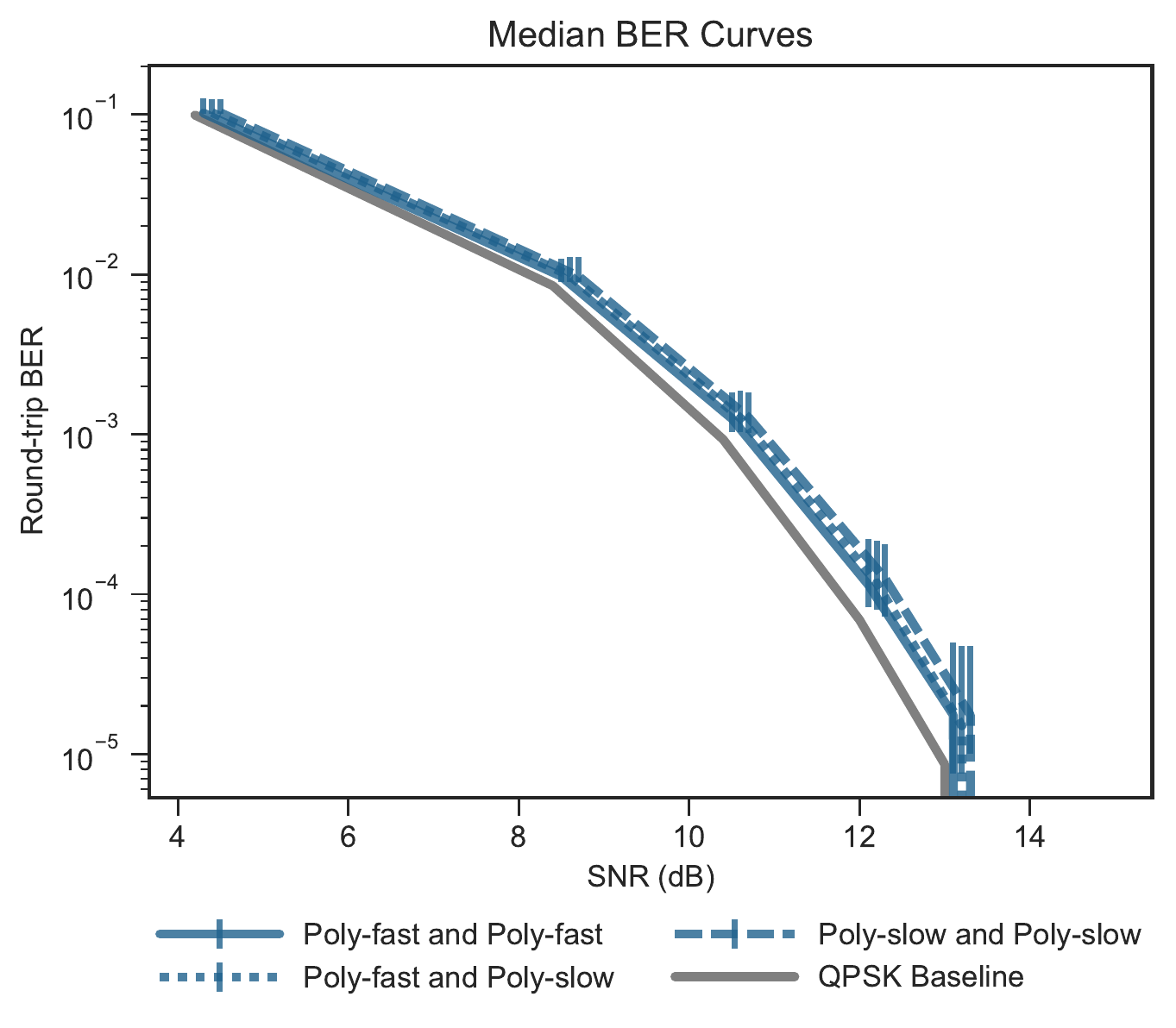}
        \label{subfig:polys-epp-ber}
    }\\
    \subfloat[Convergence of 50  trials to be within 3~dB at 8.4~dB training SNR.]{ \includegraphics[width=1.0\linewidth]{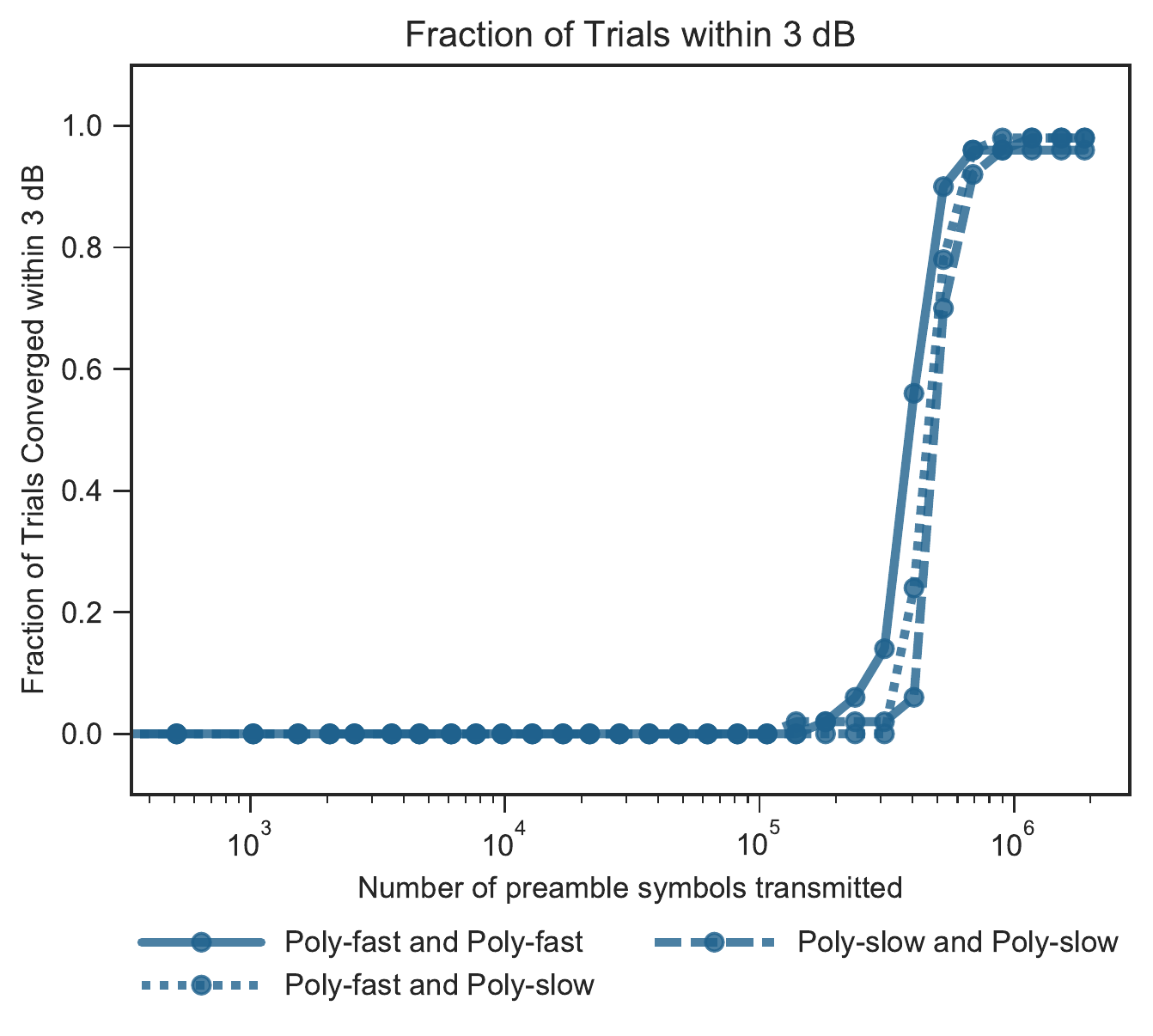}
     \label{subfig:polys-epp-db3}
    }
    \caption{Learning with clones (\PF-and-\PF, \PS-and-\PS) compared to learning with self-alien (\PF-and-\PS) under the \EPP{} protocol. The BER plot (a) shows that the round-trip BER in all cases is almost identical and  close to the QPSK baseline. From the convergence plot (b), we observe that \PF-and-\PF{} converges about twice as fast as \PS-and-\PS. The convergence time for the self-alien pairing,  \PF-and-\PS{}, is in between the clone pairings; the \PF{} agent helps the \PS{} agent learn faster when paired together. }
    \label{fig:polys-epp}
\end{figure}

\begin{figure}[h!]
     \subfloat[Round-trip median BER. The error bars reflect the \nth{10} to \nth{90} percentiles across 50 trials. All agents are evaluated at the same
SNR but error bars have been dithered for readability. ]{
 \includegraphics[width=1.0\linewidth]{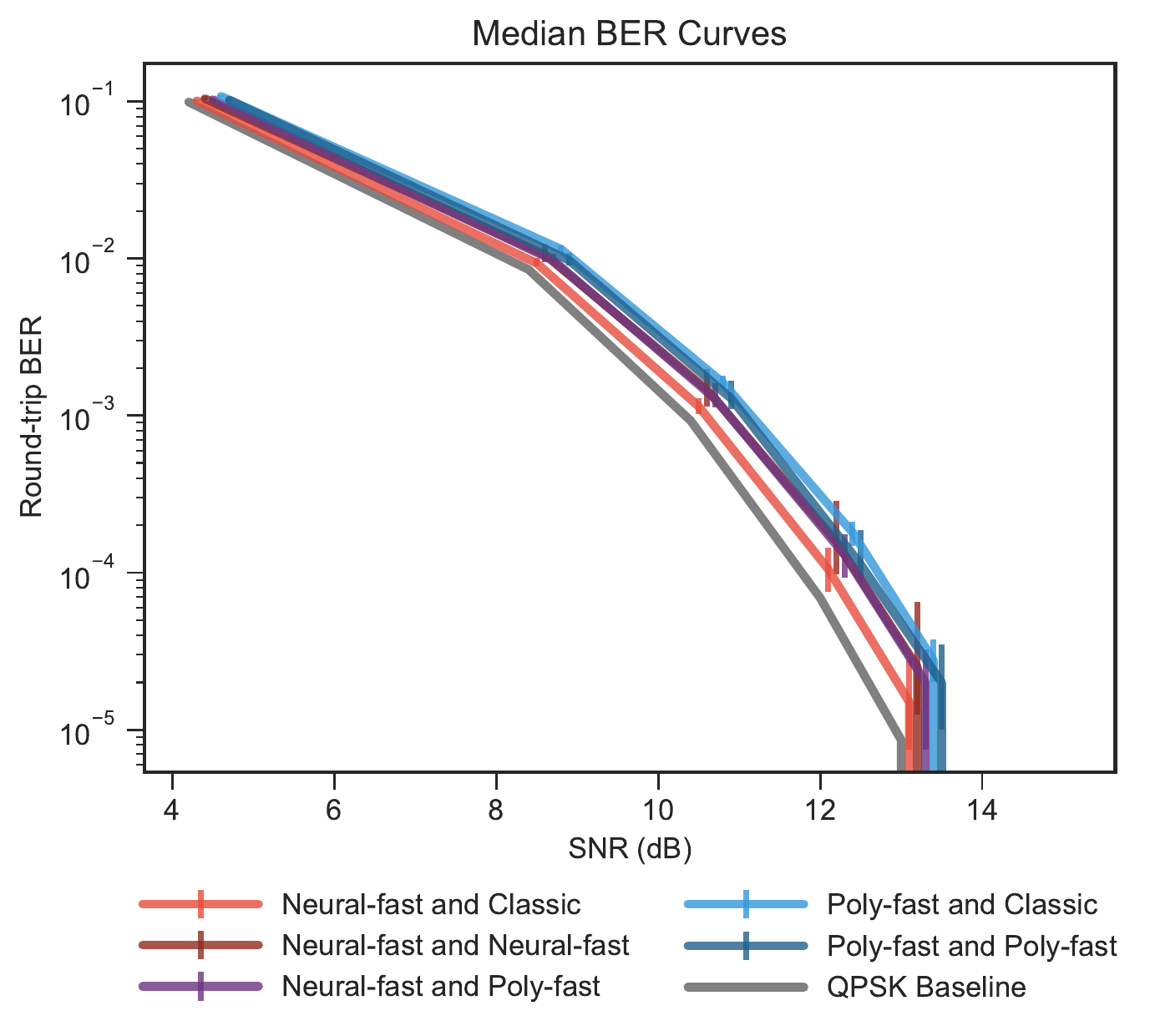}
 \label{subfig:esp-combos-ber}
 }\\
    \subfloat[Convergence of 50  trials to be within 3~dB at testing SNR 8.4~dB.]{ \includegraphics[width=1.0\linewidth]{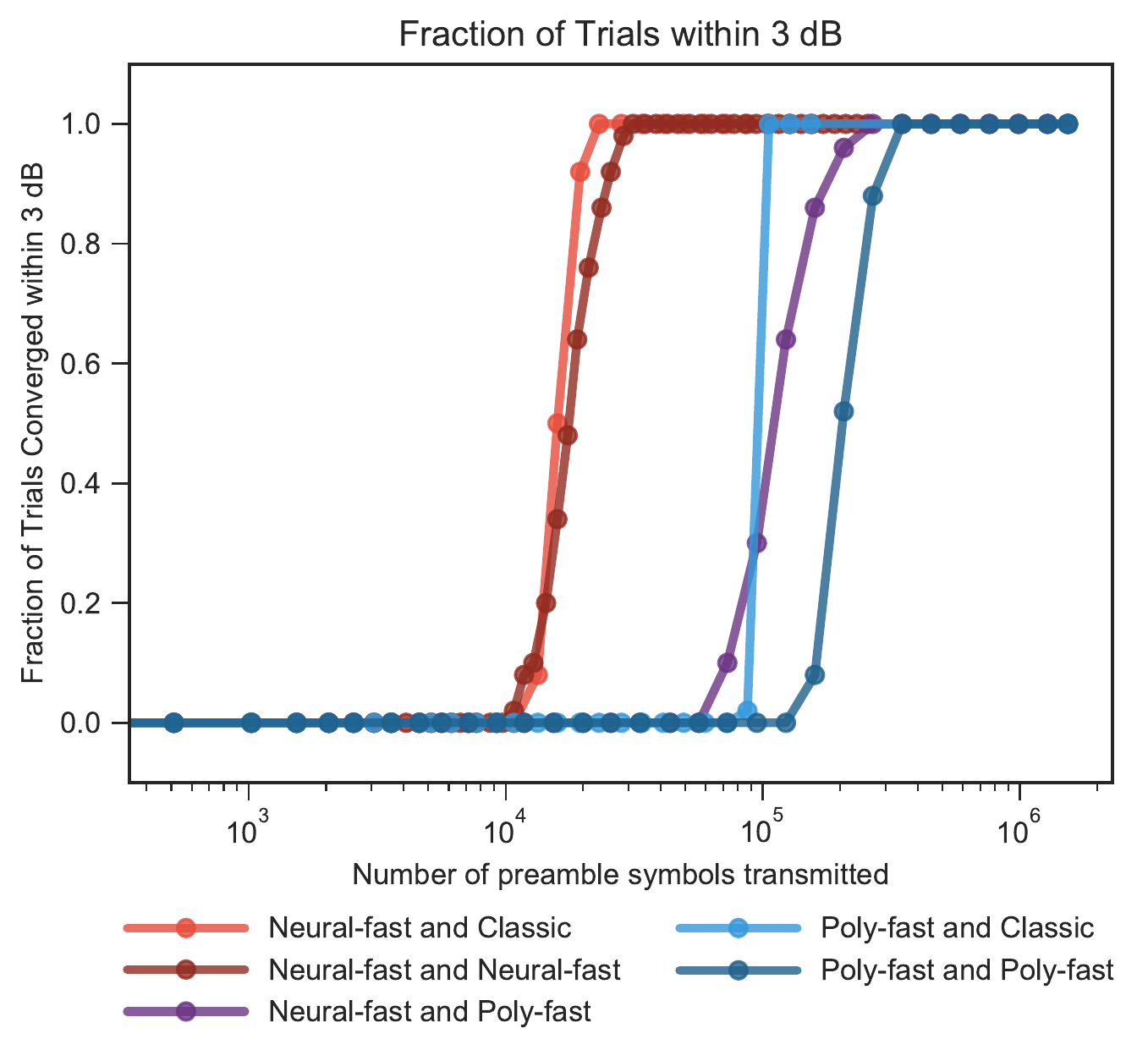}
     \label{subfig:esp-combos-db3}
    }
        \caption{Learning under \ESP{} protocol: The BER plot (a) shows that all protocols achieve BER close to that of QPSK baseline. From the convergence plot (b), we observe that learning with fixed agents is faster than learning with clones for both Neural and Poly agents. The alien pairing, Neural-fast and Poly-fast, has convergence times in between those of the individual clone pairings. The Neural agent helps the Poly agent learn faster when paired together.    }
    \label{fig:esp-combos}
\end{figure}

Here we include results using \Poly{} agents which demonstrate the same behaviors as the \Neural{} agents in Section~\ref{sec:results}. Figs.~\ref{fig:poly-fast-poly-fast-allproto}~to~\ref{fig:polys-epp} demonstrate the effects of information sharing and that the \EPP{} protocol works with fixed \Classic{} and self-alien agents using polynomial function approximators. As with \Neural{} agents, more information sharing leads to faster training. Similarly, \Classic{} agents speed up training, and two self-alien agents converge at a rate in between their individual convergence speeds. Fig.~\ref{fig:esp-combos} shows the performance of several combinations of \Classic{}, clone, and self-alien agents using the \ESP{} protocol. The relative ordering of performance is the same as when using \EPP{}, even though each combination trains faster with \ESP{}.

\FloatBarrier
\section{Unsuccessful Constellation Centering Methods}
\label{app:constellation-centering}

We investigated several methods for forcing modulator constellations to be centered to avoid the training problems caused by DC offset correction in the USRP radios. Because not all of them worked, it is important to report the results for scientific integrity. The first method we investigated was to add a function, implemented in PyTorch, which calculated the center of the means output by the Gaussian policy and subtracted that value from the modulated symbols.
\[ f\textsubscript{center}(\bm{s}) = \bm{s} - \frac{1}{2^b} \sum_{i=1}^{2^b} \mu_i, \]
where $\mu_i$ are the means of each possible constellation point and $\bm{s}$ is the set of complex symbols modulated by the current policy.

\begin{figure}[ht]
    \centering
    \includegraphics[width=0.7\linewidth]{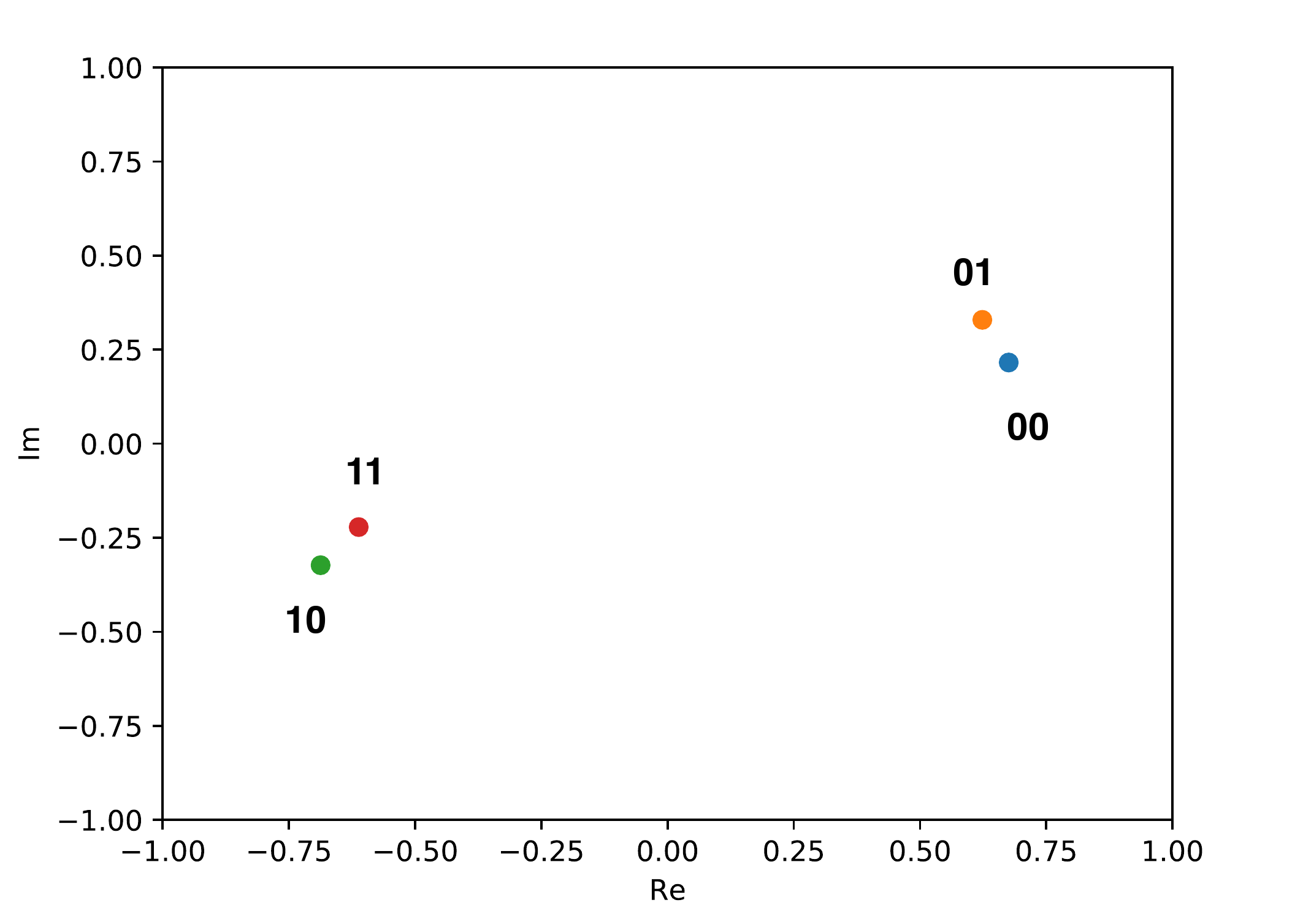}
    \caption{An example of a pseudo-BPSK constellation reached during one training run with the GNU Radio \EPP{} implementation. Two pairs of constellation points exist antipodally just like a BPSK constellation. There is not enough separation between constellation points within the pairs to reliably demodulate the correct bit sequences.}
    \label{fig:pseudo-bpsk}
\end{figure}

The rate of successfully trained trials improved while using this centering method but we discovered that QPSK constellations were often unable to split out from a pseudo-BPSK constellation, where two pairs of constellation points existed in nearly the same location. Fig.~\ref{fig:pseudo-bpsk} shows an example pseudo-BPSK constellation reached during one training run. We hypothesize that this hard centering required two constellation points to split out in tandem, which is difficult using noisy feedback.

As an alternative to the hard centering method, we applied `soft' centering by adding a term for the constellation center's distance from the origin to the loss function. With this soft centering we were able to achieve successful training rates similar to the baseline simulation results. We verified that setting the weight of the constellation center location loss term to infinity reproduced the behavior seen in the hard forcing method above, namely that modulators reach a pseudo-BPSK constellation but were unable to split into a true QPSK constellation. Similarly, reducing the weight of the loss term to zero produced results seen in the baseline method where DC offset correction caused the modulators to be unable to train.

\section{Classic Modulation Schemes}
\label{app:mod_order_mods_demods}

\begin{figure}[ht]
    \centering
    \subfloat[QPSK Classic Modulator
        \label{fig:qpsk_mod}]{
        \includegraphics[width=0.3\linewidth]{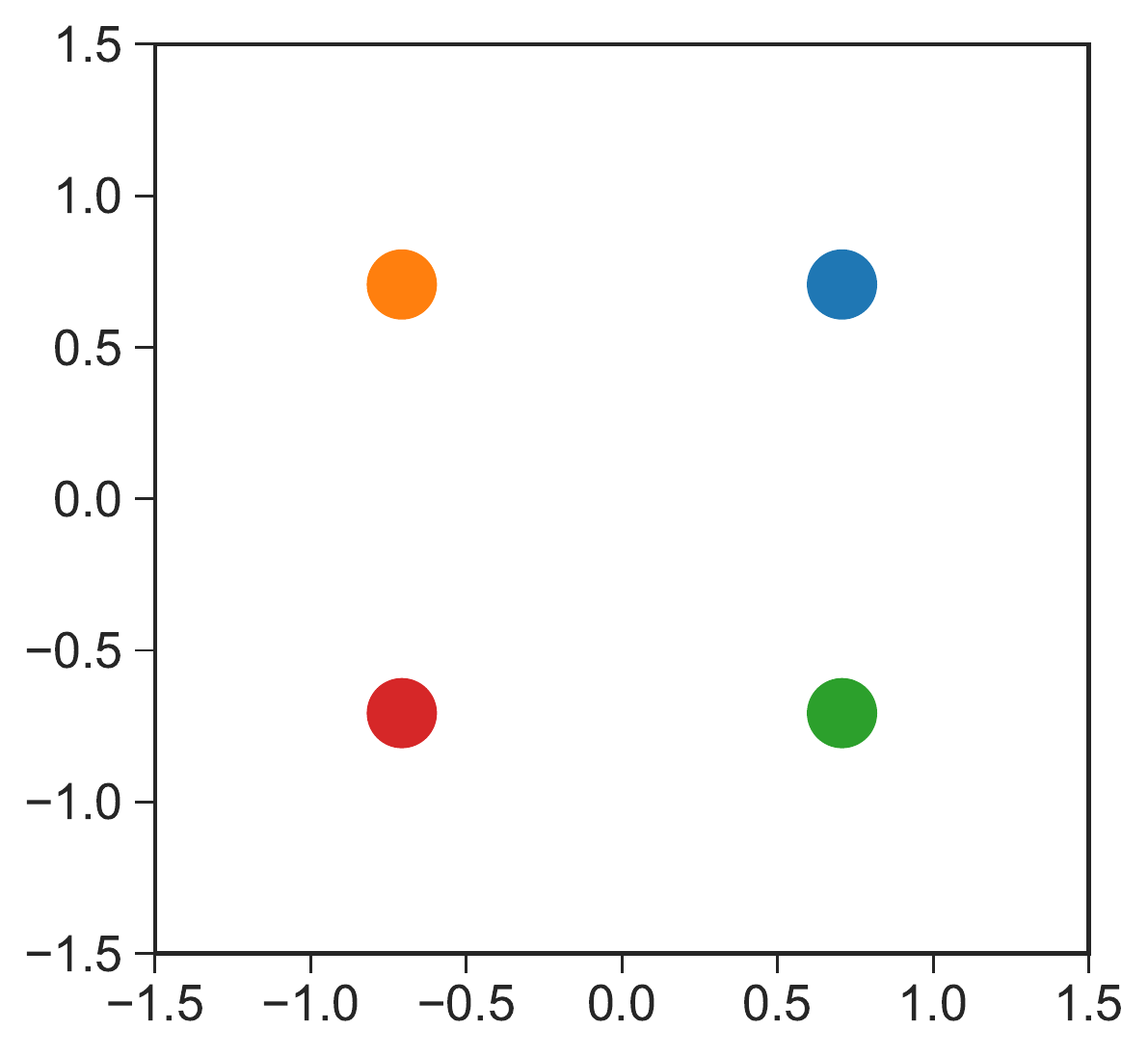}
    }
    ~
    \subfloat[8PSK Classic Modulator
        \label{fig:8psk_mod}]{
        \includegraphics[width=0.3\linewidth]{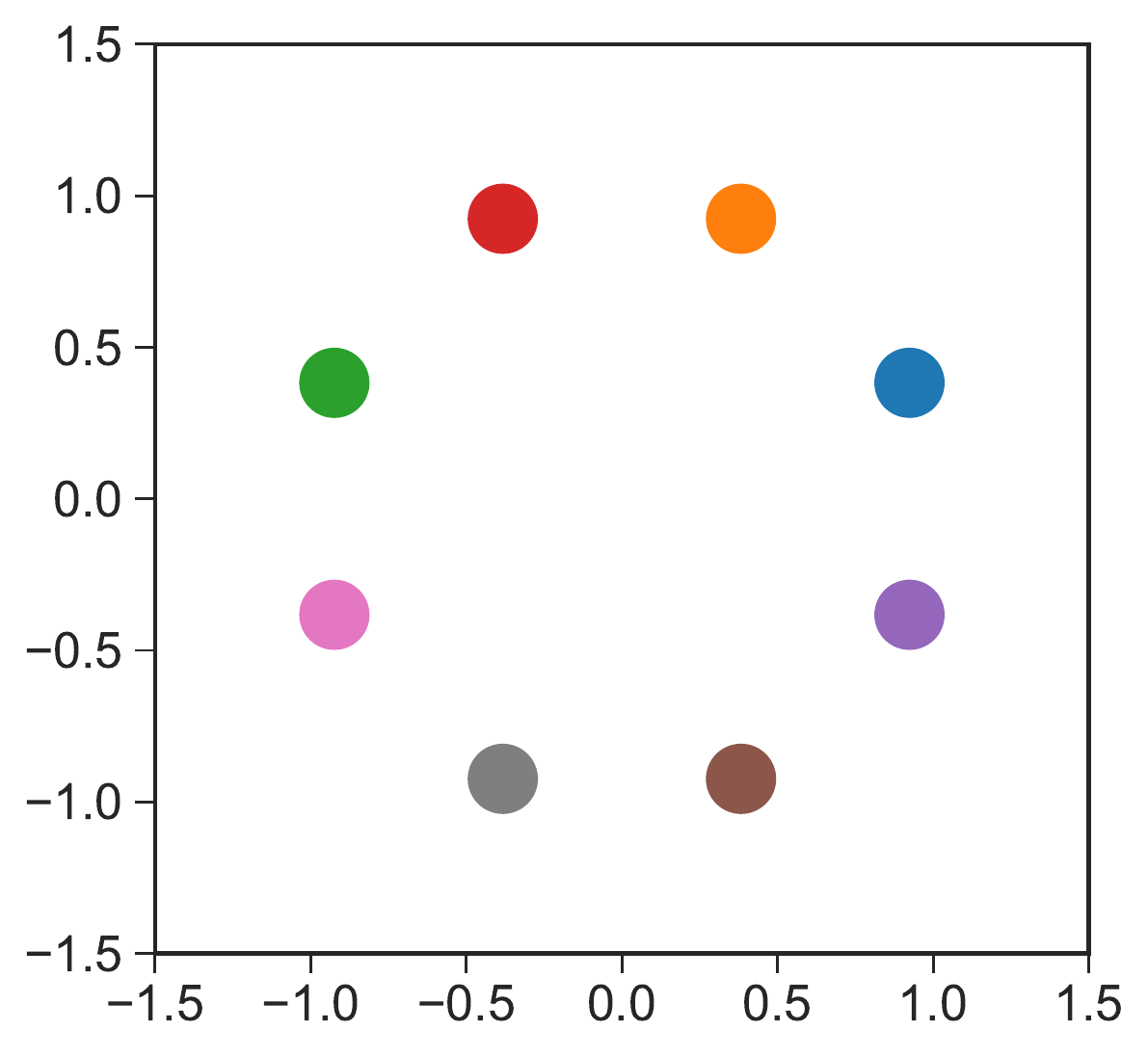}
    }
    ~
    \subfloat[16QAM Classic Modulator
        \label{fig:qam16_mod}]{
        \includegraphics[width=0.3\linewidth]{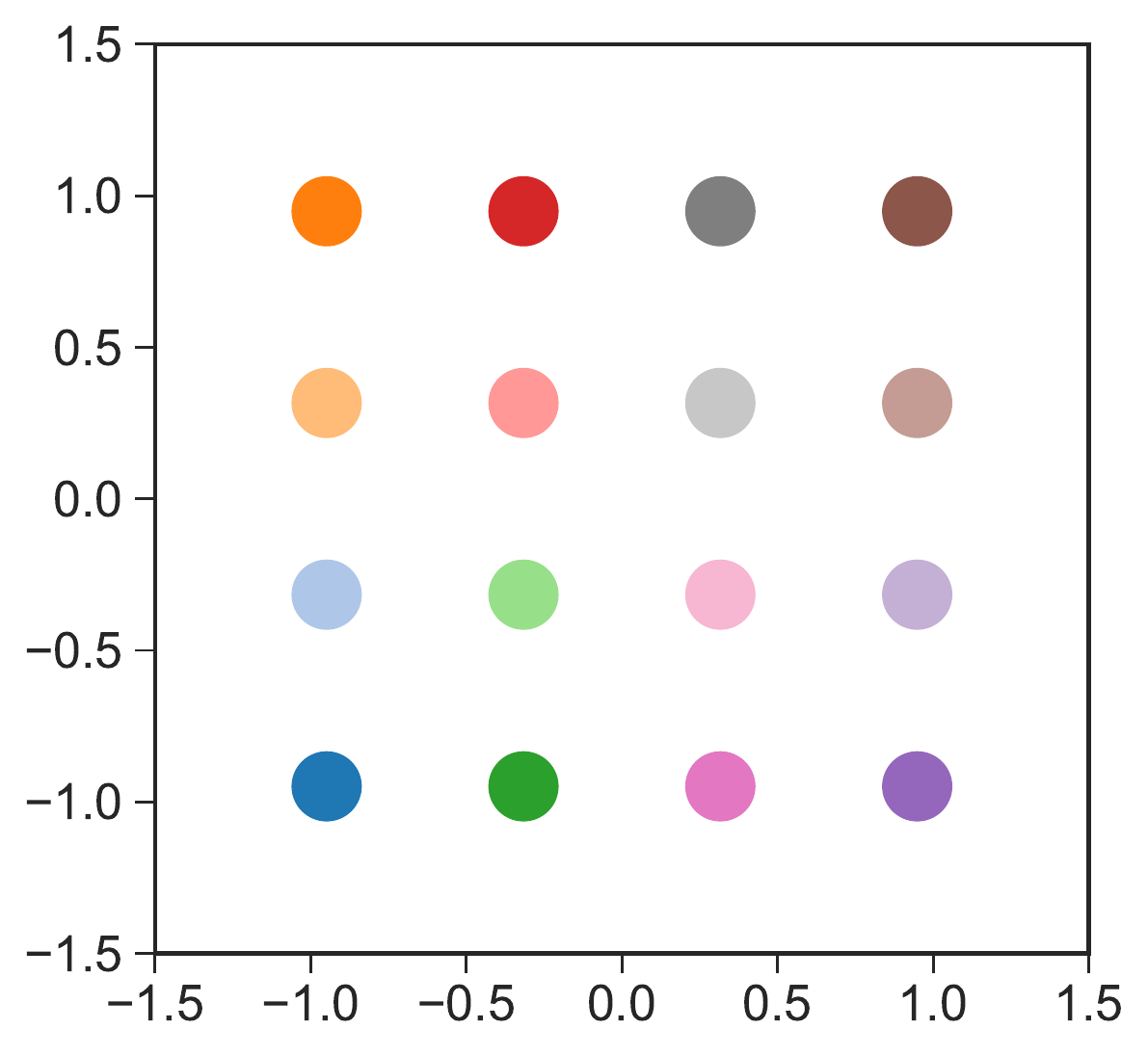}
    }
    
    \subfloat[QPSK Classic Demodulator
        \label{fig:qpsk_demod}]{
        \includegraphics[width=0.3\linewidth]{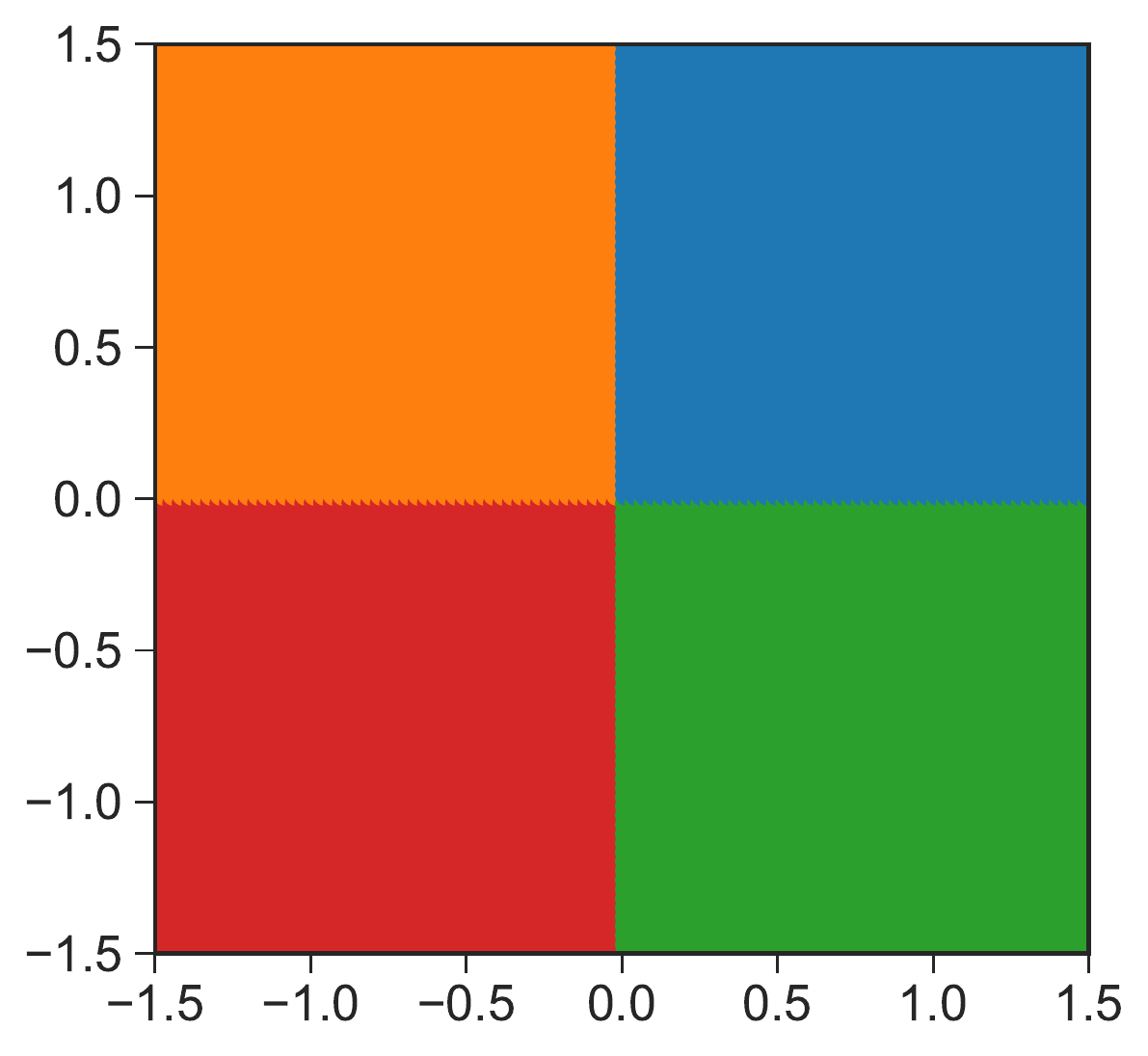}
    }
    ~
    \subfloat[8PSK Classic Demodulator
        \label{fig:8psk_demod}]{
        \includegraphics[width=0.3\linewidth]{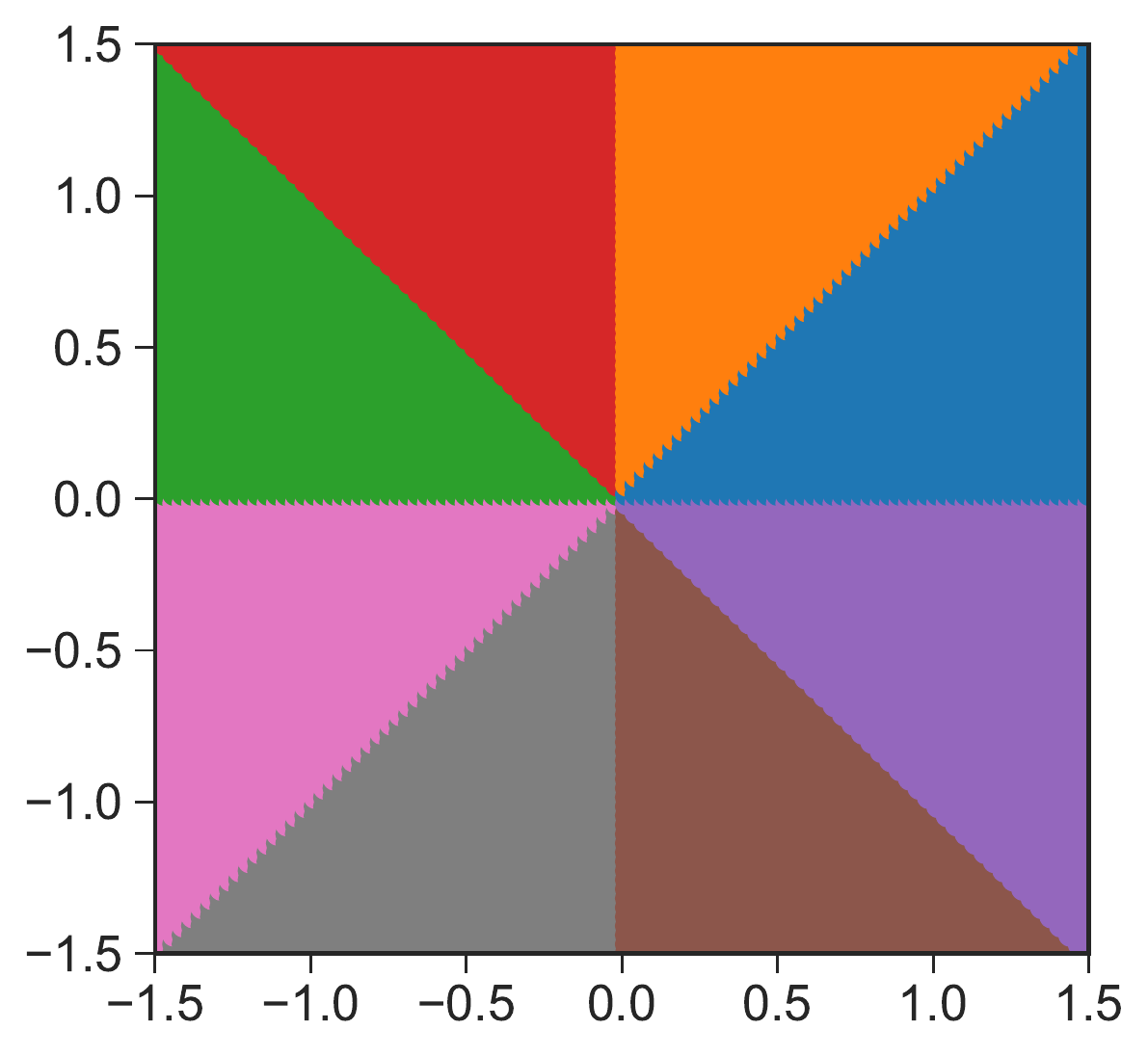}
    }
    ~
    \subfloat[16QAM Classic Demodulator
        \label{fig:qam16_demod}]{
        \includegraphics[width=0.3\linewidth]{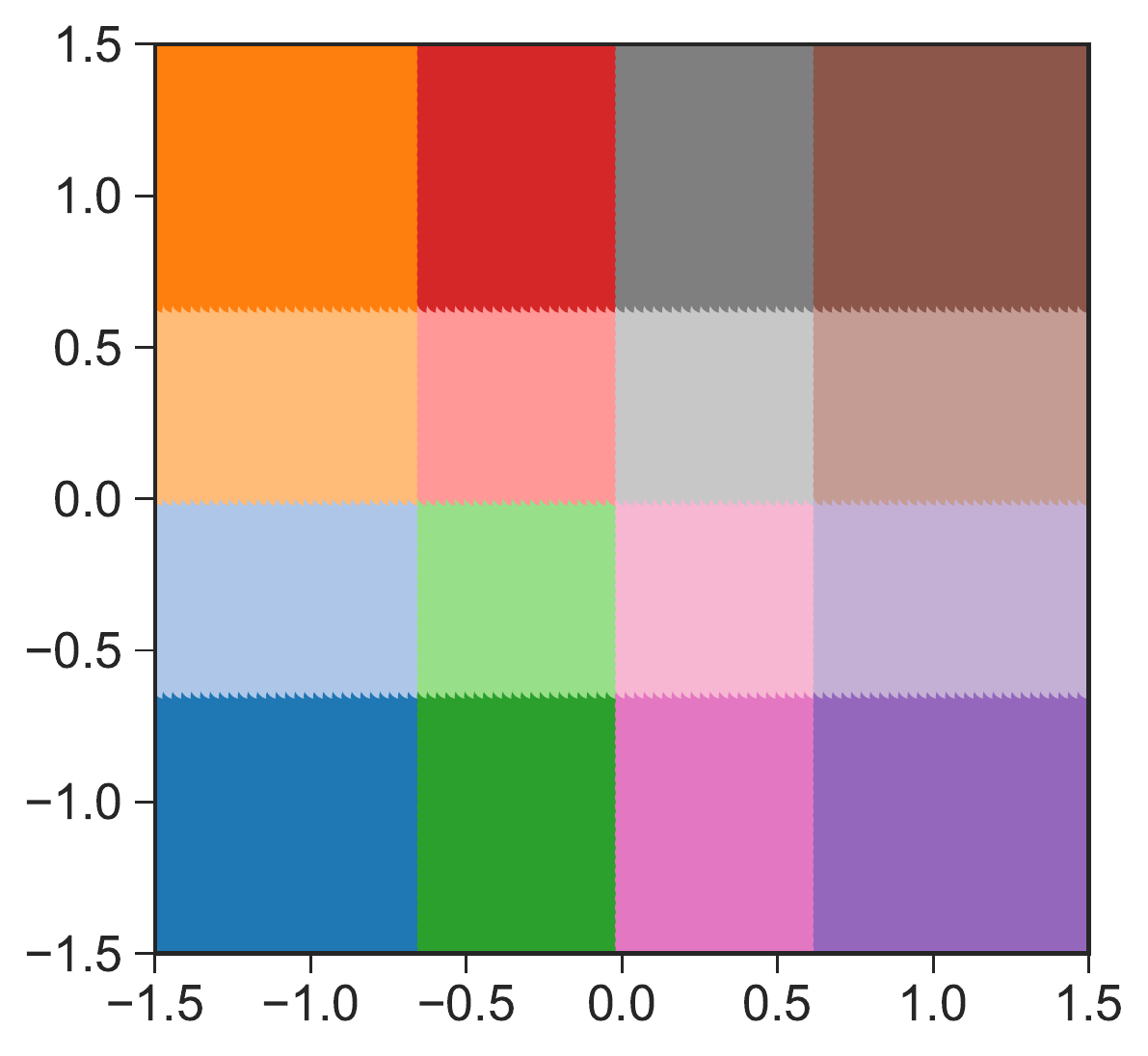}
    }
    
    \caption{Figures (a) through (c) show the fixed, optimal modulation used for Classic models; (d) through (f) show the corresponding demodulation boundaries.}
    \label{fig:mod_order_mods_demods}
\end{figure}

Figure~\ref{fig:mod_order_mods_demods} illustrates the fixed modulation schemes used by \Classic{} agents. These schemes are known to be optimal for AWGN channels \cite{qammod}.

\FloatBarrier
\section{Simulation Settings}
\label{app:simulation-settings}
Unless specified otherwise, the training SNR values default to the values in Table~\ref{tab:ber-snr-mod}. Testing SNR values default to those corresponding to 0.001\%, 0.01\%, 0.1\%, 1\%, and 10\% BER from Table~\ref{tab:ber-snr-mod}. Table \ref{tab:exp_settings} describes other simulation settings such as the number of iterations, preamble length, and testing frequency. 

\begin{table}[ht]
\begin{center}
\begin{tabularx}{\linewidth}{lrrr}
\toprule
& \makecell{preamble length\\ (symbols)} & \makecell{\# iterations} &  \makecell{testing intervals\\(spacing, \# tests)} \\
\midrule
\textbf{gradient passing} &  &  &   \\
\hspace{0.25in}\textbf{QPSK} & 256 & 100 & log, 30 \\
\textbf{loss passing} &  &  &   \\
\hspace{0.25in}\textbf{QPSK} & 256 & 600 & log, 30 \\
\textbf{echo, shared} &  &  &   \\
\hspace{0.25in}\textbf{QPSK} & 256 & 2500 & log, 30 \\
\hspace{0.25in}\textbf{8PSK} & 256 & 6000 & log, 30 \\
\hspace{0.25in}\textbf{16QAM} & 256 & 8000 & log, 30 \\ 
\textbf{echo, private} &  &  &   \\
\hspace{0.25in}\textbf{QPSK} & 256 & 3000* & log, 30 \\
\hspace{0.25in}\textbf{8PSK} & 256 & 10000 & log, 30 \\
\hspace{0.25in}\textbf{16QAM} & 256 & 20000 & log, 30 \\ 
\bottomrule
\end{tabularx}
\end{center}
\caption{Experiment settings for the different protocols and modulation orders. (* Because \NF{}-and-\NF{} converged so fast, we only trained for 500 iterations in order to adequately sample the convergence curve.)}
\label{tab:exp_settings}
\end{table}

\section{Simulation Training Hyperparameters}
\label{app:simulation-hyperparams}

The hyperparameter names and values in this appendix match the exact arguments used for running experiments with the code found at \url{https://github.com/ml4wireless/echo}. Table~\ref{tab:hyperparam-descriptions} describes the purpose of each hyperparameter.

\begin{table}[H]
\begin{tabularx}{\linewidth}{@{}ll@{}}
\toprule
\textbf{Hyperparameter} & \textbf{Description} \\ \midrule
hidden\_layers & List of hidden layer widths \\
bits\_per\_symbol & Number of bits per symbol \\
restrict\_energy & Energy constraint for modulator output  \\
activation\_fn\_hidden & Hidden layer activation function \\
optimizer & Optimizer used \\
lambda\_prob & Used for numerical stability in loss function \\
stepsize\_cross\_entropy & Learning rate for demodulator network weights \\
cross\_entropy\_weight & Weighting for cross-entropy loss terms \\
stepsize\_mu & Learning rate for modulator network weights \\
stepsize\_sigma & Learning rate for Gaussian policy stdev \\
initial\_std & Starting exploration stdev \\
max\_std & Maximum stdev for the Gaussian policy \\
min\_std & Minimum stdev for the Gaussian policy \\
max\_amplitude & Maximum average power of constellation \\
lambda\_center & Weight for constellation center offset loss \\
lambda\_l1 & Weight for L1 penalty on weights \\
degree\_polynomial & Degree of the polynomial features \\
\bottomrule
\end{tabularx}
\caption{Descriptions of hyperparameters used in experiments.}
\label{tab:hyperparam-descriptions}
\end{table}

\subsection{\GP{} Training Hyperparameters}
\label{app:passing-neural-hyperparams}

\begin{table}[H]
\begin{tabularx}{\linewidth}{lll}
\toprule
\textbf{Hyperparameter} & \textbf{\begin{tabular}[c]{@{}l@{}}Neural\\ Modulator\end{tabular}} & \textbf{\begin{tabular}[c]{@{}l@{}}Neural\\ Demodulator\end{tabular}} \\ \midrule
hidden\_layers & {[}50{]} & {[}50{]} \\
bits\_per\_symbol & 2 & 2  \\
restrict\_energy & 1 & -  \\
activation\_fn\_hidden & `tanh' & `tanh' \\
optimizer & `Adam' & `Adam' \\
stepsize\_mu & 3e-2 & - \\
max\_amplitude & 1.0 & - \\
stepsize\_cross\_entropy & - & 3e-2 \\
cross\_entropy\_weight & - & 1.0 \\ 
\bottomrule
\end{tabularx}
\caption{\Neural{} hyperparameters used for \GP{} QPSK Simulation experiments. There are no exploration-related parameters because \GP{} does not use a Gaussian policy.}
\label{tab:gradient-neural-qpsk-hyperparams}
\end{table}
\vfill

\begin{table}[H]
\begin{tabularx}{\linewidth}{lll}
\toprule
\textbf{Hyperparameter} & \textbf{\begin{tabular}[c]{@{}l@{}}Poly\\ Modulator\end{tabular}} & \textbf{\begin{tabular}[c]{@{}l@{}}Poly\\ Demodulator\end{tabular}} \\ \midrule
degree\_polynomial & *default & 1 \\
bits\_per\_symbol & 2 & 2  \\
restrict\_energy & 1 & -  \\
optimizer & `Adam' & `Adam' \\
stepsize\_mu & 1e-1 & - \\
max\_amplitude & 1.0 & - \\
stepsize\_cross\_entropy & - & 1e-1 \\
cross\_entropy\_weight & - & 1.0 \\ 
lambda\_l1 & - & 1e-3 \\
\bottomrule
\end{tabularx}
\caption{\Poly{} hyperparameters used for \GP{} QPSK Simulation experiments. There are no exploration-related parameters because \GP{} does not use a Gaussian policy. (* Due to the fact that inputs to modulators are bits, a unique max degree polynomial can be used.)}
\label{tab:gradient-poly-qpsk-hyperparams}
\end{table}

\subsection{\LP{}, \ESP{}, and \EPP{} Neural Training Hyperparameters}
\label{app:echo-neural-hyperparams}

\begin{table}[H]
\begin{tabularx}{\linewidth}{@{}lll@{}}
\toprule
\textbf{Hyperparameter} & \textbf{\begin{tabular}[c]{@{}l@{}}Neural\\ Modulator\end{tabular}} & \textbf{\begin{tabular}[c]{@{}l@{}}Neural\\ Demodulator\end{tabular}} \\ \midrule
hidden\_layers & {[}50{]} & {[}50{]} \\
bits\_per\_symbol & 2 & 2  \\
max\_std & 1 & - \\
min\_std & 1e-1 & - \\
initial\_std & 3e-1 & - \\
restrict\_energy & 1 & -  \\
activation\_fn\_hidden & `tanh' & `tanh' \\
optimizer & `Adam' & `Adam' \\
lambda\_prob & 1e-10 & - \\
stepsize\_mu & 8e-3 & - \\
stepsize\_sigma & 1e-4 & - \\
max\_amplitude & 1.0 & - \\
stepsize\_cross\_entropy & - & 5e-3 \\
cross\_entropy\_weight & - & 1.0 \\ 
\bottomrule
\end{tabularx}
\caption{\textbf{\NF{}} hyperparameters used for \LP{}, \ESP{}, and \EPP{} QPSK Simulation experiments.}
\label{tab:echo-neural-fast-qpsk-hyperparams}
\end{table}

\begin{table}[H]
\begin{tabularx}{\linewidth}{@{}lll@{}}
\toprule
\textbf{Hyperparameter} & \textbf{\begin{tabular}[c]{@{}l@{}}Neural\\ Modulator\end{tabular}} & \textbf{\begin{tabular}[c]{@{}l@{}}Neural\\ Demodulator\end{tabular}} \\ \midrule
hidden\_layers & {[}50{]} & {[}50{]} \\
bits\_per\_symbol & 2 & 2  \\
max\_std & 1 & - \\
min\_std & 1e-1 & - \\
initial\_std & 3e-1 & - \\
restrict\_energy & 1 & -  \\
activation\_fn\_hidden & `tanh' & `tanh' \\
optimizer & `Adam' & `Adam' \\
lambda\_prob & 1e-10 & - \\
stepsize\_mu & 6e-4 & - \\
stepsize\_sigma & 1e-4 & - \\
max\_amplitude & 1.0 & - \\
stepsize\_cross\_entropy & - & 1e-3 \\
cross\_entropy\_weight & - & 1.0 \\ 
\bottomrule
\end{tabularx}
\caption{\textbf{\NS{}} hyperparameters used for \ESP{} and \EPP{} QPSK Simulation experiments.}
\label{tab:echo-neural-slow-qpsk-hyperparams}
\end{table}

\begin{table}[H]
\begin{tabularx}{\linewidth}{@{}lll@{}}
\toprule
\textbf{Hyperparameter} & \textbf{\begin{tabular}[c]{@{}l@{}}Neural\\ Modulator\end{tabular}} & \textbf{\begin{tabular}[c]{@{}l@{}}Neural\\ Demodulator\end{tabular}} \\ \midrule
hidden\_layers & {[}100{]} & {[}100{]} \\
bits\_per\_symbol & 3 & 3  \\
max\_std & 1 & - \\
min\_std & 1e-2 & - \\
initial\_std & 2e-1 & - \\
restrict\_energy & 1 & -  \\
activation\_fn\_hidden & `tanh' & `tanh' \\
optimizer & `Adam' & `Adam' \\
lambda\_prob & 1e-10 & - \\
stepsize\_mu & 8e-3 & - \\
stepsize\_sigma & 4e-3 & - \\
max\_amplitude & 1.0 & - \\
stepsize\_cross\_entropy & - & 1e-2 \\
cross\_entropy\_weight & - & 1.0 \\ 
\bottomrule
\end{tabularx}
\caption{\Neural{} hyperparameters used for \ESP{} and \EPP{} 8PSK Simulation experiments}
\label{tab:echo-neural-8psk-hyperparams}
\end{table}

\begin{table}[H]
\begin{tabularx}{\linewidth}{lll}
\toprule
\textbf{Hyperparameter} & \textbf{\begin{tabular}[c]{@{}l@{}}Neural\\ Modulator\end{tabular}} & \textbf{\begin{tabular}[c]{@{}l@{}}Neural\\ Demodulator\end{tabular}} \\ \midrule
hidden\_layers & {[}200{]} & {[}200{]} \\
bits\_per\_symbol & 4 & 4  \\
max\_std & 1 & - \\
min\_std & 1e-2 & - \\
initial\_std & 1e-1 & - \\
restrict\_energy & 1 & -  \\
activation\_fn\_hidden & `tanh' & `tanh' \\
optimizer & `Adam' & `Adam' \\
lambda\_prob & 1e-10 & - \\
stepsize\_mu & 7e-4 & - \\
stepsize\_sigma & 5e-4 & - \\
max\_amplitude & 1.0 & - \\
stepsize\_cross\_entropy & - & 1e-3 \\
cross\_entropy\_weight & - & 1.0 \\ 
\bottomrule
\end{tabularx}
\caption{\Neural{} hyperparameters used for \ESP{} and \EPP{} 16QAM Simulation experiments}
\label{tab:echo-neural-qam16-hyperparams}
\end{table}

\begin{table}[H]
\begin{tabularx}{\linewidth}{@{}lll@{}}
\toprule
\textbf{Hyperparameter} & \textbf{\begin{tabular}[c]{@{}l@{}}Neural\\ Modulator\end{tabular}} & \textbf{\begin{tabular}[c]{@{}l@{}}Neural\\ Demodulator\end{tabular}} \\ \midrule
hidden\_layers & {[}50{]} & {[}50{]} \\
bits\_per\_symbol & 2 & 2  \\
max\_std & 1 & - \\
min\_std & 1e-1 & - \\
initial\_std & 3e-1 & - \\
restrict\_energy & 1 & -  \\
activation\_fn\_hidden & `tanh' & `tanh' \\
optimizer & `Adam' & `Adam' \\
lambda\_prob & 1e-10 & - \\
stepsize\_mu & 1e-3 & - \\
stepsize\_sigma & 1e-4 & - \\
max\_amplitude & 1.0 & - \\
stepsize\_cross\_entropy & - & 1e-3 \\
cross\_entropy\_weight & - & 1.0 \\ 
\bottomrule
\end{tabularx}
\caption{\Neural{} hyperparameters used for \EPP{} QPSK Training SNR Simulation experiments.}
\label{tab:echo-neural-snr-qpsk-hyperparams}
\end{table}

\subsection{\LP{}, \ESP{}, and \EPP{} Polynomial Training Hyperparameters}
\label{app:echo-poly-hyperparams}

\begin{table}[H]
\begin{tabularx}{\linewidth}{@{}lll@{}}
\toprule
\textbf{Hyperparameter} & \textbf{\begin{tabular}[c]{@{}l@{}}Poly\\ Modulator\end{tabular}} & \textbf{\begin{tabular}[c]{@{}l@{}}Poly\\ Demodulator\end{tabular}} \\ \midrule
degree\_polynomial & *default & 1 \\
bits\_per\_symbol & 2 & 2  \\
max\_std & 2 & - \\
min\_std & 2e-1 & - \\
initial\_std & 1.0 & - \\
restrict\_energy & 1 & - \\
optimizer & `Adam' & `Adam' \\
stepsize\_mu & 4e-2 & - \\
stepsize\_sigma & 4e-3 & - \\
max\_amplitude & 1.0 & - \\
stepsize\_cross\_entropy & - & 1e-2 \\
lambda\_l1 & - & 1e-3 \\
\bottomrule
\end{tabularx}
\caption{\textbf{\PF{}} hyperparameters used for \LP{}, \ESP{}, and \EPP{} QPSK Simulation experiments. (* Due to the fact that inputs to modulators are bits, a unique max degree polynomial can be used.)}
\label{tab:echo-poly-fast-qpsk-hyperparams}
\end{table}

\begin{table}[H]
\begin{tabularx}{\linewidth}{@{}lll@{}}
\toprule
\textbf{Hyperparameter} &
\textbf{\begin{tabular}[c]{@{}l@{}}Polynomial\\ Modulator\end{tabular}} & \textbf{\begin{tabular}[c]{@{}l@{}}Polynomial\\ Demodulator\end{tabular}} \\ \midrule
degree\_polynomial & *default & 1 \\
bits\_per\_symbol & 2 & 2  \\
max\_std & 2 & - \\
min\_std & 2e-1 & - \\
initial\_std & 1.0 & - \\
restrict\_energy & 1 & - \\
optimizer & `Adam' & `Adam' \\
stepsize\_mu & 3e-2 & - \\
stepsize\_sigma & 3e-3 & - \\
max\_amplitude & 1.0 & - \\
stepsize\_cross\_entropy & - & 1e-2 \\
lambda\_l1 & - & 1e-3 \\
\bottomrule
\end{tabularx}
\caption{\textbf{\PS{}} hyperparameters used for \EPP{} QPSK Alien Simulation experiments. (* Due to the fact that inputs to modulators are bits, a unique max degree polynomial can be used.)}
\label{tab:echo-poly-slow-qpsk-hyperparams}
\end{table}

\section{GNU Radio Training Hyperparameters}
\label{app:gnr-hyperparams}
The hyperparameter names and values in Table~\ref{tab:gnr-hyperparams} match the exact arguments used for running experiments with the code found at \url{https://github.com/ml4wireless/gr-echo}. Because of the extra lambda\_center term in the loss function, the hyperparameters were tuned on the radio rather than reusing the hyperparameters from the rest of the simulations. These hyperparameters were used for both trials on the radio and simulations used for comparison with radio results.

\begin{table}[H]
\begin{tabularx}{\linewidth}{@{}lll@{}}
\toprule
\textbf{Hyperparameter} & \textbf{\begin{tabular}[c]{@{}l@{}}Neural\\ Modulator\end{tabular}} & \textbf{\begin{tabular}[c]{@{}l@{}}Neural\\ Demodulator\end{tabular}} \\ \midrule
hidden\_layers & {[}50{]} & {[}50{]} \\
bits\_per\_symbol & 2 & 2  \\
max\_std & 100 & - \\
min\_std & 1e-1 & - \\
initial\_std & 2e-1 & - \\
restrict\_energy & 1 & -  \\
activation\_fn\_hidden & `tanh' & `tanh' \\
optimizer & `Adam' & `Adam' \\
lambda\_prob & 1e-10 & - \\
stepsize\_mu & 1e-3 & - \\
stepsize\_sigma & 1e-4 & - \\
max\_amplitude & 0.5 & - \\
lambda\_center & 125 & - \\
stepsize\_cross\_entropy & - & 1e-2 \\
\end{tabularx}
\caption{\Neural{} hyperparameters used for GNU Radio experiments}
\label{tab:gnr-hyperparams}
\end{table}

\end{appendices}



\section*{Acknowledgments}
This work is a continuation of work reported in \cite{wcm:de2018cooperative}, begun with students Colin de Vrieze, Shane Barratt, and Daniel Tsai, and which grew out of a special topics research course offered in Spring 2017 at UC Berkeley. We thank all the students in that course for the productive discussions in this area. We also thank Laura Brink and Sameer Reddy for their contributions to discussions that led to this work. Finally, we thank the reviewers for their helpful comments.

\nocite{tange_ole_2018_1146014}
\bibliographystyle{IEEEtran}
\bibliography{IEEEabrv,echo}

\begin{IEEEbiography}[{\includegraphics[width=1in,height=1.25in,clip,keepaspectratio]{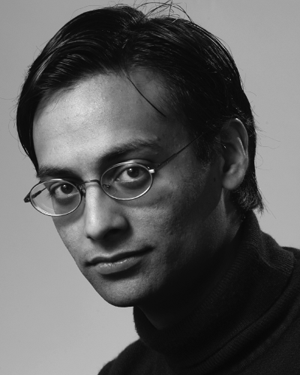}}]{Anant Sahai} Anant Sahai received the B.S. degree from the University of California (UC) at Berkeley, Berkeley, CA, USA, in 1994, and the S.M. and Ph.D. degrees from the Massachusetts Institute of Technology, Cambridge, MA, USA, in 1996 and 2001, respectively. Prior to joining the UC at Berkeley as a Faculty Member in 2002, he spent a year as a Research Scientist at the wireless startup Enuvis, South San Francisco, where he was involved in developing software-radio signal-processing algorithms to enable very sensitive GPS receivers for indoor operation. His current research interests are the intersection of information theory, wireless communication, and decentralized control, including dynamic spectrum sharing and the role of machine learning in it. From 2007 to 2009, he was the Treasurer for the IEEE Information Theory Society.
\end{IEEEbiography}

\begin{IEEEbiography}[{\includegraphics[width=1in,height=1.25in,clip,keepaspectratio]{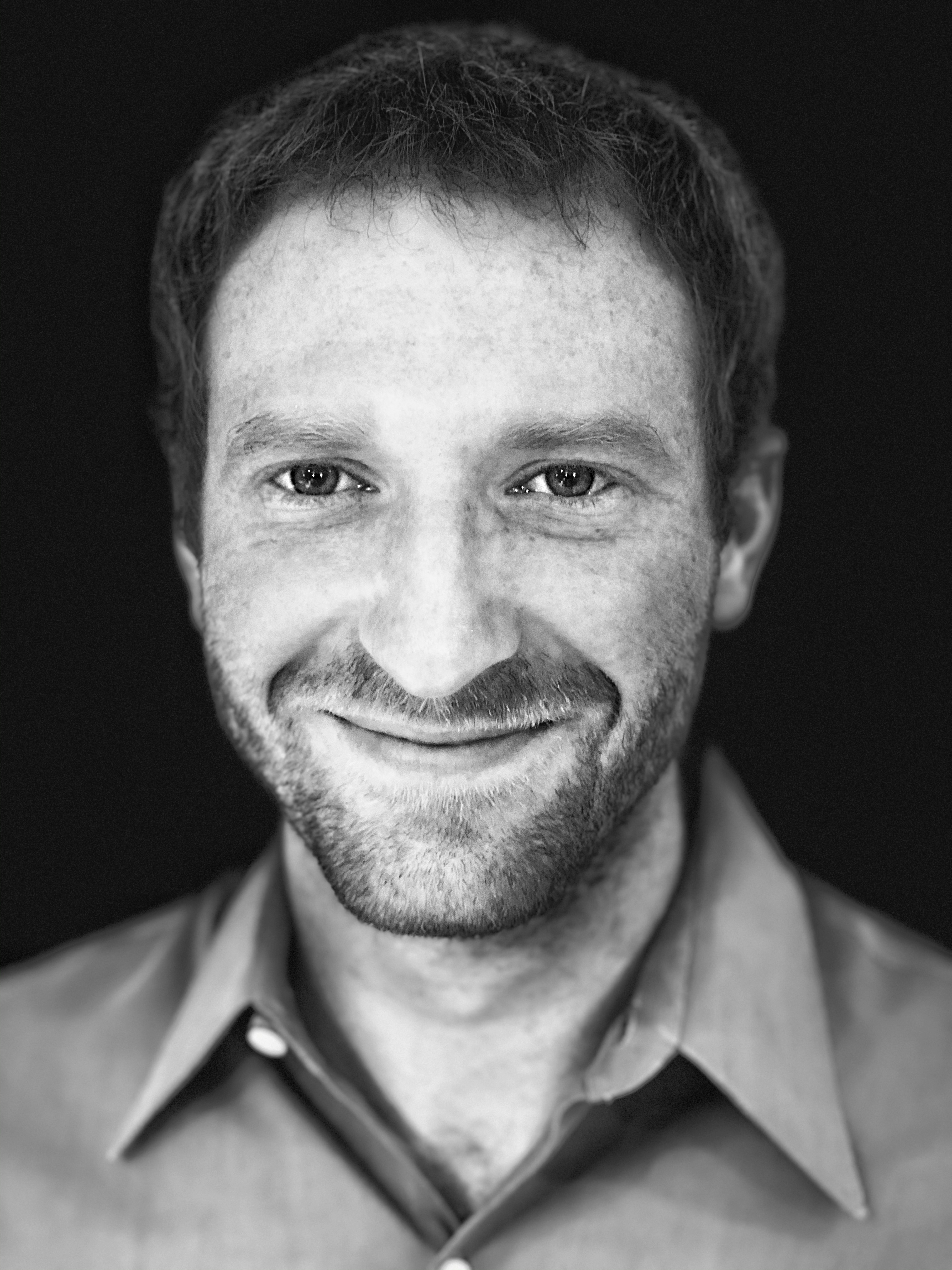}}]{Joshua Sanz} Joshua Sanz received the B.S. degree in engineering from Harvey Mudd College, Claremont, CA, USA, in 2015. He is currently pursing the Ph.D. degree in electrical engineering and computer science at the University of California at Berkeley, Berkeley, CA, USA. 
He worked at MIT Lincoln Laboratory as Associate Technical Staff from 2015 to 2018. His current research interests are in the applications of machine learning to wireless communications.
\end{IEEEbiography}

\begin{IEEEbiography}[{\includegraphics[width=1in,height=1.25in,clip,keepaspectratio]{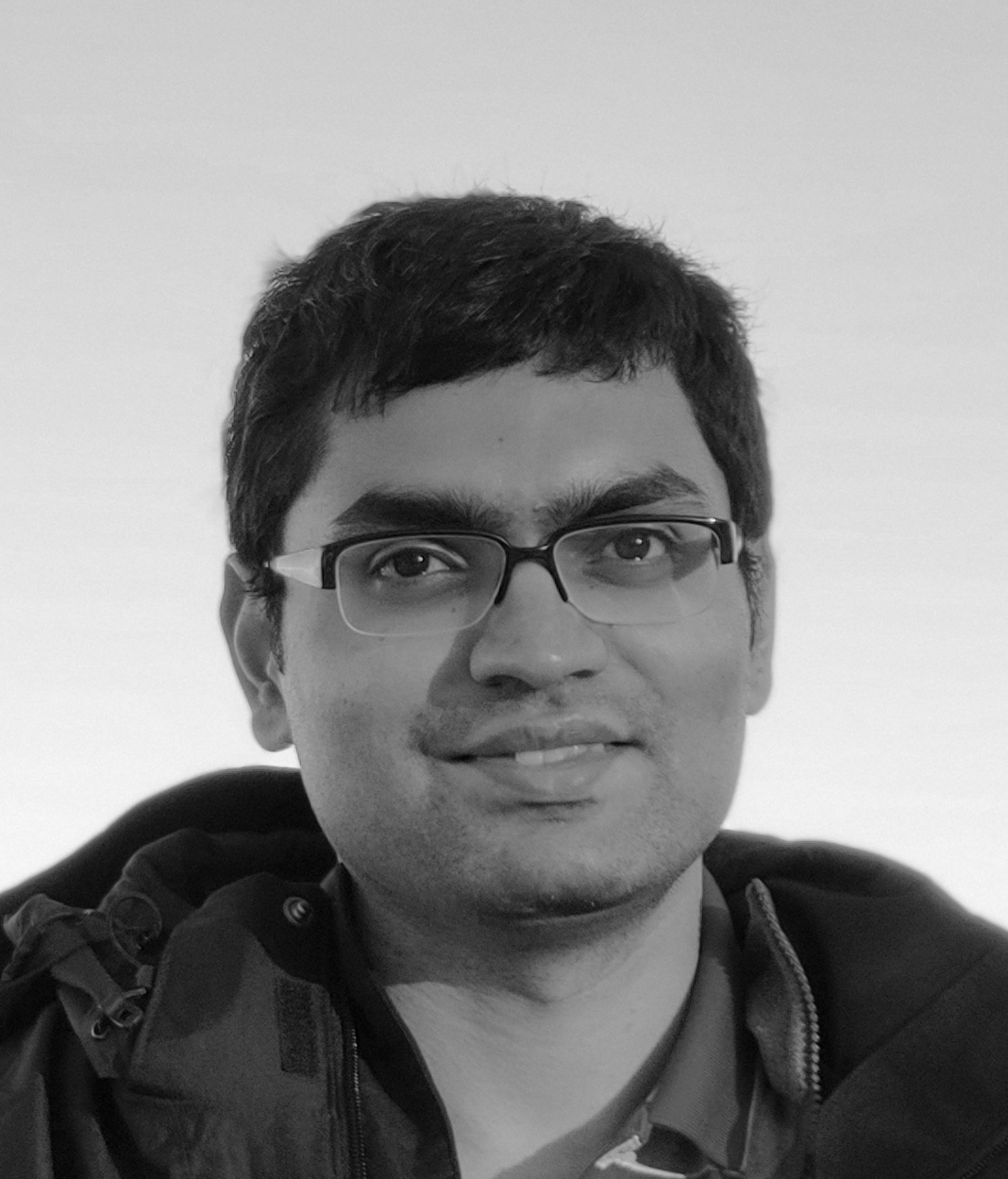}}]{Vignesh Subramanian} Vignesh Subramanian received his Bachelor's (BTech) and Masters (MTech) degrees in Electrical Engineering from Indian Institute of Technology, Bombay, India. He is currently pursuing a Ph.D. in Electrical Engineering and Computer Science from the University of California at Berkeley, Berkeley, CA, USA.
His research interests include application of machine learning techniques to wireless communication and control and theoretical analysis of machine learning techniques in overparameterized regimes. 
His awards include the Institute Gold Medal and Institute Silver Medal from IIT Bombay in 2015. 
\end{IEEEbiography}

\begin{IEEEbiography}[{\includegraphics[width=1in,height=1.25in,clip,keepaspectratio]{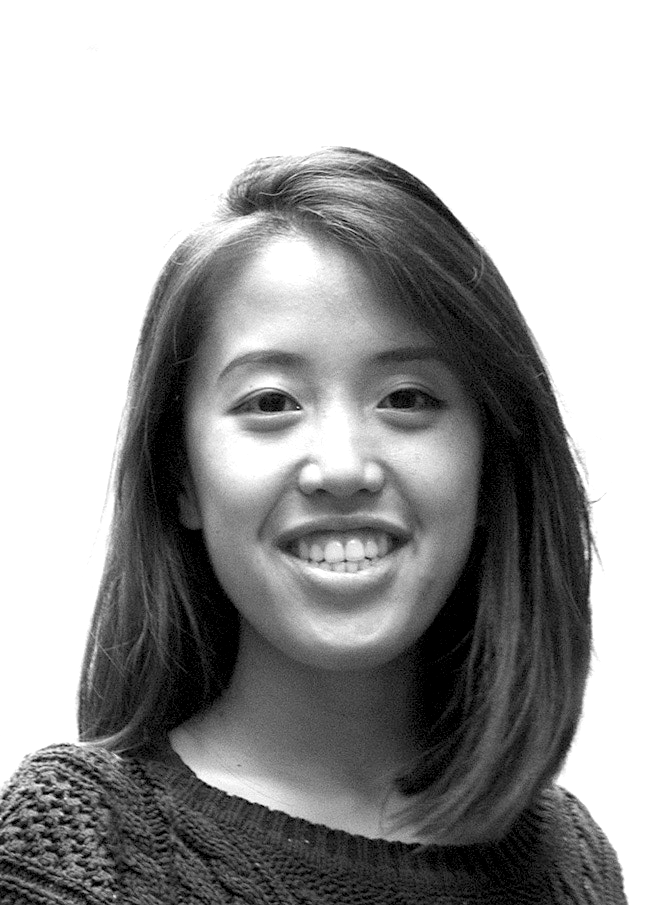}}]{Caryn Tran} Caryn T. N. Tran was born in California, USA in 1995. She received a B.A. in Computer Science in 2017 and a M.S. in Electrical Engineering and Computer Science in 2019 from the University of California (UC) at Berkeley, Berkeley, CA, USA. Her academic interests are in computer science education.
\end{IEEEbiography}

\begin{IEEEbiography}[{\includegraphics[width=1in,height=1.25in,clip,keepaspectratio]{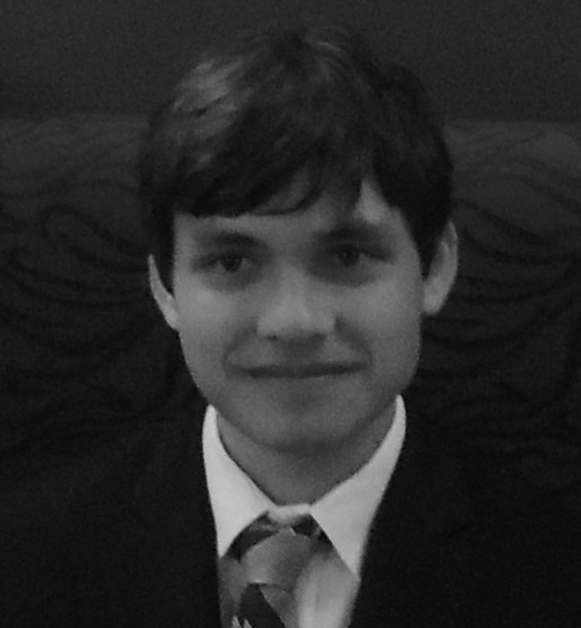}}]{Kailas Vodrahalli} Kailas Vodrahalli received the B.S. degree in electrical engineering and computer sciences from the University of California, Berkeley, in 2019. He is currently completing the Ph.D. degree in electrical engineering from Stanford University, Stanford, CA where he is supported by a Stanford Graduate Fellowship.
He was a summer intern at Intel Corporation in 2017 and at Starkey Hearing Technologies in 2018. He is interested in theoretical and practical applications of machine learning, with a current interest in the medical field. 
\end{IEEEbiography}

\EOD
\end{document}